\documentclass[useAMS,usenatbib,usegraphicx]{mn2e}
\usepackage{times}
\usepackage{morefloats}
\usepackage[total={17.8cm,24.0cm},centering]{geometry}

\newcommand{\ha}{H$\alpha$}
\newcommand{\hb}{H$\beta$}
\newcommand{\hg}{H$\gamma$}
\newcommand{\mgb}{Mg$\,b$}

\newcommand{\zh}{[Z/H]}
\newcommand{\afe}{[$\alpha$/Fe]}

\newcommand{\oiii}{[O{\small III}]}
\newcommand{\nii}{[N{\small I}]}
\newcommand{\niii}{[N{\small II}]}
\newcommand{\kms}{$\mbox{km s}^{-1}$}
\newcommand{\re}{$R_e$}
\newcommand{\lsim}{\mathrel{\hbox{\rlap{\hbox{\lower4pt\hbox{$\sim$}}}\hbox{$<$
}}}}
\newcommand{\gsim}{\mathrel{\hbox{\rlap{\hbox{\lower4pt\hbox{$\sim$}}}\hbox{$>$
}}}}
\newcommand{\sauron}{{\texttt {SAURON}}}
\newcommand{\oasis}{{\texttt {OASIS}}}

\newcommand{\XOasis}{{\small {XOASIS}}}
\newcommand{\farcsec}{\hbox{$.\!\!^{\prime\prime}$}}  
\newcommand{\degree}{$^{\circ}\!$}  
\newcommand{\msun}{$M_{\odot}$}
\newcommand{\kinpa}{PA$_{\mathrm{kin}}$}
\newcommand{\ki}{k$_1$}

\title[The SAURON project VIII: OASIS/CFHT spectroscopy of \sauron\/
       E/S0 galaxy centres]
      {The SAURON project - VIII. OASIS/CFHT\,\,integral-field spectroscopy of elliptical and lenticular galaxy centres\thanks{Based on observations obtained at the Canada-France-Hawaii Telescope which is operated by the National Research Council of Canada, the Institut National des Sciences de l'Univers of the Centre National de la Recherche Scientifique of France, and the University of Hawaii.}\looseness-2}
\author[R.M.\ McDermid, et al.]
       {Richard M.\ McDermid$^{1}$\thanks{Email: mcdermid@strw.leidenuniv.nl}, Eric Emsellem$^{2}$, Kristen L.\ Shapiro$^{3}$, Roland Bacon$^{2}$, \newauthor Martin Bureau$^{4}$, Michele Cappellari$^{1}$, Roger L.\ Davies$^{4}$, Tim\ de Zeeuw$^{1}$, \newauthor Jes{\'u}s Falc{\'o}n-Barroso$^{1,5}$, Davor Krajnovi{\'c}$^{4}$, Harald Kuntschner$^{6}$, Reynier F.\ Peletier$^{7}$,
       \newauthor and Marc Sarzi$^{8}$\\
~\\
  $^1$ Leiden Observatory, Postbus 9513, 2300 RA Leiden, The Netherlands\\
  $^2$ Université de Lyon 1, CRAL, Observatoire de Lyon, 9 av. Charles André, F-69230 Saint-Genis Laval; CNRS, UMR 5574 ; ENS de Lyon, France\\
  $^3$ UC Berkeley Department of Astronomy, Berkeley, CA 94720, USA\\
  $^4$ Denys Wilkinson Building, University of Oxford, Keble Road, Oxford, United Kingdom\\
  $^5$ European Space and Technology Centre (ESTEC), Keplerlaan 1, Postbus 299, 2200 AG Noordwijk, The Netherlands\\
  $^6$ Space Telescope European Coordinating Facility, European Southern Observatory, Karl-Schwarzschild-Str 2, 85748 Garching, Germany\\
  $^7$ Kapteyn Astronomical Institute, Postbus 800, 9700 AV Groningen, The Netherlands\\
  $^8$ Centre for Astrophysics Research, Science \& Technology Research Institute, University of Hertfordshire, Hatfield, United Kingdom\\
}

\date{Accepted by MNRAS, 13th September, 2006}

\pagerange{\pageref{firstpage}--\pageref{lastpage}}
\pubyear{}

\begin{document}

\maketitle

\label{firstpage}

\begin{abstract}
We present high spatial resolution integral-field spectroscopy of 28 elliptical (E) and lenticular (S0) galaxies from the \sauron\/ representative survey obtained with the \oasis\/ spectrograph during its operation at the CFHT. These seeing-limited observations explore the central 8\arcsec$\times$10\arcsec\/ (typically one kiloparsec diameter) regions of these galaxies using a spatial sampling four times higher than \sauron\/ (0\farcsec27 vs. 0\farcsec94 spatial elements), resulting in almost a factor of two improvement in the median PSF. These data allow accurate study of the central regions to complement the large-scale view provided by \sauron. Here we present the stellar and gas kinematics, stellar absorption-line strengths and nebular emission-line strengths for this sample. We also characterise the stellar velocity maps using the `kinemetry' technique, and derive maps of the luminosity-weighted stellar age, metallicity and abundance ratio via stellar population models. We give a brief review of the structures found in our maps, linking also to larger-scale structures measured with \sauron. We present two previously unreported kinematically-decoupled components (KDCs) in the centres of NGC\,3032 and NGC\,4382. We compare the intrinsic size and luminosity-weighted stellar age of all the visible KDCs in the full \sauron\/ sample, and find two types of components: kiloparsec-scale KDCs, which are older than 8~Gyr, and are found in galaxies with little net rotation; and compact KDCs, which have intrinsic diameters of less than a few hundred parsec, show a range of stellar ages from 0.5 - 15~Gyr (with 5/6 younger than 5~Gyr), are found exclusively in fast-rotating galaxies, and are close to counter-rotating around the same axis as their host. Of the 7 galaxies in the \sauron\/ sample with integrated luminosity-weighted ages less than 5~Gyr, 5 show such compact KDCs, suggesting a link between counter-rotation and recent star-formation. We show that this may be due to a combination of small sample size at young ages, and an observational bias, since young KDCs are easier to detect than their older and/or co-rotating counterparts.\looseness-2

\end{abstract}



\section{Introduction}

In the era of the {\it Hubble Space Telescope} ({\it HST}), early-type galaxy nuclei (central 10-50~pc) have become a key component in the modern picture of galaxy evolution. Empirical relationships between the central luminosity profile and numerous large-scale galaxy properties, such as isophotal shape, rotational support, and stellar populations, strongly suggest that the formation and evolution of the central regions is inextricably linked to that of the galaxy as a whole \citep[e.g.][]{faber97}. The ubiquitous presence of super-massive black holes (SMBHs) at the centres of galaxies \citep[e.g.][]{ferrarese00,gebhardt00} is thought to be an important aspect of this relationship, controlling global star-formation and dictating the central density distribution \citep[e.g.][]{silk98,granato04,binney04,dimatteo05}.

Most observational studies of early-type galaxy nuclei have so far been based only on {\it HST} imaging, with a small number of objects also observed spectroscopically from space, either with a single aperture or a long slit. The limitations of these data become apparent when considering the complexity of structures found in the centres of these otherwise `boring' objects. Spectroscopy is required to establish dynamical black hole mass estimates, and it has been shown that a single long slit is grossly insufficient to constrain fully general dynamical models \citep{verolme02,cappellari05}. To understand the stellar orbital structure of galaxy nuclei therefore requires information across a sizeable two-dimensional field, obtained with the highest possible spatial resolution. Additionally, {\it HST} imaging reveals highly complex gas and dust distributions \citep[e.g.][]{vdokkum95, tomita00,lauer05}, which are sensitive indicators of dynamical torques that are not detectable from the stellar distribution alone. Understanding the kinematic and chemical properties of such irregular gas distributions requires two-dimensional spectroscopy. {\it HST} imaging studies also find broad-band colour gradients, indicative of distinct star-formation histories towards the very central regions \citep[e.g.][]{carollo94b,carollo97a,lauer05,cote06}. However, to accurately quantify the populations requires spectroscopy on comparable spatial scales, of which there are rather few studies of early-type galaxy samples, due to the high signal-to-noise ratio required \citep[e.g.][]{krajnovic04,sarzi05}. Moreover, these and comparable ground-based studies looking at central properties \citep[e.g.][]{carollo94b,trager2000,thomas05} are restricted to either aperture measurements or long-slit profiles, making the association between population changes and other structural variations difficult to determine.

In response to these issues, we have initiated a program of high spatial resolution integral-field spectroscopy of early-type galaxy nuclei. We select our sample from galaxies observed as part of the \sauron\/ survey: a study of 72 representative nearby early-type galaxies and spiral bulges observed with \sauron, our custom-built panoramic integral-field spectrograph mounted at the William Herschel Telescope (WHT) on La Palma \citep[Paper I]{bacon01}. The aims of this survey and the observed galaxy sample are described in \citet[Paper II]{dezeeuw02}. For the sub-sample of 48 E/S0s we present the stellar kinematic maps in \citet[Paper III]{emsellem04}; the two-dimensional ionised gas properties in \citet[Paper V]{sarzi06}; and maps of the stellar absorption line strengths in \citet[Paper VI]{kuntschner06}. We refer the reader to these papers (and references therein) for discussion of the global properties of our sample galaxies.

To maximise the field of view, the spatial sampling of \sauron\/ was set to 0\farcsec94$\times$0\farcsec94 (per lenslet) for the survey, therefore often undersampling the typical seeing at La Palma (0\farcsec8). This does not affect the results for the main body of early-type galaxies, as they generally exhibit smoothly-varying structure on scales larger than the seeing. However, the central regions are poorly resolved in the \sauron\/ data, preventing accurate measurements of the central properties.

For this reason we have observed the centres of a subset of the \sauron\/ sample using the \oasis\/ integral-field spectrograph, mounted on the 3.6-m Canada-France-Hawaii Telescope (CFHT), Hawaii. This instrument allows optimal sampling of the natural seeing on Mauna Kea, resulting in a significant improvement in spatial resolution. To ensure a high degree of complementarity between the \oasis\/ data presented here and our \sauron\/ observations, we configured \oasis\/ to have an almost identical wavelength domain and spectral resolution. The \oasis\/ data can therefore be easily combined with \sauron\/ data, e.g. for constraining dynamical models \citep[e.g.][]{shapiro06}. All data presented in this paper will be made available through the \sauron\/ web page.{\footnote{http://www.strw.leidenuniv.nl/sauron}\looseness=-2

The \oasis\/ observations themselves reveal a number of previously unresolved features in the centres of our sample galaxies. In particular, we find several galaxies which exhibit very young central stellar populations that can be directly associated with peculiar (misaligned) components in the stellar rotation fields. We explore this further by considering all objects in the \sauron\/ sample with clear kinematically decoupled components, and find a link between the size of these components and their stellar populations.

The layout of the paper is as follows: Section~\ref{sec:obs} describes the instrument, the observations and the basic data reduction; Section~\ref{sec:analysis} describes the application of our analysis tools used to derive the stellar kinematics, gas properties and line-strength measurements (including the application of stellar population models), and presents detailed comparisons with equivalent quantities from \sauron; Section~\ref{sec:results} presents the two-dimensional maps of all measured quantities, and discusses their main features; Section~\ref{sec:discussion} presents a discussion of young stellar populations in our sample, and the connection with KDCs; and Section \ref{sec:conclusions} concludes.


\section{Observations and Data Reduction}
\label{sec:obs}

\subsection{The \oasis\/ Spectrograph}

\oasis\/ is an integral-field spectrograph based on the {\tt TIGER} concept \citep{bacon95} and is designed for high spatial resolution observations. It is a multi-mode instrument, with both imaging and spectroscopic capabilities, and can be assisted by an adaptive optics (AO) system.  \oasis\/ operated at the Cassegrain focus of the CFHT between 1997 and 2002 (optionally with the {\tt PUEO} AO system), and was transferred to the Nasmyth focus of the WHT behind the {\tt NAOMI} AO system in March 2003. All observations presented here were obtained at the CFHT, hence in the following we solely refer to the \oasis\/ CFHT configurations.

Since most of the objects in the \sauron\/ sample do not have a bright, nearby guiding source ($m_V < 16$ within $\sim$~30\arcsec), the $f/8$ (non-AO) mode of \oasis\/ was used. A spatial scale of 0\farcsec27$\times$0\farcsec27 per lenslet was selected for the observations to properly sample the generally excellent seeing at Mauna Kea, providing 1009 individual spectra in a 10\arcsec$\times$8\arcsec\ field-of-view. \oasis\/ was configured to give similar spectral coverage and resolution as \sauron, resulting in a wavelength range of $4760$--$5558$~\AA, with a resolution of 5.4~\AA\/ full-width half-maximum (FWHM), sampled at $1.95$~\AA~pix$^{-1}$. Given the generally large velocity broadening in the central regions of early-type galaxies, this configuration is suitable for measuring stellar kinematics in these objects, and also covers key absorption/emission features such as Mg\,$b$, a number of Fe lines, H$\beta$, [O{\small III}]$\lambda\lambda$4959,5007, and [N{\small I}]$\lambda\lambda$5198,5200.

\subsection{Observed Sample}

In the chosen configuration, \oasis\/ has a spectral resolution around 20\%\/ larger than that of \sauron, effectively increasing the minimum velocity dispersion that can be reliably measured. For this reason, we restricted our sample to only the E/S0 galaxies from the \sauron\/ survey (which tend to have higher dispersions than the Sa bulges), and gave priority to objects with a central velocity dispersion (as measured from Paper III) larger than the instrumental resolution of \oasis\/ ($\sigma_{\mathrm{OAS}} = 135$~\kms). We observed a total of 28 galaxies from the parent sample of 48 during three runs between March 2001 and April 2002 (Table~\ref{tab:runs}). Table \ref{tab:observations} lists the galaxies and gives a summary of the observations. Fig. \ref{fig:sample} shows how these objects populate the plane of ellipticity versus absolute $B$-magnitude (used to select the \sauron\/ survey galaxies, see Paper II).

\begin{figure}
 \begin{center}
  \includegraphics[height=8cm, angle=90]{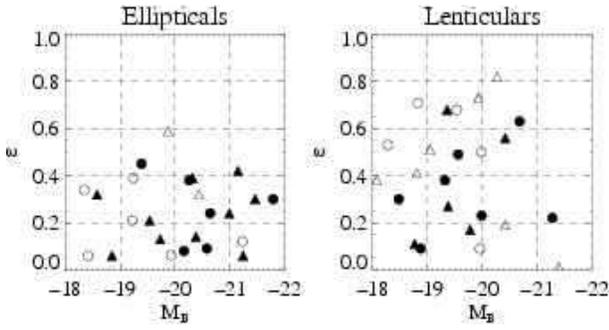}
 \end{center}
 \caption{\sauron\/ survey sample of ellipticals and lenticulars.
 Objects are separated as `field' (circular symbols) and `cluster'
 (triangular symbols) as defined in Paper II. Filled symbols indicate
 the subset of objects observed with \oasis\/ at the CFHT. }
 \label{fig:sample}
\end{figure}

\begin{table}
 \begin{center}
  \caption{Summary of the three \oasis\ observing runs.}
  \begin{tabular}{ccc}
  \hline
   Run  &Dates  &Clear    \\
   \hline
    1 & 24--27/03/2001    &3/3   \\
    2 &  8--13/01/2002    &2.5/3.5   \\
    3 &  6--10/04/2002    &4.5/5   \\
   \hline
  \label{tab:runs}
  \end{tabular}
 \end{center}
  Note: The dates indicate the duration of the run, although some of
  the nights were shared with other programs. The last column
  indicates the effective number of clear nights from the total number
  of nights dedicated to this program.
\end{table}

\begin{table}
\begin{center}
\caption{Summary of observations and PSF estimates.}
\begin{tabular}{llcclr}
\hline
NGC & Type & Run    & T$_{\mathrm exp}$ &Source & Seeing \\
~(1) & ~(2)  & ~(3) &  ~(4)         & ~~(5)   & (6)~~ \\
\hline
1023 & SB0$^{-}$(rs) &  2 & $2 \times 3600$ &F555W & 1.41 \\ 
2549 & S0$^{0}$(r)sp &  1 & $1800,3600$ &F702W & 0.91 \\ 
2695 & SAB0$^{0}$(s) &  2 & $2 \times 3600$ &Acq. & 0.91 \\
2699 & E:            &  2 & $2 \times 3600$ &F702W & 1.67 \\ 
2768 & E6:           &  3 & $2 \times 3600$ &F555W & 0.89 \\
2974 & E4            &2,3 & $2 \times 3600$ &F555W & 0.99 \\ 
3032 & SAB0$^0$(r)   &  2 & $2 \times 3600$ &F606W & 0.92 \\
3379 & E1            &  1 & $2 \times 3600$ &F555W & 0.91 \\ 
3384 & SB0$^-$(s):   &  1 & $1 \times 2700$ &F555W & 0.72 \\ 
3414 & S0~pec        &  3 & $2 \times 3600$ &F814W & 0.86 \\
3489 & SAB0$^+$(rs)  &  3 & $2 \times 3600$ &F814W & 0.69 \\
3608 & E2            &  1 & $1800, 2 \times 2700$ &F555W & 1.13 \\ 
4150 & S0$^0$(r)?    &  2 & $3 \times 3600$ &F814W & 2.15 \\ 
4262 & SB0$^-$(s)    &  3 & $2 \times 3600$ &F475W & 0.60 \\
4382 & S0$^+$(s)pec  &  2 & $3 \times 3600$ &F555W & 1.11 \\ 
4459 & S0$^+$(r)     &  3 & $2 \times 3600$ &F814W & 1.53 \\
4473 & E5            &  3 & $2 \times 3600$ &F555W & 0.80 \\
4486 & E0-1$^+$pec   &  1 & $3 \times 2700$ &F555W & 0.57 \\ 
4526 & SAB0$^0$(s)   &  1 & $1800, 3600$ &F555W & 0.94 \\ 
4552 & E0-1          &  1 & $2 \times 2700$ &F555W & 0.67 \\ 
4564 & E             &  3 & $2 \times 3600$ &F702W & 0.70 \\
4621 & E5            &  3 & $2 \times 3600$ &F555W & 0.86 \\
5198 & E1-2:         &  3 & $2 \times 3600$ &F702W & 0.84 \\
5813 & E1-2          &  1 & $3 \times 2700$ &F555W & 0.87 \\ 
5831 & E3            &  3 & $3600, 1200$ &F702W & 0.95 \\
5845 & E:            &  3 & $2 \times 3600$ &F555W & 0.91 \\
5846 & E0-1          &  3 & $2 \times 3600$ &F555W & 0.58 \\
5982 & E3            &  3 & $2 \times 3600$ &F555W & 0.77 \\
\hline
\label{tab:observations}
\end{tabular}
\end{center}
Notes:
(1)~NGC number.
(2)~Hubble type (RC3: \citeauthor{RC3} \citeyear{RC3}).
(3)~Run number corresponding to Table~\ref{tab:runs}.
(4)~Exposure time, in seconds.
(5)~Source for the seeing determination: either {\it HST} filter used or
    acquisition image (Acq.).
(6)~Seeing, full width at half maximum in arcseconds.
\end{table}

A number of spectroscopic standard stars were observed to allow calibration of the stellar absorption line strengths to the Lick/IDS system (see Section \ref{sec:line-strengths}). Several photometric standards were also observed, giving an estimate of the absolute throughput and total spectroscopic response of the instrument and telescope over the three runs.\looseness=-2

\subsection{Basic Data Reduction}

The data were reduced using the publicly available\footnote{http://www-obs.univ-lyon1.fr/ $\!\!\sim$oasis/download} \XOasis\/ software V6.2 \citep{rousset92} developed at CRAL (Lyon). Pre-processing of the images included bias and dark subtraction. Dark frames were acquired for each exposure time used, including for calibrations and standard stars, to correct for non-linear `hot' pixels on the CCD. The spectra were then extracted from the two-dimensional CCD image into a three dimensional data-cube by means of an instrument model, or `mask'. This mask is essentially a map of the wavelengths falling on the different pixels for each lens, based on a numerical model of the instrument and telescope. The mask includes a model of the pupil for each lens, and simulates the curvature and chromatic variations of each spectrum on the CCD. Neighbouring spectra on the CCD are de-blended based on this model, and the spectra apertures are summed in the cross-dispersion direction, applying optimal weights \citep{horne86}.\looseness=-2

Although the extraction process imparts first-order wavelength
information on the spectra from the instrument model, a more accurate
wavelength calibration was performed in the conventional way using neon
arc-lamp exposures: this removes higher-order variations in the dispersion
solution or inaccuracies in the mask. \oasis\/ was mounted at the
Cassegrain focus of the CFHT, and so flexures were accounted for by
obtaining arc-lamp exposures bracketing the science
exposures. Both neon exposures were used as a reference for the mask,
giving the average position of the spectra on the CCD during the exposure. The dispersion
solution itself is computed on a single arc-lamp exposure,
generally using around 15 lines fitted with a second-order polynomial,
with a mean dispersion of around a tenth of a pixel ($\sim 0.2$~\AA).

Flat-fielding of the data-cube was performed simultaneously in the three dimensions $(x, y, \lambda)$. Exposures of a continuum source were used for the spectral domain; exposures of the twilight sky were used to provide an illumination correction, or spatial flat-field. These two calibration exposures were combined to produce a single flat-field data-cube, which was divided through the science cubes.

Cosmic ray hits affect neighbouring pixels on the CCD, which are
effectively well-separated within the extracted data-cube, due to the
staggered layout of spectra on the detector. This means that comparing
spectra of neighbouring (on the sky) lenses (which should be rather
similar due to PSF effects and the general nature of these objects)
gives high contrast to cosmic ray events for a given spectral
element. Pixels affected by cosmic rays were replaced by the median value of the neighbouring lenses at the corresponding wavelength.\looseness=-2

No sky subtraction was performed on these data since in all cases, the
contribution from night sky lines was insignificant at the adopted
wavelengths and within the small field of view centred on the bright
galaxy nucleus.

The spectral resolution can vary slightly within the field of view, which may introduce an irregular broadening of the lines when separate exposures are combined. We therefore homogenised the resolution to a common value. This was done by comparing twilight exposures with a high-spectral resolution solar reference spectrum\footnote{http://bass2000.obspm.fr/solar\_spect.php} to measure the effective spectral resolution of each lens. These values were then smoothed with a circular spatial boxcar filter of 0\farcsec5 in radius to reduce noise. These `maps' of spectral resolution were used to establish a representative instrumental broadening, taken as 5.4~\AA\/ (FWHM), the value for which 90\% of the lenses (averaged over all available twilight frames of different nights and runs) were on or below that resolution. The twilight closest in time to the science exposure was used as a reference to smooth the data-cube with the appropriate Gaussian to obtain the adopted resolution.

\subsection{Fringe Pattern}

The data obtained during the three \oasis\/ runs suffered from an interference pattern similar to fringing within the spectral range of our data. This added a periodic pattern with amplitude of about $10$~\kms\/ to the mean absorption-line velocity fields derived from the spectra. The pattern was found to vary between exposures due to instrument flexure, and was therefore difficult to remove accurately. In principle, one would prefer to correct for this effect in the pre-extracted images, rather than on the extracted data-cubes, in which several adjacent pixels are combined at each wavelength. The interference pattern was not well understood during runs 1 and 2, however, and so appropriate calibration frames (high-quality dome flats) were not obtained. For these runs, we were therefore required to make a correction on the extracted data-cubes. This was done by removing the average twilight spectrum (in which the effects of the fringing were largely randomised) from the twilight data-cube, leaving a residual fringe pattern, unique to each lens. Dividing the science data-cubes by this frame removed much of the fringe pattern, although some signature was left due to flexures between the twilight and science exposures. For run 3, we obtained the appropriate calibration frames to allow the fringe correction to be made before extraction. A median combination of six dome flats was used to generate a correction frame. This frame was obtained by dividing each column of the master dome flat image by a high-order spline, fitted in such a way as to follow the complex light distribution, but not fit into the fringe pattern itself. Before extraction, we divided all data by the resulting correction frame to remove the fringe pattern. Small linear offsets between the correction image and the data image (due to flexure) were accounted for by comparing the associated arc-lamp exposures of the two frames, and shifting the correction frame accordingly.\looseness-2

Fig.~\ref{fig:fringe} demonstrates the effect of the fringing and fringe-correction on the
mean stellar velocity field derived from a run 3 twilight sky exposure ($S/N > 100$), which should show a flat velocity field. Velocity fields of the same twilight exposure are shown after correction with both methods. The standard deviation of velocity values decreases from 7~\kms\/ in the uncorrected case to 4~\kms\/ with either correction. A residual gradient remains in the corrected velocity fields, with an amplitude of $\sim 5$~\kms\/ across the field. This is within the fundamental uncertainties of our wavelength calibration, and below the typical measurement error of our galaxy observations (see Section \ref{sec:stellar_kin}). Although the correction using dome flats is preferred, the solution using twilight exposures yields comparable results.

\begin{figure}
 \begin{center}
  \includegraphics[width=8.0cm, angle=0]{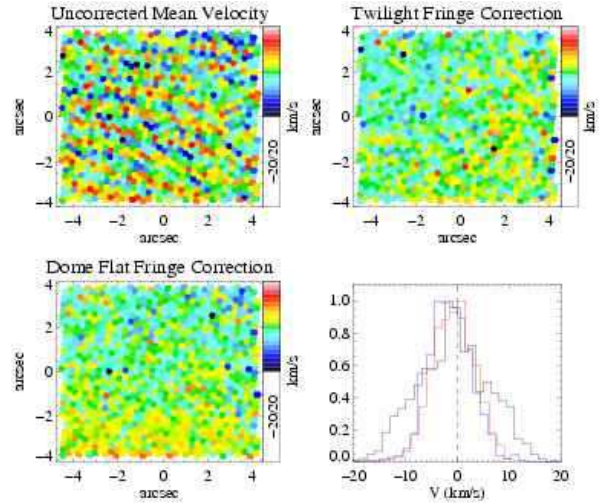}
 \end{center}
 \caption{{\it Top left:} mean velocity field of a twilight sky exposure derived using a solar reference spectrum and the pPXF method described in Section \ref{sec:stellar_kin}. A periodic pattern can be seen in the velocity field, resulting from a spatially-varying interference pattern in the spectra. {\it Top right:} mean velocity of the same twilight frame, corrected {\em after} extraction using a correction frame constructed from a twilight exposure from a different night. This method was used to correct runs 1 and 2. {\it Bottom left:} mean velocity of the same twilight frame, corrected {\em before} extraction, using a median-combination of several dome-flat exposures. This method was used to correct data from run 3. {\it Bottom right:} histogram of velocities from each field: black line = uncorrected field, blue line = dome-flat correction, magenta line = twilight correction. After the correction, the periodic pattern is removed, and the dispersion in the recovered velocities decreases from 7~\kms\/ in the uncorrected case, to 4~\kms\/ for both correction methods. (See text for details.)}
 \label{fig:fringe}
\end{figure}

\subsection{Flux Calibration}

During each night of the three observing runs, photometric standard stars were observed, primarily in order to calibrate chromatic variations in throughput and instrument response, but also to give an absolute flux calibration for the data. The observed stars are listed in Table \ref{tab:flux_stars}, and include several repeat measurements within and between runs. A single spectrum was extracted from each reduced data-cube by integrating all spectra within a circular aperture centred on the star. To ensure that the flux star spectrum was representative of the total flux in our wavelength range without simply adding noise, the PSF was measured from the reconstructed image of the star, and a summation of the flux within an aperture three times larger than the PSF FWHM was made, weighted by the Gaussian PSF model.

\begin{table}
 \begin{center}
  \caption{Observed photometric standard stars.}
  \begin{tabular}{ccc}
  \hline
   Name & Run(\#obs)  & Calibration \\
   \hline
   HD093521   & 2(3),3(4) & STIS+Oke \\
   G191-B2B   & 1(2),2(2) & Model \\
   Feige 66   & 1(1),3(1) & FOS+Oke \\
   BD+33 2642 & 1(2)      & FOS+Oke \\
   HZ 44      & 1(1)      & STIS \\
   \hline
  \label{tab:flux_stars}
  \end{tabular}
 \end{center}
 Notes: The second column indicates the run number, and in parentheses, the number of observations during that run. Only one star was observed per night. The third column indicates the source of the reference calibration spectrum, as described in \citet{bohlin2001}.
\end{table}

The observed photometric standard stars were chosen to have as few spectral features as possible, thus making them suitable for calibrating residual spectral response variations. Over the relatively small spectral domain of \oasis, such variations are expected to be low-order. However, imperfections in the calibration of our data (e.g. due to the fringe pattern) may give rise to fluctuations on scales of tens of Angstroms. Such residuals are difficult to detect, given the typically low spectral resolution of available spectrophotometric standard star reference tables \citep[e.g.][]{oke90}. One solution is to compare with higher-resolution theoretical models. Such a model is available for the well-known star G191-B2B from \cite{bohlin2000}, which provides flux measurements at varying wavelength intervals, from around 20~\AA\/ at the red end of our wavelength domain (where the spectrum is varying only slowly) to $<1$~\AA\/ inside absorption features.

Fig.~\ref{fig:g191b2b} shows a comparison of this model with our four \oasis\/ observations {\em before} any flux calibration correction was applied. The model is interpolated to have the same spectral sampling as our observations, and is fitted to the observations using the penalized pixel-fitting method described in Section \ref{sec:stellar_kin}. As well as applying a Gaussian velocity broadening (analogous to spectral resolution) and Doppler shift, the fitting process allows the inclusion of polynomial terms. The (average) fit shown in the top panel of Fig.~\ref{fig:g191b2b} is achieved using a second-order multiplicative polynomial term. The bottom panel of Fig.~\ref{fig:g191b2b} shows the residuals of the four individual fits to the \oasis\/ observations, showing that the second-order polynomial is a good description of the relation between our observations and the theoretical model of G191-B2B. Deviations from the parabolic relation, although systematic across the two runs, are small, with a standard deviation of less than half a percent. This shows that structures on intermediate wavelength scales which may be present in our spectra due to e.g., filter throughput, are well removed by our spectral flat-fielding process.

To evaluate whether there are global changes in the chromatic response of the observational setup, Fig.~\ref{fig:flux} compares the correction polynomials derived from all observed standard stars. In general, the curves agree well, with a maximum 1$\sigma$ spread of 2\%\/ at the spectra edges. Within a run, the scatter is significantly less ($\sim 0.5$\%), however, and there appear to be some small systematic effects between runs. We therefore employ a single flux calibration curve per run, rather than a single correction for all data. \looseness-2

\begin{figure}
 \begin{center}
  \includegraphics[height=8cm, angle=90]{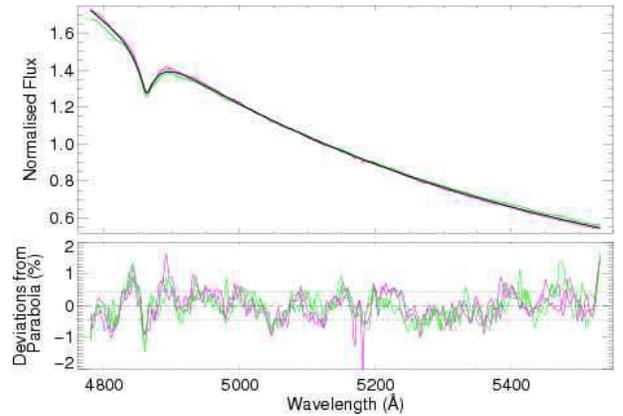}
 \end{center}
 \caption{{\it Top panel:} The four normalised \oasis\/ observations of the photometric standard star G191-B2B (thin lines: magenta = run 1, green = run2) overplotted with the (normalised) average best-fit of the \citet{bohlin2000} theoretical model (thick black line) using penalised pixel-fitting (pPXF, see Section \ref{sec:stellar_kin}). A multiplicative second-order polynomial term is included in the fit, as well as a varying shift and broadening of the model spectrum. {\it Bottom panel:} ratio of the \oasis\/ observations and individual pPXF fits expressed as percentage deviations around the parabolic polynomial. The line colours indicate different runs as in the upper panel. Dotted horizontal lines indicate the (robust) standard deviation of the residuals with respect to zero, having an amplitude of $0.44$\%.}
 \label{fig:g191b2b}
\end{figure}

We perform the flux calibration of our data comparing our observed stellar spectra with the calibrated spectra of \citet{bohlin2001}, available from the ``CALSPEC'' database\footnote{http://www.stsci.edu/hst/observatory/cdbs/calspec.html} of standard flux calibration spectra used for {\it HST}. Variations in the {\em absolute} flux calibration are as large as 20\%. However, to improve observing efficiency, no effort was made to obtain an accurate absolute photometric calibration by observing standard stars close (in time) to each galaxy pointing.

\begin{figure}
 \begin{center}
  \includegraphics[height=8cm, angle=90]{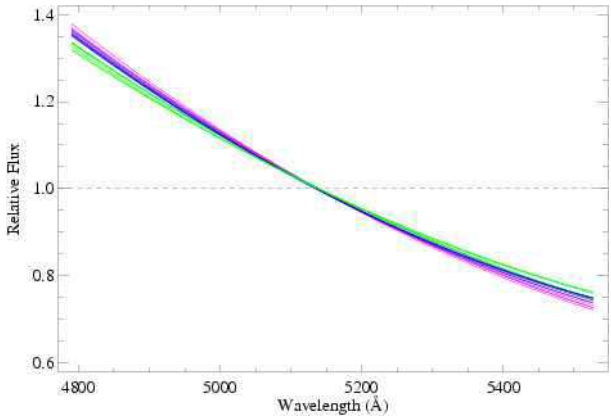}
 \end{center}
 \caption{Normalised flux correction curves determined from each observed photometric standard star for all three observing runs (magenta = run 1, green = run2, blue = run 3). From the stellar spectra, and from the comparison of G191-B2B shown in Fig.~\ref{fig:g191b2b}, a second-order polynomial was deemed sufficient to describe the chromatic response of the observational setup. Variations between the curves have a maximum 1$\sigma$ spread of 2\%\/ (with a range of 3\%), although variations {\em within} a single run are smaller ($\sim 0.5$\%). The systematic differences lead us to apply a single flux correction within a run, rather than a unique correction for all data.}
 \label{fig:flux}
\end{figure}

\subsection{Merging and Spatial Binning}

Two or more exposures were obtained for each galaxy (Table~\ref{tab:observations}), each one offset by a small, non-integer number of lenslets to provide oversampling and avoid systematic effects due to bad CCD regions. Multiple exposures were combined by first truncating to a common wavelength region and centring the spatial coordinates on the galaxy nucleus using images reconstructed from the data-cubes by summing flux in the spectral direction. Exposures were then re-normalised to account for transparency variations, and resampled from the original hexagonal spatial grid (set by the lens array) onto a common spatial grid of 0\farcsec2$\times$0\farcsec2. Co-spatial spectra were combined via an optimal summing routine, taking into account the error spectra which are propagated through the reduction. For the spectroscopic standard stars, single spectra were extracted by summing all the data-cube spectra within a 2\arcsec\/ circular aperture.

At the stage of merging multiple pointings, a correction was made for differential refraction by the atmosphere, which causes the image to change position on the lens array as a function of wavelength. We applied the theoretical model of \cite{filippenko82}, which uses atmospheric parameters to estimate the shift of the image as a function of wavelength and zenith angle. We adopted global values of 620~mb and 4$^\circ$C for the air pressure and temperature, respectively, for all runs. Most observations were taken at airmass $\lsim 1.3$, ensuring that the typical relative shift due to atmospheric dispersion across the short wavelength range of our \oasis\/ observations is $<0$\farcsec2.  No correction was made for atmospheric dispersion for the standard stars, as the flux is combined over a relatively large aperture.

It is necessary to analyse the galaxy spectra with a minimum signal-to-noise ratio ($S/N$) to ensure accurate and unbiased measurements. This was achieved using the binning method developed by \cite{cappellari03}, in which spectra are coadded starting from the highest $S/N$ lenslet, and accreting additional lenslets closest to the current bin centroid. When the target $S/N$ is reached, a new bin is begun. The resulting bin centroids are then used as the starting points for a centroidal Voronoi tessellation, generating compact, non-overlapping bins with little scatter around the imposed minimum $S/N$. We conservatively bin all data to a minimum mean $S/N$ of 60 per spectral resolution element for our analysis.\looseness-2

\subsection{Point Spread Function}

By comparing images constructed by integrating the flux in each spectrum with imaging from {\it HST} (where available), an estimate of the PSF can be made. We do this by parameterising the \oasis\/ PSF with two concentric circular Gaussians, and minimizing the difference between the {\it HST} image convolved with this PSF, and the reconstructed image. In practice, {\it HST} images in the appropriate band are not available for all galaxies, in which case features such as dust may introduce small inaccuracies in our recovered values. For those galaxies with archival {\it HST} images, the resulting PSF estimates, characterised by the FWHM of the best-fitting single Gaussian to our PSF model, are presented in column 6 of Table \ref{tab:observations}. Fig.~\ref{fig:psf_comp} compares these values with the equivalent PSF measurements for the \sauron\/ observations of these galaxies (Table 3 of Paper III).\looseness-1

Although the spatial resolution of the \sauron\/ and \oasis\/ data sets are limited by atmospheric seeing conditions, the \oasis\/ data has the advantage of fully sampling even the best conditions on Mauna Kea within this wavelength range. For this reason, the \oasis\/ data has generally better effective spatial resolution, and in some cases is significantly better, with more than a quarter of our sample having a PSF $\lsim $0\farcsec8. The median PSF of the \oasis\/ observations (0\farcsec91) is almost half that of \sauron\/ (1\farcsec7), with up to a factor three improvement in some cases. For four galaxies the measured PSF of \oasis\/ is similar to that of \sauron: namely NGC\,1023, NGC\,2699, NGC\,4459 and NGC\,4150. We nevertheless present the data in this paper, as the spectral domain of \oasis\/ includes several Lick iron indices which are not obtained with \sauron.

\begin{figure}
 \begin{center}
  \includegraphics[width=8cm]{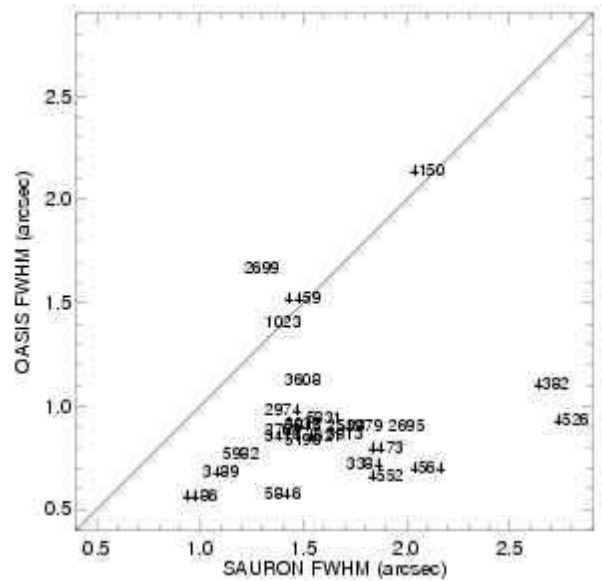}
 \end{center}
 \caption{Comparison of \oasis\/ and \sauron\/ PSF estimates for the
          galaxies in Table \ref{tab:observations}. The solid line
          indicates the 1:1 relation, showing that the PSFs of the
          \oasis\/ observations are generally significantly smaller
          than those of \sauron. }
 \label{fig:psf_comp}
\end{figure}


\section{Data Analysis}
\label{sec:analysis}

\subsection{Stellar Kinematics}
\label{sec:stellar_kin}

Stellar absorption-line kinematics were derived for each galaxy by
directly fitting the spectra in pixel-space, using the penalised
pixel-fitting method (hereafter pPXF) of \citet{cappellari04}, as in
Paper III.  This method was chosen over Fourier-based methods due to
its robustness to contamination by nebular emission lines (emission
lines are simply excluded from the fitting process), which can be
strong in our spectral domain around the central regions of early-type
galaxies (e.g., Paper V).

Although robust to emission-line contamination, pixel-fitting methods
are sensitive to template mismatch effects, which can significantly
bias the resulting kinematics \citep[e.g.,][]{rix92}. This problem was
minimised by selecting an `optimal template' at each iteration of the
fitting process. For each trial set of kinematic parameters, an
optimal linear combination of appropriately convolved absorption
spectra were fitted to the data. We use the same library of absorption
spectra as in Paper III, which was taken from the single-burst stellar
population (SSP) models of \cite{vazdekis99}, with the addition of
several individual stellar spectra with strong Mg\,$b$ from the
\cite{jones97} library (from which the SSP models are built) to
compensate for the near-solar abundance ratio inherent in the SSP
models. It was found that this library was able to reproduce the
galaxy spectra very well, minimizing biases caused by template
mismatch.

After rebinning the wavelength axis of the data-cube to a logarithmic
scale (equivalent to pixels of constant velocity scale), we used the pPXF
method to determine the line-of-sight velocity distribution (hereafter
LOSVD) parameterised by the truncated Gauss-Hermite series
\citep{vdmarel93,gerhard93}. Only the first four terms (mean velocity $V$;
velocity dispersion $\sigma$; and the higher-order terms $h_3$ and
$h_4$, related to the skewness and kurtosis respectively) were measured, as these are generally sufficient to describe the significant features
in early-type galaxy LOSVDs. Our data are generally of adequate
quality to also measure $h_5$ and $h_6$, which may be useful for some
specific purposes (for example in the fitting of dynamical models),
however they are not reported here.\looseness -2

The penalty term of the pPXF method, quantified by the near-unity parameter, $\lambda_p$, suppresses the large uncertainties inherent in measuring the higher-order terms $h_3$ and $h_4$ when the LOSVD is undersampled by biasing the solution towards a simple Gaussian. In general, the objects in our sample have sufficiently large velocity dispersions such that this biasing is not important. Several objects, however, have velocity dispersions close to or below the instrumental dispersion. Following the procedure described in \citet{cappellari04}, we therefore optimised $\lambda_p$ via Monte Carlo simulations to suppress the measurement scatter to a level at which the true value (known in the simulation) lies well within the standard deviation of the biased value. This was found to be $\lambda_p = 0.5$ for our adopted $S/N$ of 60, differing slightly from the value used for the \sauron\/ data in Paper III due to the different spectral resolution and wavelength coverage of the two instruments. From these same simulations, errors on the kinematic parameters ($V$, $\sigma$, $h_3$, $h_4$) for $S/N \sim 60$ were found to be 9~\kms, 15~\kms, 0.06, 0.06 respectively at a velocity dispersion of 110~\kms; and 12~\kms, 12~\kms, 0.04, 0.04 respectively at a velocity dispersion of 250~\kms.

Final errors on the derived galaxy kinematic parameters were determined via 100 Monte-Carlo realisations of each galaxy spectrum, in which a representative spectrum of simulated white noise was added {\it on top of} the noisy galaxy spectrum for each realisation. This approach, in contrast to adding noise to a noise-free broadened template spectrum, has the advantage that biases in the random error due to template mismatch are included (although the systematic uncertainty caused by the mismatch is not). The difficulty in quantifying systematic uncertainties due to template mismatch is well known, and the reader is directed to Appendix B3 of Paper III for relevant discussion. 

Spectral regions which could possibly be affected by ionised gas emission were excluded from the fit by imposing `windows' around the main features (\hb, \oiii\/ and \nii). The half-width of these windows was conservatively taken as 600~\kms. In cases where the velocity dispersion of the detected gas $\sigma_\mathrm{gas}$ was very high (taken as when $\sigma_\mathrm{gas} > 400$~\kms), the windows were adjusted such that the half-width was given by $2 \times \sigma_\mathrm{gas}$, and the fit was repeated. In practice, only NGC\,4486 (M87) exhibits such strong emission, and on the second iteration of pPXF, the stellar continuum fit was generally greatly improved.

\subsection{Comparison with \sauron\/ stellar kinematics}
\label{sec:sau_comp_kin}

Due to the relatively small field of view of \oasis, the conventional
comparison with literature values of velocity dispersion is not
straight-forward, since these values are often integrated within,
e.g., half an effective radius: generally much larger than the
\oasis\/ field of view.  Rather than introducing large and uncertain
aperture corrections, the most direct comparison can be made with the
\sauron\/ measurements themselves, which were already shown to be
consistent with independent studies in Paper III.

To reduce the effects of atmospheric seeing on the dispersion values, we integrate the spectra from both instruments within a 4\arcsec\/ circular aperture, and derive the velocity dispersion from this high $S/N$ spectrum using pPXF, fitting only for the mean velocity and velocity dispersion. Fig.~\ref{fig:sigma_comp} shows the resulting comparison of \oasis\/ and \sauron\/ velocity dispersion measurements ({\it top panel}) and the resulting scatter around the unity relation ({\it bottom panel}). The values are consistent between the two studies, with a mean offset of 0.84$\pm$7.8~\kms. The spectra used in this comparison are also used to compare the gas properties and line-strengths below. In Appendix~\ref{appendix:c}, we present a comparison of the \oasis\/ and \sauron\/ spectra themselves.

\begin{figure}
 \begin{center}
  \includegraphics[width=8cm]{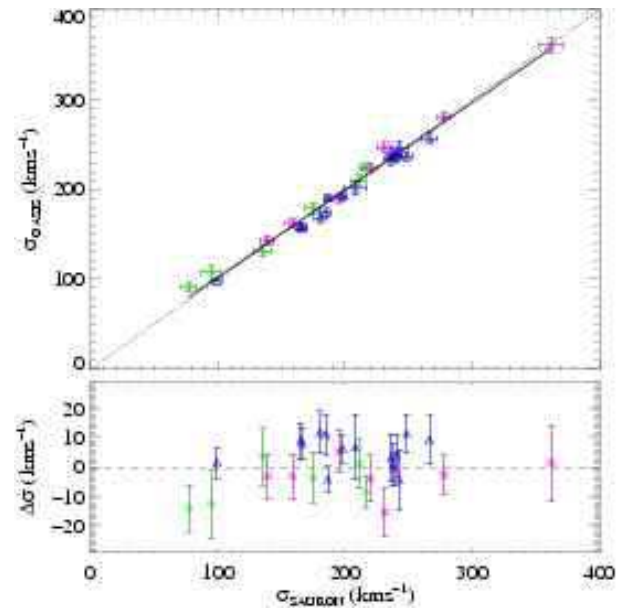}
 \end{center}
 \caption{{\it Top:} Comparison of \oasis\/ and \sauron\/ velocity dispersions measured on spectra integrated within equivalent circular apertures of 4\arcsec\/ radius. Different coloured symbols indicate galaxies observed on different runs: magenta asterisk = Run 1, green diamonds = Run 2, blue triangles = Run 3. The dotted line indicates the unity relation, and the thick solid line is a robust straight-line fit to the data, with slope 0.97$\pm$0.02. {\it Bottom:} Difference between the \sauron\/ and \oasis\/ velocity dispersion measurements, with quadratically summed errors.}
 \label{fig:sigma_comp}
\end{figure}

\subsection{Kinemetry}
\label{sec:kinemetry}

To quantify the structures found in our stellar velocity fields, we employ the so-called `kinemetry' method of \citet{krajnovic06}. This technique is an extension of the classic Fourier analysis of photometry \citep[e.g.][]{jedrzejewski87,franx89}, which can be used to describe the radial behaviour of two-dimensional kinematic maps. Annuli with increasing radius are considered, and for each annulus, a Fourier expansion is performed to fit the changing amplitude as a function of polar angle. 

By determining the flattened (elliptical) annulus along which the higher moments of the expansion are minimized, one can measure the characteristic `flattening' of the rotation field, equivalent to approximating the rotation to that of a thin disk. The amplitude of the first term, \ki, gives a robust estimate of the peak rotation as a function of radius, equivalent to the circular velocity in the limit of a true thin disk. By also fitting the phase of the first moment of the expansion, a robust measurement of the kinematic position angle, \kinpa, is obtained, defined here as the angular position of the peak rotation. This parameter traces the rotation axis, and is an important diagnostic of the kinematic structure of the object, since changes in the rotation axis are a sensitive indicator of deviations from axisymmetry \cite{krajnovic05}.\looseness-2

Profiles of \kinpa\/ and \ki\/ are presented for all galaxies in Appendix~\ref{appendix:b}, and are used to support out interpretation of the structures found in our velocity maps. For example, an abrupt change of \kinpa\/ combined with a minimum in \ki\/ indicates the presence of a kinematically decoupled component in the stellar velocity field that is counter-rotating to the outer parts, or has a different rotation axis. Detailed analysis of the full set of kinematic moments of the complete \sauron\/ sample of E and S0 galaxies, using both \sauron\/ and \oasis\/ data where available, will be given in a future paper of this series.\looseness-2

\subsection{Gas Properties}
\label{sec:gas}


The spectral region covered by our data includes a number of possible nebular emission lines, namely \hb, \oiii, and \nii. It is necessary to separate these emission lines, where present, from the underlying stellar absorption continuum in order to obtain accurate line-strength indices, as well as to study the emission-line distribution and kinematics themselves. We do this using the method of Paper V, which includes a model of the emission-line spectrum as part of an optimal template fit. The stellar kinematics are fixed at the values determined by excluding the emission regions from pPXF as above. With the stellar kinematics fixed, the emission-line spectrum, modelled as a series of independent single Gaussians for each emission line, is included in the template library. The velocity shift and broadening of the Gaussians are determined via non-linear optimisation, and the best-fitting non-negative linear combination of stellar spectra and emission-line models is found at each iteration. This method has the advantage that both absorption-line templates and emission-line model are fitted simultaneously, avoiding the need for `windowing' possible emission line regions. This makes optimal use of the information in the spectrum, resulting in accurate absorption and emission line measurements. Paper V gives a full discussion of this method.

We adopt the approach of first fitting the \oiii\/ lines alone, and then repeating the fit including the \hb\/ and \nii\/ lines in the emission spectrum, but with their kinematics fixed to the values measured from \oiii. This approach was chosen since measuring the \oiii\/ lines is less sensitive to template mismatch, and furthermore, all the objects with detected \hb\/ emission show signs of \oiii\/ emission, which is usually stronger. For this reason, we use \oiii\/ as a robust indicator of emission, before considering that other lines may be present. In Paper V we discuss the merits and pitfalls of fitting \hb\/ and \oiii\/ independently. Here we adopt the conservative approach of always assigning the kinematics derived from \oiii\/ to the \hb\/ line. Although this may differ slightly from the true \hb\/ kinematics, in practice these differences are small compared to our measurement errors, and in any case do not significantly bias the fluxes we measure.

The detection or non-detection of gas is based on the ratio of the amplitude of the fitted gas line and the scatter of the residual spectrum after subtraction of the best-fitting stellar and gas templates. By using the residuals of the template fit, this `amplitude-to-noise' ratio (hereafter $A/N$) includes both random noise and systematic errors (e.g. due to template mismatch) in the noise estimate. The limiting values of $A/N$ below which our measurements become strongly biased or uncertain were determined from simulations in Paper V. For the unconstrained fit of \oiii, emission lines with $A/N \ge 4$ are considered reliable. For \hb, the fitting process has less freedom, being constrained by the \oiii\/ kinematics. In this case, emission lines with $A/N \ge 3$ are considered reliable.\looseness-2

The detection of \nii\/ emission is complicated by the fact that the library of templates used to describe the stellar continuum are of near-solar abundance ratio. The effects of template mismatch are therefore particularly acute around the \mgb\/ feature ($\lambda = 5177$~\AA), slightly blueward of the \nii\/ doublet. To incorporate this into our detection threshold, when computing the $A/N$ for fitted \nii\/ features, we base the residual noise estimate $N$ specifically on a wavelength region extending across both the \nii\/ and \mgb\/ features. Following Paper V, we impose that $A/N \ge 4$ for this feature. In addition, \nii\/ is only considered if both \hb\/ and (by proxy) \oiii\/ emission is also present. We further constrain the fit of \nii\/ by imposing the same kinematics as \hb\/ and \oiii.

Based on these detection thresholds, we have shown in Paper V that it is possible to derive typical sensitivity limits, in terms of the equivalent width $EW$ of the (Gaussian) emission line. Specifically, from eq. (1) of Paper V:

\begin{equation}
EW = \frac{A/N \times \sigma_{\mathrm{obs}}\sqrt{2\pi}}{S/N} \mathrm{,}
\end{equation}

\noindent where $\sigma_{\mathrm{obs}}$ is the effective dispersion of the fitted emission line {\em including} the instrumental broadening. For an emission line with intrinsic velocity broadening $\sigma_{\mathrm{int}} = 50$~\kms, $\sigma_{\mathrm{obs}} = \sqrt{\sigma_{\mathrm{int}}^2 + \sigma_{\mathrm{oas}}^2} = 144$~\kms $\equiv 2.4$~\AA\/ for \oiii\/ (2.3~\AA\/ for \hb). For our imposed threshold of $S/N = 60$ for the continuum, this gives an estimate for the limiting $EW$ of \oiii\/ of 0.40~\AA\/ for the chosen limit of $A/N = 4$ (0.29~\AA\/ for \hb). For given $A/N$ thresholds, the limiting $EW$ for \oasis\/ compared with \sauron\/ (within the same aperture) scales as the ratio of the instrumental broadening and $S/N$, such that

\begin{equation}
\frac{EW_{\mathrm{oas}}}{EW_{\mathrm{sau}}} =
 \frac{\sigma_{\mathrm{oas}}}{\sigma_{\mathrm{sau}}} \times
 \frac{S/N_{\mathrm{sau}}}{S/N_{\mathrm{oas}}} =
 1.25 \times \frac{S/N_{\mathrm{sau}}}{S/N_{\mathrm{oas}}} \mathrm{.}
\end{equation}

In the central few arcseconds of our galaxies, the \sauron\/ data are generally not spatially binned, and have typical $S/N$ far in excess of the binning threshold. The \oasis\/ lenses are smaller on the sky than those of \sauron, with more than a factor 10 less collecting area. For this reason, the \oasis\/ spectra are usually close to or below the binning threshold in the same regions. The sensitivity in these regions is consequently less with \oasis\/ than with \sauron, although the effective spatial resolution is higher (see also Appendix~\ref{appendix:d}).

Errors on the emission line parameters are estimated via 100 Monte Carlo realisations of the full analysis process. The galaxy spectra (including the emission) with added noise used in the simulations to compute the kinematic errors (see Section \ref{sec:stellar_kin}) are used as input to the gas-fitting method. In this way, we propagate errors in the template-fitting process, as well as the emission line fitting itself. The amplitude of the errors varies with the $A/N$ of the emission line, which can vary significantly within and between galaxies. For $A/N = 60$, typical errors in the derived flux, velocity and dispersion are 3\%, 2\% and 10\% respectively. For $A/N = 10$, these values increase to 10\%, 4\% and 25\% respectively.

\subsection{Comparison with \sauron\/ gas measurements}
\label{sec:sau_comp_gas}

Given the complexities of separating often faint emission lines from the stellar continuum, and the high-quality data required to make reliable measurements, making a meaningful comparison of our measured gas properties with literature values is a difficult task. In Paper V we established that there is a satisfactory agreement between \sauron\/ gas measurements and those of the Palomar spectroscopic survey \citep{ho95}, given the differences in techniques and data quality. We therefore again adopt the \sauron\/ measurements as our reference, since the data quality are similar to our observations, and the apertures used can be accurately matched.

Fig.~\ref{fig:gas_comp_ratio} presents a comparison of the \oiii/\hb\/ ratio measured with \sauron\/ and \oasis. This demonstrates our ability to measure emission line fluxes, independent of the absolute flux calibration of the two data sets, which are both approximate. This comparison shows reasonable agreement, with a mean \sauron:\oasis\/ ratio of $1.19\pm0.44$. There is some evidence of a systematic trend, with a fitted slope of $1.08\pm0.03$ from the error-weighted line fit. Fig.~\ref{fig:gas_comp_sigma} compares the velocity dispersion of the \oiii\/ doublet. Again, there is reasonable agreement, with a mean ratio of $0.91\pm0.24$, and a fitted slope of $0.99\pm0.05$ from the error-weighted line fit.

\begin{figure}
 \begin{center}
  \includegraphics[width=8cm]{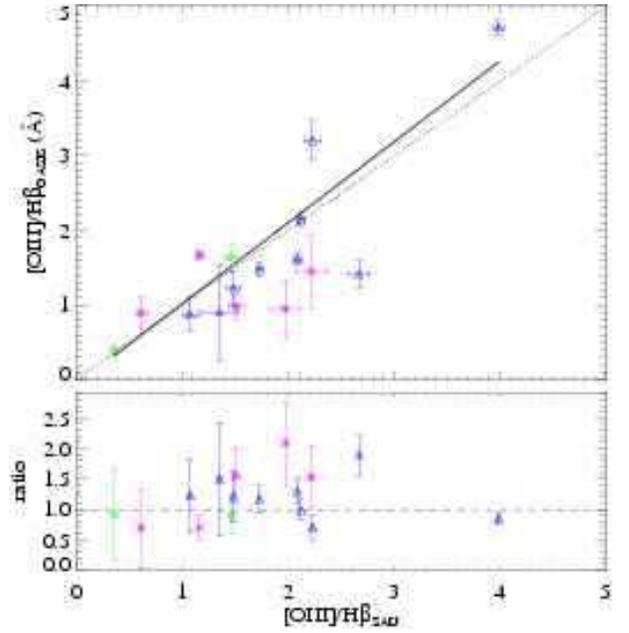}
 \end{center}
 \caption{{\it Top:} Comparison of \oasis\/ and \sauron\/ \oiii/\hb\/ measurements within a circular aperture of 4\arcsec\/ radius. The dotted line shows the unity relation, and the thick black line shows a straight-line fit weighted by the errors in both axes. {\it Bottom:} Ratio of the \sauron\/ and \oasis\/ measurements, with quadratically summed errors. Different coloured symbols indicate galaxies observed on different runs as in Fig.~\ref{fig:sigma_comp}.}
 \label{fig:gas_comp_ratio}
\end{figure}

\begin{figure}
 \begin{center}
  \includegraphics[width=8cm]{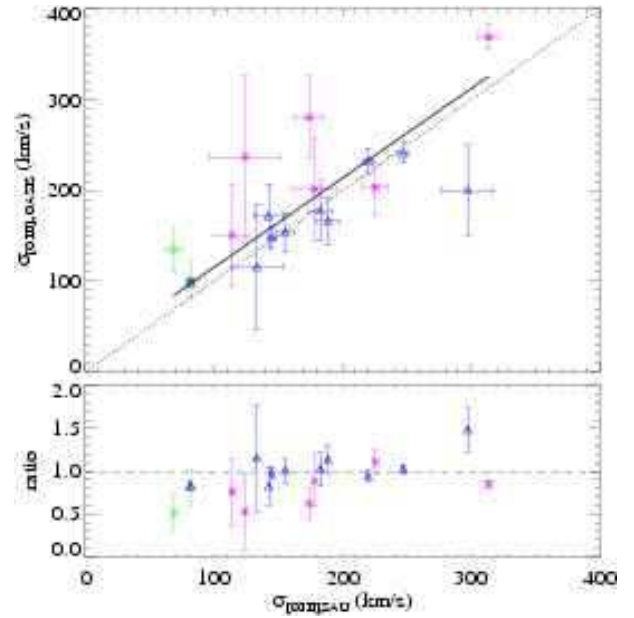}
 \end{center}
 \caption{{\it Top:} Comparison of \oasis\/ and \sauron\/ \oiii\/ velocity dispersion measurements within a circular aperture of 4\arcsec\/ radius. The overplotted lines are as in Fig.~\ref{fig:gas_comp_ratio}. {\it Bottom:} Ratio of the \sauron\/ and \oasis\/ measurements, with quadratically combined errors. Different coloured symbols indicate galaxies observed on different runs as in Fig.~\ref{fig:sigma_comp}.}
 \label{fig:gas_comp_sigma}
\end{figure}

\subsection{Line Strengths}
\label{sec:line-strengths}

The spectral range of our \oasis\/ data contains a number of absorption features which can be used as diagnostic tools to determine the distribution of stellar populations within a galaxy, in terms of the luminosity-weighted population parameters: age, metallicity and element abundance ratio. We quantify the depth of these absorption features through the use of atomic line-strength indices,  taken from the Lick/IDS system \citep{burstein84,worthey94,trager98}, the bandpass definitions of which are given in Table \ref{tab:lick_bands}.\looseness-2

\begin{table}
 \begin{center}
  \caption{Bandpass definitions of Lick indices within our \oasis\/ spectral range.}
  \begin{tabular}{lccc}
  \hline
   Index & Units & Central bandpass & Pseudo-continuum \\
   \hline
\hb    & \AA & 4847.875 - 4876.625 & 4827.875 - 4847.875 \\  
       &     &                     & 4876.625 - 4891.625 \\
Fe5015 & \AA & 4977.750 - 5054.000 & 4946.500 - 4977.750 \\ 
       &     &                     & 5054.000 - 5065.250 \\
\mgb   & \AA & 5160.125 - 5192.625 & 5142.625 - 5161.375 \\ 
       &     &                     & 5191.375 - 5206.375 \\
Fe5270 & \AA & 5245.650 - 5285.650 & 5233.150 - 5248.150 \\ 
       &     &                     & 5285.650 - 5318.150 \\
Fe5335 & \AA & 5312.125 - 5352.125 & 5304.625 - 5315.875 \\ 
       &     &                     & 5353.375 - 5363.375 \\
Fe5406 & \AA & 5387.500 - 5415.000 & 5376.250 - 5387.500 \\ 
       &     &                     & 5415.000 - 5425.000 \\
   \hline
  \label{tab:lick_bands}
  \end{tabular}
 \end{center}
\end{table}

Three absorption features in our spectral range, namely \hb, Fe5015 and \mgb, can be significantly altered due to `infilling' by possible \hb, \oiii\/ and \nii\/ emission features, respectively. Before measuring the absorption-line strengths, we therefore subtract the emission spectrum as derived in Section \ref{sec:gas} from the original data. Finally, the absorption-line strengths are calibrated to the classical Lick/IDS system \citep[e.g.][]{trager98}. Errors are estimated via 100 Monte Carlo simulations of our complete line-strength analysis, using the output `cleaned' spectrum from the Monte Carlo simulations of the emission line analysis (see Sections \ref{sec:stellar_kin} and \ref{sec:gas}). In this way, we include the uncertainty in our emission line correction and stellar kinematics directly. For $S/N \sim 60$, typical errors for the indices \hb, Fe5015, \mgb, Fe5270, Fe5335 and Fe5406 are 0.15~\AA, 0.3~\AA, 0.15~\AA, 0.15~\AA, 0.17~\AA, and 0.13~\AA\/ respectively.

Below we describe the steps taken to calibrate our line-strength measurements to the Lick system. To account for the difference in spectral resolution of our observations and the Lick/IDS system, and to facilitate comparison with other authors, we convolve our spectra using a wavelength-dependent Gaussian kernel based on a linear interpolation of the resolution values given in the appendix of \citet{worthey97}.

\subsubsection{Velocity Broadening Correction}

The absorption lines of integrated galaxy spectra are broadened by the collective velocity distribution of the stars along the line of sight, usually approximated by the velocity dispersion, $\sigma$. The effect of this broadening is generally to weaken an index by lowering the surrounding continuum bands, as well as contaminating the central pass-band with contributions from neighbouring absorption features. The conventional correction for this effect involves broadening a collection of observed stellar spectra and measuring the Lick indices as a function of $\sigma$ only. The size of this correction can vary strongly as a function of $\sigma$ and, for some indices, with the intrinsic strength of the absorption feature. In addition, deviations from a Gaussian velocity distribution can also introduce a small systematic error following this approach \citep{kuntschner04}.\looseness-2

We therefore refine this method by using the optimal template derived for the stellar kinematics to determine the velocity broadening correction of each spectrum (as in Paper VI). The correction is given by comparing the index values of the unbroadened optimal template with those of the template broadened by the derived velocity distribution, including higher-order moments. The optimal templates derived by pPXF provide very good representations of the galaxy spectra in our sample, thus the method accounts directly for dependence on intrinsic line strength, as well as the detailed shape of the velocity distribution, without relying on average corrections determined from a range of stellar or model templates.\looseness-2

\subsubsection{Lick Offsets}

The Lick system is based on non-flux-calibrated spectra, and it is therefore necessary to account for offsets in the measured line indices caused by small differences in the local continuum shape of our spectra when compared with those of the (non-flux calibrated) Lick system. This is done by observing a number of stars in common with the Lick library, and comparing the measured indices with those in the appendix of \citet{worthey94}.

During the \oasis\/ observing campaign, a total of 31 Lick standard stars were observed, including 8 repeat observations between one run or more, and 3 repeat observations within runs. In total, there were 45 observations of stars in common with the Lick library. Fig.~\ref{fig:lick_offsets} presents the differences between the reference Lick values and our observations, using all measured stars (repeat measurements are included as separate points). Unfortunately there are too few repeat observations to fully evaluate night-to-night and run-to-run variations. However, the maximum difference between observations of the same star are consistent with the scatter of all measurements. From the distribution of points in Fig.~\ref{fig:lick_offsets}, the offsets measured from each run individually are self-consistent, suggesting minimal run-to-run biases. The final offsets are given in Table \ref{tab:lick_offsets}, which have been added to the measured indices presented here unless otherwise stated.

In general, the determined offsets are small, and consistent within the 1$\sigma$ errors of those found in Paper VI, where many more stars were used. The exception to this is Fe5015, which shows an offset larger than expected from \sauron, consistent only at the 2$\sigma$ level. For all indices, a number of stars are clear outliers of the distribution. Often these are M-type stars, where small errors in the velocity measurement (required to place the bandpasses at the appropriate rest wavelengths) can give rise to large changes in the measured index. The mean offset is determined with a biweight estimator \citep{hoaglin83}, and so is robust to these deviant points.

\begin{table}
 \begin{center}
  \caption{Estimate of Lick offsets derived from all stars.}
  \begin{tabular}{lc}
  \hline
   Index  & Offset    \\
   \hline
\hb    & -0.10 $\pm$  0.03~\AA \\
Fe5015 & ~0.45 $\pm$  0.07~\AA \\
\mgb   & -0.02 $\pm$  0.03~\AA \\
Fe5270 & -0.04 $\pm$  0.03~\AA \\
Fe5335 & -0.03 $\pm$  0.04~\AA \\
Fe5406 & ~0.05 $\pm$  0.04~\AA \\
   \hline
  \label{tab:lick_offsets}
  \end{tabular}
 \end{center}
 Note: For each index, the biweight mean and dispersion estimate was used to reduce the influence of outlying values. The error on the mean offset derived from the $N$ observations is given as the dispersion scaled by $1/\sqrt{N}$.
\end{table}

\begin{figure*}
 \begin{center}
  \includegraphics[width=12cm, angle=90]{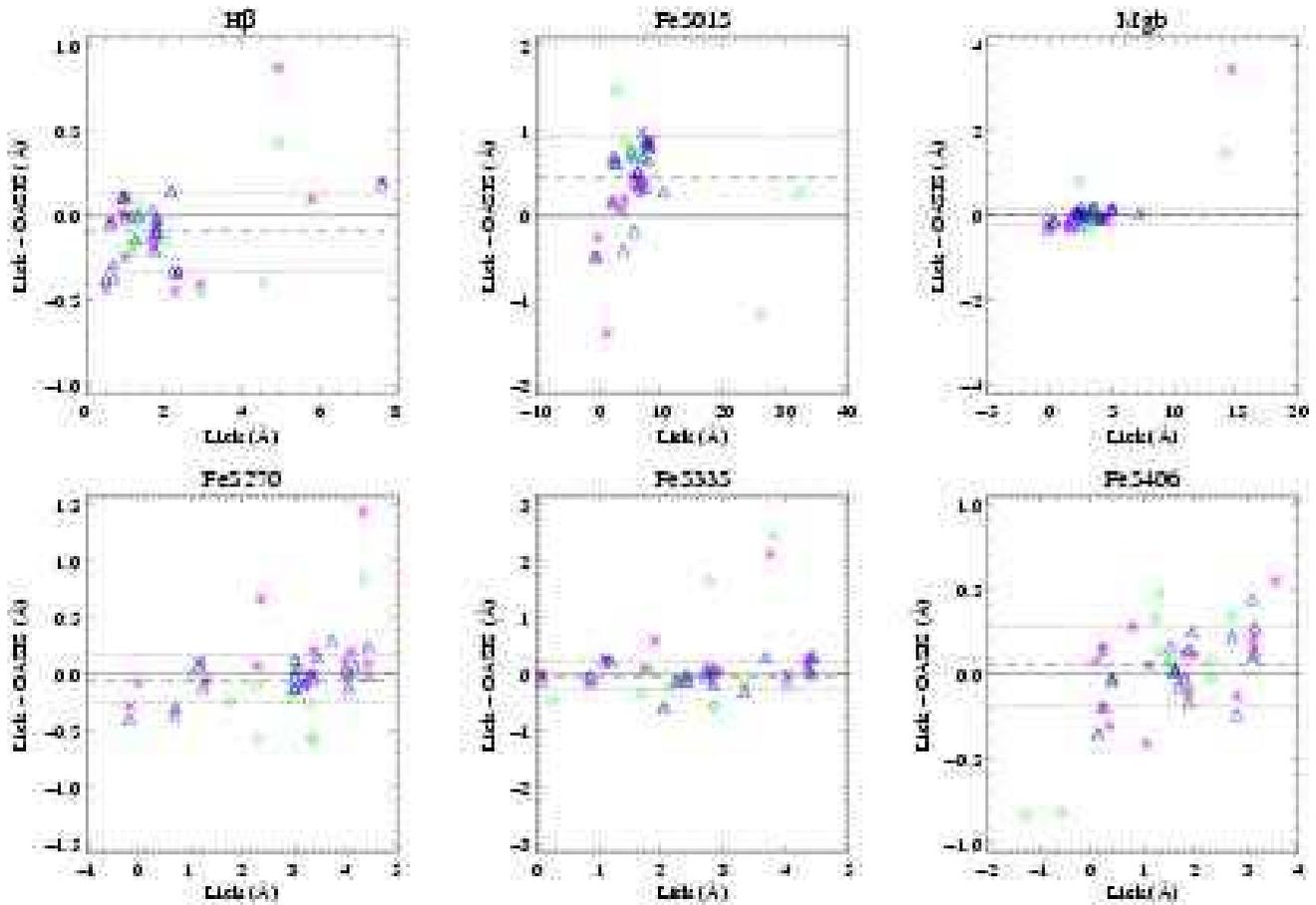}
 \end{center}
 \caption{Measured differences between Lick index reference values \citep{worthey94} and \oasis\/ observations for the six indices in our spectral range. Different coloured symbols indicate stars observed on different runs as in Fig.~\ref{fig:sigma_comp}. Repeat measurements of the same star are included as separate points. The biweight mean and robust dispersion estimates based on all measurements are indicated by the dashed and dotted lines respectively. Values of the biweight mean and the error on this mean are given in Table~\ref{tab:lick_offsets}.}
 \label{fig:lick_offsets}
\end{figure*}

\subsection{Comparison with \sauron\/ line strengths}
\label{sec:sau_comp_ls}

As for the comparisons with \sauron\/ data presented in Sections \ref{sec:sau_comp_kin} and \ref{sec:sau_comp_gas}, the line-strengths are best compared when derived from spectra within the same circular aperture from both data sets. Fig.~\ref{fig:lick_index_comp} presents a comparison of \oasis\/ and \sauron\/ line-strength measurements within the same 4\arcsec\/ aperture used before. Due to truncation of the \sauron\/ wavelength range by the instrument filter, only three lines can be measured within this aperture for all galaxies: \hb, \mgb\/ and Fe5015. The measurements are made after the emission lines have been removed as described in Section \ref{sec:gas}. Given the sensitivity of line-strength measurements to systematic errors, as well as the inherent difficulty of separating the emission lines from the stellar absorption features, the agreement between the two data sets is remarkably tight. For \hb\/ and \mgb, the (biweight) mean difference and standard error on the mean are $-0.09 \pm 0.03$~\AA\/ and $0.05 \pm 0.04$~\AA\/ respectively. For Fe5015, the agreement is less good, with a mean difference of $-0.36\pm 0.05$~\AA, and evidence for some systematic behaviour.\looseness-2

\begin{figure*}
 \begin{center}
  \includegraphics[width=5.8cm, angle=0]{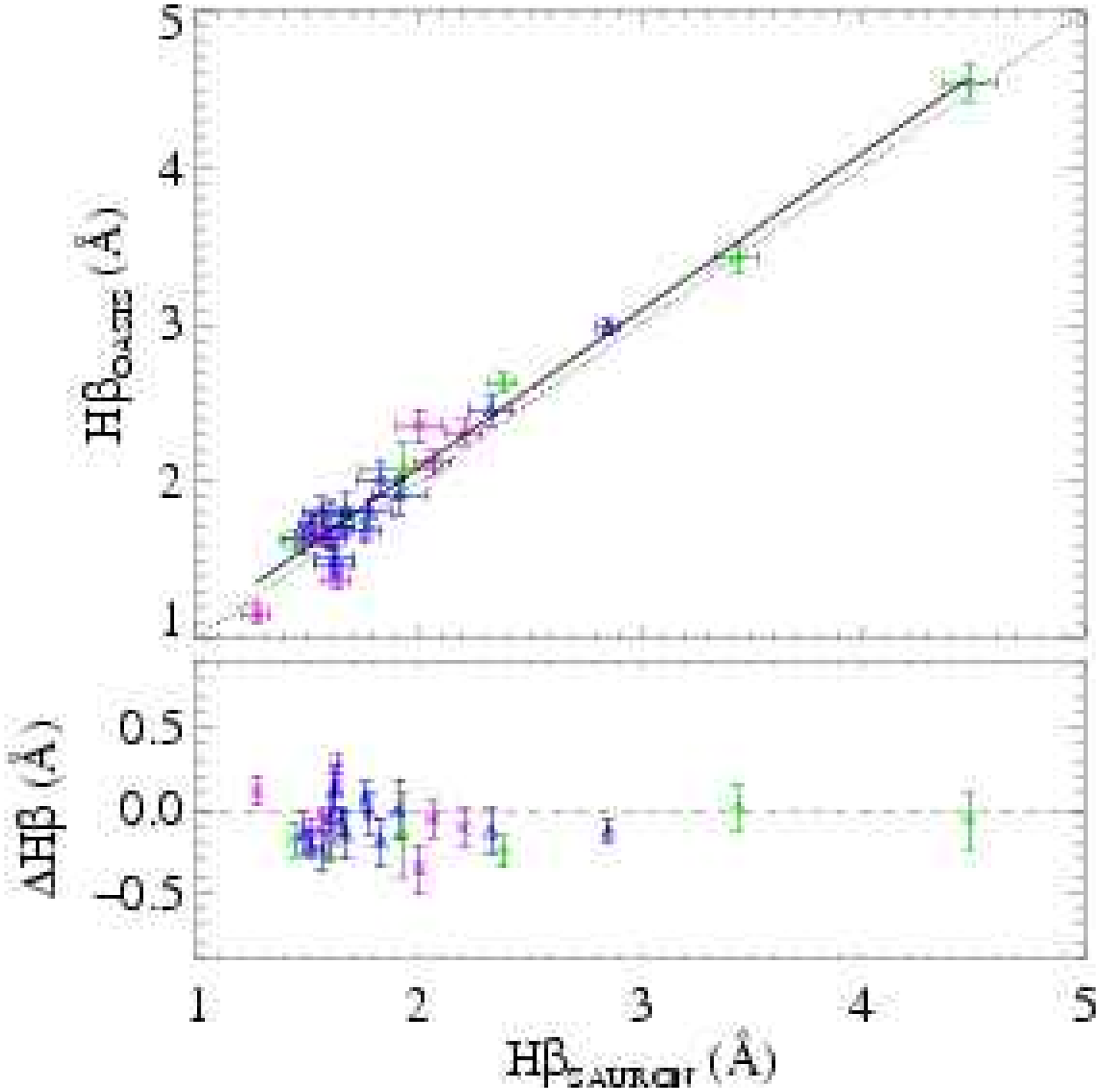}
  \includegraphics[width=5.9cm, angle=0]{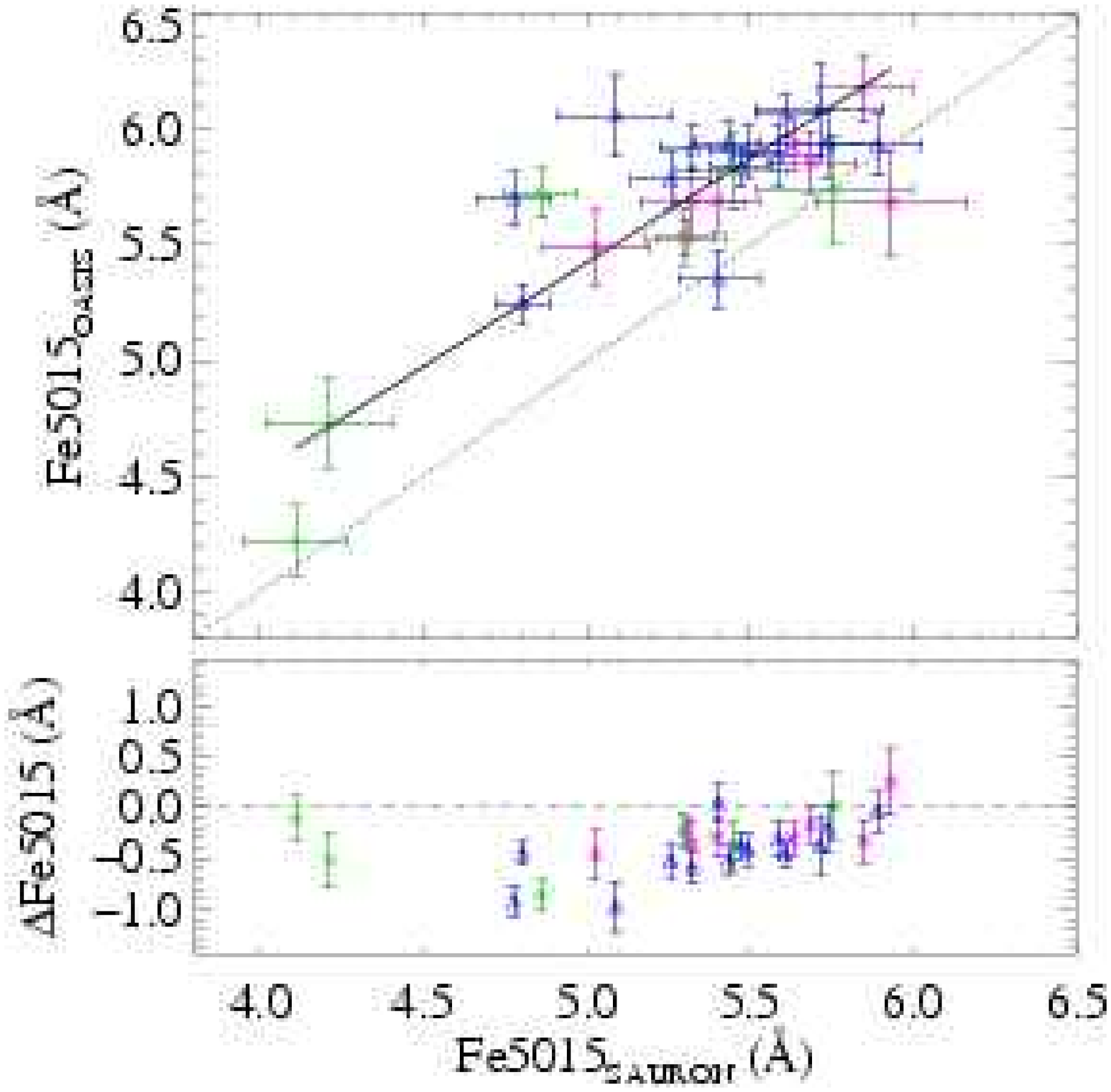}
  \includegraphics[width=5.8cm, angle=0]{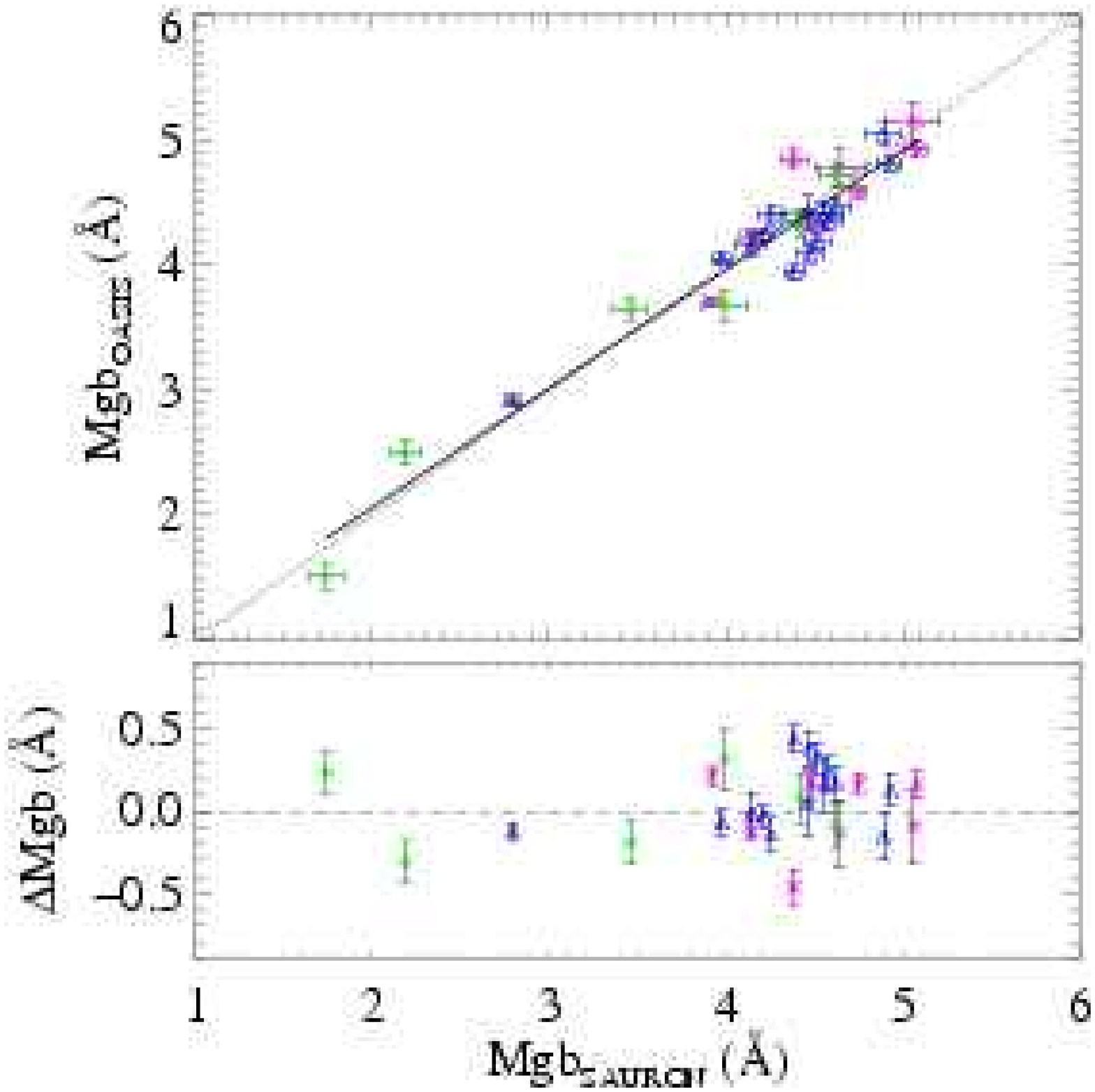}
 \end{center}
 \caption{Comparison of \sauron\/ and \oasis\/ line-strength measurements. Different coloured symbols indicate galaxies observed on different runs as in Fig.~\ref{fig:sigma_comp}. Dotted lines show the identity relation; solid lines indicate robust straight-line fits, from which slopes are derived. Since both data sets are flux calibrated, no Lick offsets are applied to the values in this comparison. For both \hb\/ and \mgb, the mean differences are consistent with zero, with the fitted slopes consistent with no measurable systematic trends ($1.0\pm0.04$ and $0.95\pm0.05$ respectively). The agreement is less good for Fe5015, which gives a fitted slope of $1.0\pm0.09$, but with a constant mean offset of -0.36$\pm$0.05\AA.}
 \label{fig:lick_index_comp}
\end{figure*}

\subsection{Age, Metallicity and Abundance Ratio}
\label{sec:ssp}

The purpose of measuring the line-strength indices is ultimately to understand the distribution of stellar ages, metallicities and element abundances in these galaxies. Although certain features can be more sensitive to these quantities than others, there is always a certain degree of degeneracy between the population parameters and single absorption features. By using a combination of line-strength indices, it is possible to constrain stellar population models, which in turn allow us to interpret the distribution of line-strengths as one of age, metallicity and element abundance.

In order to determine these physical quantities, we use the single-burst stellar population models of \citet{thomas03}. We interpolate the original grid of models to produce a cube of $40 \times 40 \times 40 = 64000$ individual models, spanning [Z/H] $= -0.33$ to $+0.67$ (in steps of 0.026 dex), age $= 0.1$ to 15~Gyr (logarithmically, in steps of 0.056 dex), and \afe\/ $= -0.2$ to 0.5 (in steps of 0.018 dex). We locate the model which lies closest to our six measured Lick indices simultaneously for each bin in our data using the $\chi^2$ technique, similar to the method described in \cite{proctor04}. We estimate the confidence levels on the derived parameters using the $\Delta \chi^2$ values, which were checked to be consistent with those found from our Monte-Carlo simulations. For measurements which lie outside of the model population parameter-space, values on the closest boundary are chosen.

Fig.~\ref{fig:pop_grid} shows an example of the $\Delta\chi^2$ contours obtained with this method, indicating the minimum and confidence levels. This shows that the typical 1$\sigma$ uncertainties we obtain on the parameters of metallicity, abundance ratio and age are of the order 0.1 dex. The effect of the age-metallicity degeneracy \cite[e.g.,][]{worthey94} can clearly be seen as a tilt in the contours of the right-most panel of Fig.~\ref{fig:pop_grid}.

We note here that the approach of using SSP models is intrinsically limited to providing only luminosity-weighted properties. Moreover, this weighting does not affect all indices equally: for example, \hb\/ is strongly enhanced at younger ages (0.5 - 5~Gyr), while other diagnostic features like Mg and Fe are less so. As a result, it is not certain that the different SSP parameters derived (i.e. age, [Z/H], \afe) are weighted in the same way from the different populations that may be present. One must therefore be cautious in interpreting the maps of population parameters, since the different maps may be sensitive to different populations.

\begin{figure*}
 \begin{center}
  \includegraphics[height=16cm, angle=90]{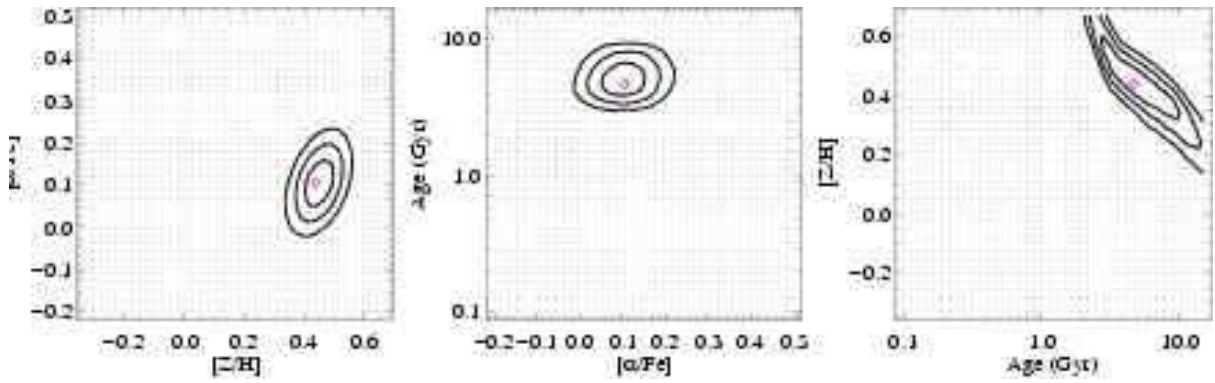}
 \end{center}
 \caption{Example of $\Delta\chi^2$ space for finding the best-fit model to the lines strengths of NGC\,4564, as measured from the 4\arcsec\/ aperture spectrum described in previous sections. The best-fit model is found by minimizing the difference between the model predictions and the six Lick indices from our data, measured using a $\chi^2$ approach. The three panels show different projections of the three-dimensional solution space (age, metallicity [Z/H], abundance ratio \afe), with one of the parameters fixed at the best-fit value. The thick contours indicate the 1$\sigma$, 2$\sigma$ and 3$\sigma$ confidence levels, based on $\Delta \chi^2$ for two degrees of freedom. We note that in the plane of age and metallicity (far right panel), the effect of the age-metallicity degeneracy can be seen as the tilt of the contours with respect to the plot axes.}
 \label{fig:pop_grid}
\end{figure*}

\section{Results}
\label{sec:results}

Maps for all measured quantities (stellar kinematics, ionised gas fluxes and kinematics, Lick indices and stellar population parameters) are presented in Fig. \ref{fig:ngc1023}. To help give an impression of our measurement errors, which can be difficult to infer only from the scatter in the parameter maps themselves, we present a simulated major-axis `slit' of NGC\,3489 in Appendix~\ref{appendix:a}, showing each of the measured and derived quantities and their individual error bars. There is a wealth of structure in the parameter maps, much of which is unseen or poorly resolved in our \sauron\/ data. For a comprehensive description of the global properties and bibliographic references of the galaxies presented here, the reader is directed to Papers III, V and VI of this series. In this section, we present an overview of the central structures found in the various parameter maps, considering the stellar kinematics, gas properties, and stellar populations in turn.

\subsection{Stellar Kinematics}
\label{sec:results_kin}

The variety of stellar motions observed in the \sauron\/ sample can be broadly separated into two categories: objects that exhibit rotation (and possible counter-rotation) around a {\em single} axis; and objects that have a {\em varying} axis of rotation. Based on the large (kpc) scales probed by the \sauron\/ observations, Emsellem et al. (Paper IX, in preparation) refine this picture, and quantify the behaviour in terms of the specific angular momentum of the galaxy, finding that fast-rotating galaxies exhibit a rotation axis which generally does not vary; whereas slowly-rotating objects tend to show misaligned components. The differentiation between fast and slow rotating galaxies is visually apparent from the velocity maps in Paper III. On the scales probed by \oasis, we find that a number of objects, while kinematically well-aligned on large-scales, may still exhibit misaligned central components. For this reason, we divide the following qualitative discussion between galaxies exhibiting misaligned and aligned rotation within the \oasis\/ maps, rather than between fast and slow rotating objects.\looseness-2

In addition to the maps, we present results from our kinemetric analysis in Appendix~\ref{appendix:b}, giving for each galaxy the kinematic position angle and maximum absolute rotation as a function of galactocentric radius. We use the kinemetric profiles of kinematic position angle (\kinpa) to help determine the velocity fields which show misaligned and aligned rotation. The kinemetric profiles are a useful tool to help quantify structures which can be approximately inferred from the maps, although, as with photometric profiles, one must also bear in mind the effects of seeing and spatial sampling when interpreting them. When the velocity gradient is low, the error in \kinpa\/ is also larger. With these caveats in mind, we consider galaxies as being aligned if the variations in \kinpa\/ are around 10\degree\/ or less within the \oasis\/ field of view. For $\Delta$\kinpa $> 10$\degree, the velocity field is considered misaligned, with strong misalignments ($\Delta$\kinpa\/ $\gsim 30$\degree) indicating a decoupled component. A more complete definition of kinematic structures found in our \sauron\/ and \oasis\/ maps will be given in a future paper of this series.\looseness-2

\subsubsection{Misaligned Rotation}

\noindent Several of the velocity maps presented in Fig. \ref{fig:ngc1023} show complex velocity fields, where the axis of rotation changes as a function of radius. This is illustrated by the variation in the \kinpa\/ profiles in Appendix~\ref{appendix:b}, which range from smooth variations of a few degrees, to an abrupt change of many tens of degrees. In addition, some galaxies rotate around an axis which shows little variation in orientation, but which is misaligned with respect to the local photometric axes.  Such features may indicate the presence of a distinct dynamical sub-component, such as a bar, or may be indicative of a triaxial potential \citep[e.g.][]{statler91}.

Differentiating between these alternatives is not always straightforward, especially when probing the limits of spatial and spectral resolution. The orientation of the system also plays an important role in determining the projected quantities we observe. With this in mind, we identify several manifestations of kinematic misalignment present in our data, varying from subtle `twists' in the zero-velocity contour (ZVC), to spatially distinct components with an apparently different orbital composition from that of the rest of the galaxy. Moreover, some objects appear regularly rotating in the \oasis\/ field of view, but are misaligned at the larger scales observed with \sauron. These objects are included in this discussion, as the \oasis\/ observations can provide a more detailed perspective on the nature of the central component.\looseness-2

The following galaxies show mild twists (10\degree\/ $< \Delta$\kinpa\/ $< 20$\degree) in their ZVC within the \oasis\/ field of view: NGC\,1023, NGC\,3489, NGC\,4459 and NGC\,4552. In the case of NGC\,1023, the axis of rotation appears to change gradually as a function of radius even to the edge of the \sauron\/ field (Paper III). In the other cases, the twist is more localised to the central regions. The strongly barred galaxies NGC\,2699 and NGC\,4262 show a single axis of rotation that is constantly {\em misaligned} with the local photometric principle axes, though the central kinematic axes align quite well with the large scale photometry.

More pronounced twisting of the ZVC is found in both NGC\,2768 ($\Delta$\kinpa $\sim 40$\degree) and NGC\,5982 ($\Delta$\kinpa $\sim 30$\degree). The central regions of these objects show a change in rotation axis from minor- to major-axis. This change is clearer in NGC\,5982, where the central component is quite distinct from the outer body, showing a peak in the profile of maximum rotation (\ki) in Appendix~\ref{appendix:b}. In NGC\,2768, the rotation increases smoothly as a function of radius, with no distinct rotation component in the centre. The velocity dispersion, however, shows a clear dip within the region of the twisted ZVC, showing that the central kinematics are different from the outer parts.\looseness-2

These two objects mark something of a transition between twisted velocity fields and decoupled components. In the case of mild twists, the point at which the galaxy exhibits a dynamically distinct component is almost impossible to identify with any degree of certainty, at least from our data alone. In NGC\,2768, the twist is stronger and localised within 1\arcsec\/ of the centre. The link with a sharp dip in the velocity dispersion gives the strong impression that the orbital distribution is indeed different from the rest of the galaxy. In NGC\,5982, the case for a distinct central component is stronger, given that there is a peak in the rotation curve, coincident with some lowering of the velocity dispersion, suggesting a possible dynamically cold component. For this specific galaxy, \citet{statler91} highlighted the important issue that a triaxial system with a smooth distribution function, when seen in projection, can naturally give rise to a velocity field very similar to NGC\,5982 \citep[see also][]{oosterloo94}. Whether or not components are actually decoupled in phase-space requires the application of dynamical models \citep[such as][]{cappellari02,verolme03,statler04}, which is beyond the scope of this paper.\looseness-2

Galaxies which have clearly distinguished misaligned components (i.e. $\Delta$\kinpa $\geq 40$\degree), which we will term kinematically decoupled components (hereafter KDCs), include NGC\,3032, NGC\,4382, NGC\,4621, and NGC\,5198. Of these, the misaligned components in NGC\,3032 and NGC\,4382 were not resolved in our \sauron\/ data (Paper III) and are reported here for the first time. Photometrically, NGC\,4382 exhibits a peculiar core region, showing a central drop in surface brightness that is not explained by dust \citep{lauer05}. Figure~\ref{fig:ngc4382} shows this galaxy has a residual rectangular central structure in the unsharp-masked {\it HST} image, which is slightly misaligned with the outer isophotes (by $\sim 10$\degree). This boxy structure is spatially coincident with the KDC and oriented in a similar direction, suggesting that the two components are directly related.

NGC\,4621 is also visible as a KDC in our data. It was first discovered by \citet{wernli02} using calcium II triplet data taken with \oasis\/ (CFHT) using adaptive optics. NGC\,4150 shows evidence of a KDC in the \sauron\/ data of Paper III, but the poor seeing conditions during our \oasis\/ observations preclude an accurate measurement of this component in our data. With the exception of NGC\,5198, each of these apparently small KDCs reside in galaxies with otherwise regular, aligned rotation at larger radii. The flattest of these objects, NGC\,4621, shows clear evidence that the KDC is embedded in a dynamically cold disk, visible in the residuals of the unsharp-masked {\it HST} image in Fig.~\ref{fig:ngc4621} \citep[see also][]{krajnovic04}.

NGC\,3414, NGC\,3608, NGC\,5813, and NGC\,5831 contain large KDCs, clearly visible in the \sauron\/ observations (Paper III). These decoupled components fill the \oasis\/ field of view, and generally show regular rotation, with some possible twisting of the ZVC as the field encroaches on the outer component beyond the \oasis\/ field of view. Both NGC\,3414 and NGC\,5813 exhibit strong rotation and anti-correlated $h_3$ fields, as well as depressed velocity dispersion where the rotation is highest, indicating that these KDCs are disk-like.

The variety of objects with misaligned rotation emphasises the fact that such behaviour can result from a number of different intrinsic orbital structures and possible sub-components. In some cases, a large-scale bar structure gives rise to the twisted velocity field; in others, the decoupled component takes the form of a disk embedded in a slowly-rotating spheroid; and in some galaxies, the decoupled component cannot be resolved even at sub-arcsecond scales, making the intrinsic structure unclear. Grouping these objects together belies the different origins of the misalignment (see also \ref{sec:discussion}), and also highlights the difficulty in separating galaxies containing distinct dynamically decoupled components from those having a complex, but smoothly varying orbital structure. With the application of new dynamical modelling techniques to study galaxies with twists and KDCs, we hope to obtain a clearer understanding of these objects' orbital structure \citep[e.g.][van den Bosch et al., in preparation]{vdven04}.\looseness-2

\subsubsection{Aligned Rotation}
\label{sec:res_stellar_kin_align}

The remaining objects in our sample show little or no evidence of misaligned stellar rotation in the \oasis\/ field of view. The two giant elliptical galaxies, NGC\,4486 and NGC\,5846, are included here, as they show almost no rotation. The strongly barred galaxy, NGC\,3384, is also included here, as the rotation is very symmetric around the local photometric minor axis. Due to the morphology of the bar, the photometric PA shows some rapid changes as a function of radius, but becomes aligned with the central PA again in the outer parts \citep{busarello96}. Within our \oasis\/ and \sauron\/ fields, the rotation axis remains quite constant. In the \oasis\/ data, this galaxy exhibits evidence of a stellar disk, with strong rotation and anti-correlated $h_3$ fields, and a remarkable depressed velocity dispersion along the major axis. This is consistent with the disk-like residuals in the unsharp-masked {\it HST} image.

The remaining objects (NGC\,2549, NGC\,2695, NGC\,2974, NGC\,3379, NGC\,4473, NGC\,4526, NGC\,4564 and NGC\,5845) appear consistent with rotation around a unique axis ($\Delta$\kinpa\/ $\lsim 10$\degree), even on the larger scales probed by \sauron. The galaxies which show the most flattened isophotes within the \oasis\/ field - namely NGC\,4473, NGC\,4526 and NGC\,5845 - show strong evidence of harbouring central stellar disks. This evidence comes from disk-like residuals in the unsharp-masked {\it HST} image, strong rotation fields with `pinched' isovelocity contours, significant $h_3$ amplitude anti-correlated with the velocity, and lowered velocity dispersion in the region occupied by the disk. In this last characteristic, NGC \,4473 is peculiar, showing a {\em rising} velocity dispersion along the major-axis. Detailed dynamical modelling of this galaxy indicates that it is composed of two flattened, almost equal-mass, components rotating in opposite directions \citep{ cappellari05}. The central drop in the dispersion indicates where the contribution from the counter-rotating component becomes less significant.

\begin{figure*}
 \begin{center}
  \includegraphics[width=15cm]{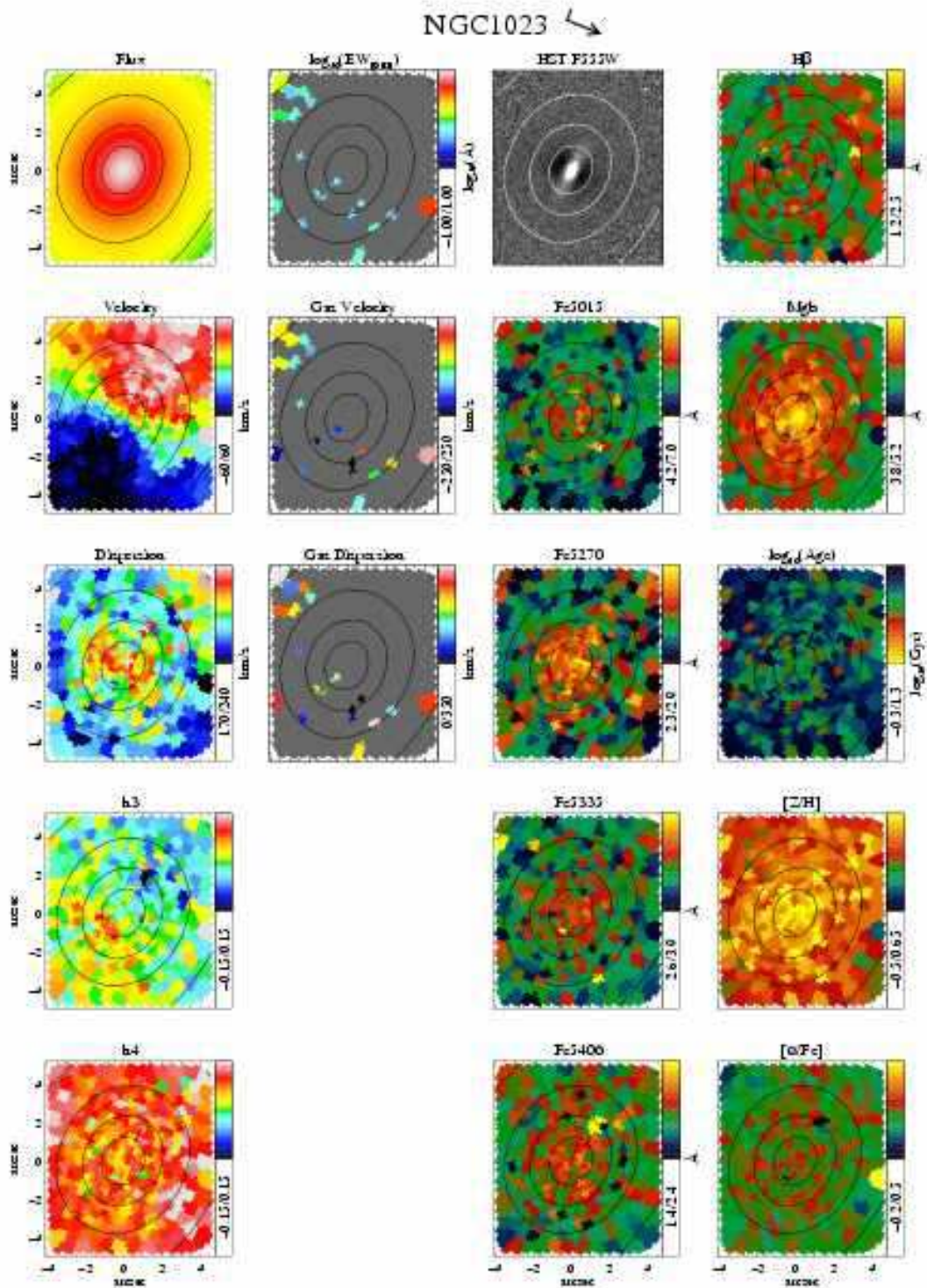}
 \end{center}
 \caption{The following figures show \oasis\/ maps of (from top to bottom, left to right): {\it first column - } reconstructed intensity, stellar velocity, stellar velocity dispersion, Gauss-Hermite coefficients $h_3$ and $h_4$, {\it second column - } \oiii\/ emission-line equivalent width, \oiii\/ gas velocity, \oiii\/ gas velocity dispersion, \hb\/ emission-line equivalent width, \oiii/\hb\/ emission line ratio, {\it third column - } unsharp-masked {\it HST} image (when available), Fe5015, Fe5270, Fe5335, Fe5406 Lick absorption-line indices, {\it fourth column - } \hb, and \mgb\ Lick absorption-line indices, luminosity-weighted stellar age, metallicity and abundance ratio. An arrow beside the figure title indicates north-east. The {\it HST} images are given as the ratio of the original image and the image convolved by a Gaussian kernel of 10 pixels. The overplotted isophotes are in half-magnitude steps. Grey bins in the emission-line maps indicate non-detections.}
 \label{fig:ngc1023}
\end{figure*}

\addtocounter{figure}{-1}

\begin{figure*}
 \begin{center}
  \includegraphics[width=15cm]{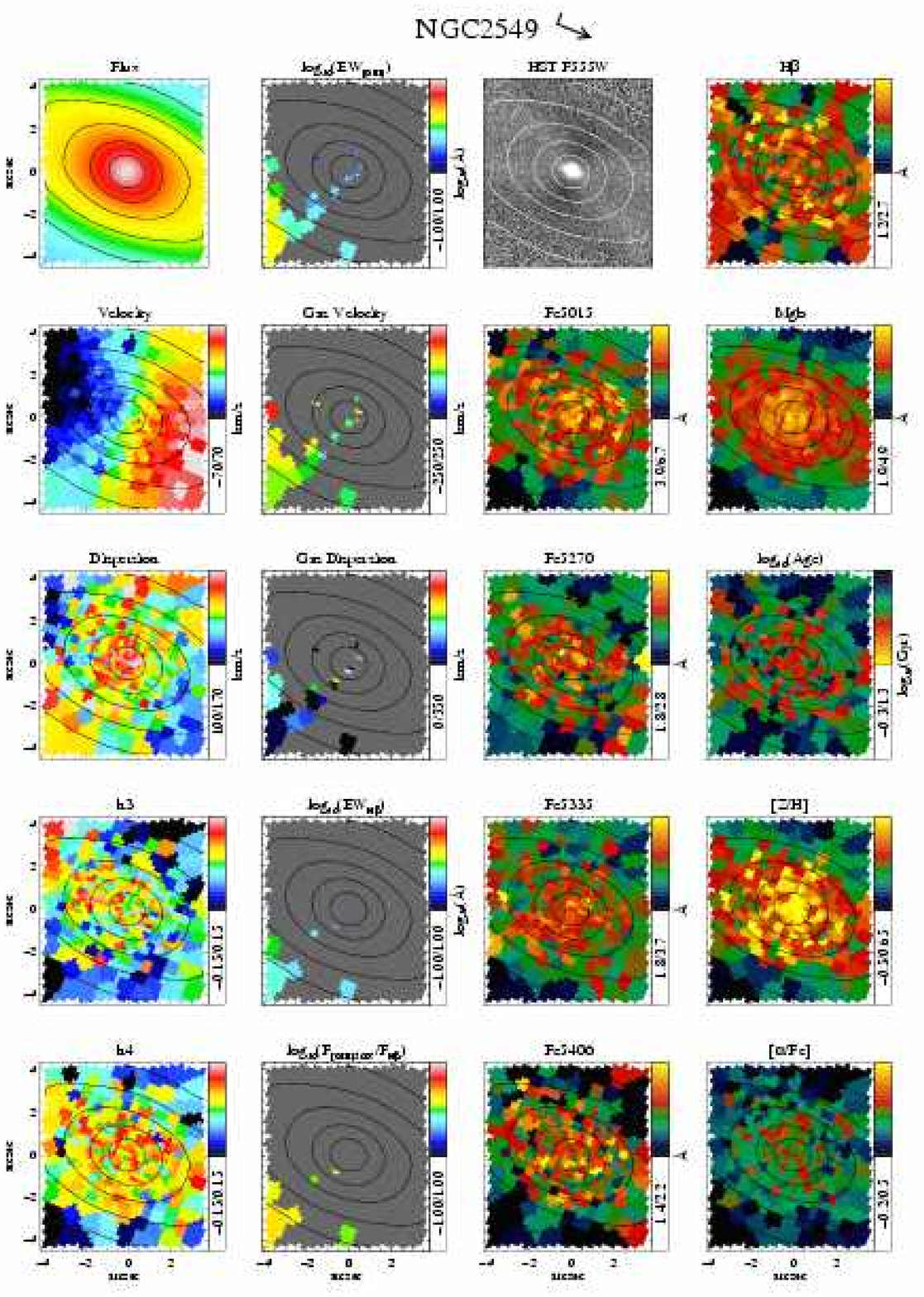}
 \end{center}
 \caption{{\it continued}}
 \label{fig:ngc2549}
\end{figure*}

\addtocounter{figure}{-1}

\begin{figure*}
 \begin{center}
  \includegraphics[width=16cm]{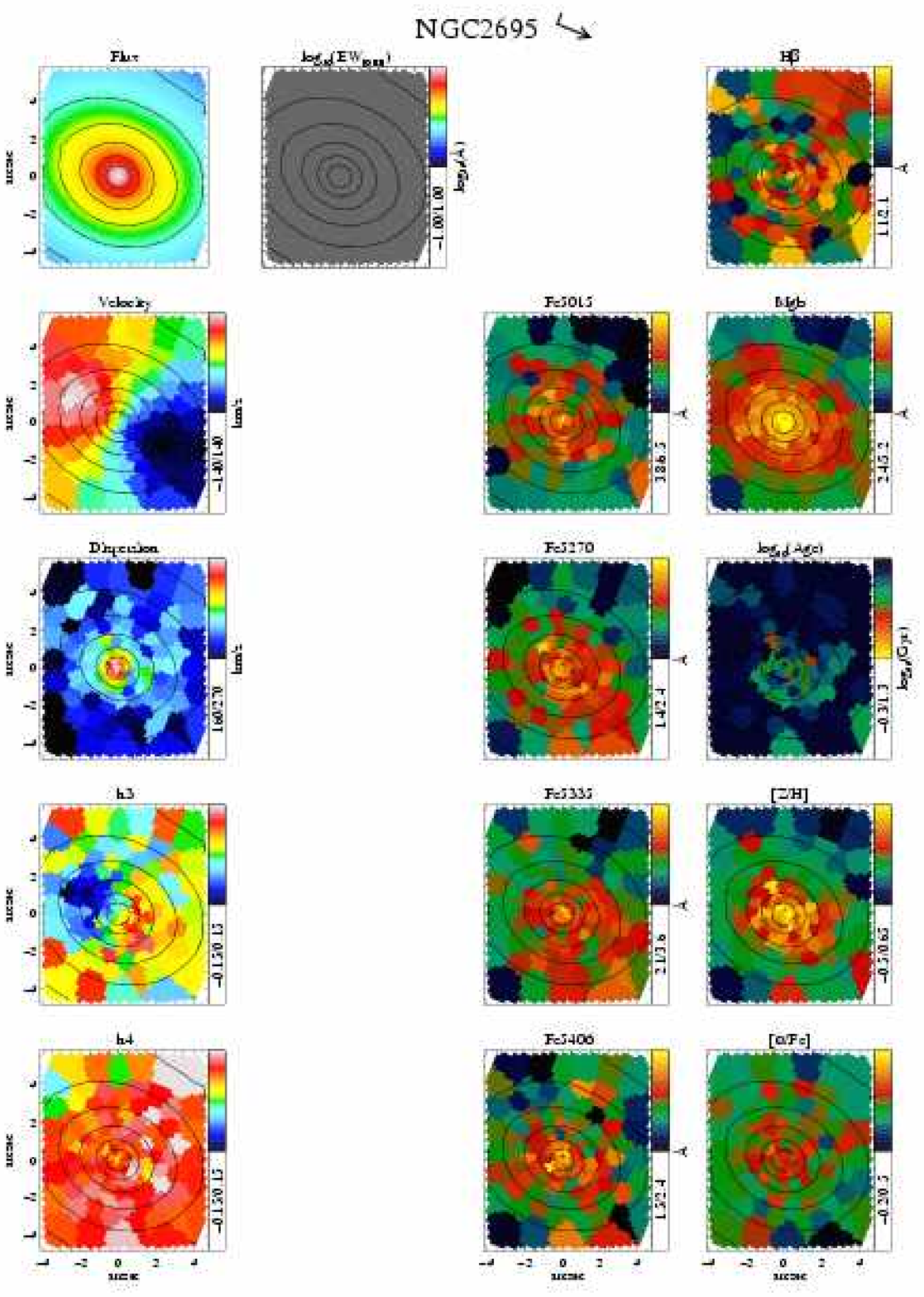}
 \end{center}
 \caption{{\it continued}}
 \label{fig:ngc2695}
\end{figure*}

\addtocounter{figure}{-1}

\begin{figure*}
 \begin{center}
  \includegraphics[width=16cm]{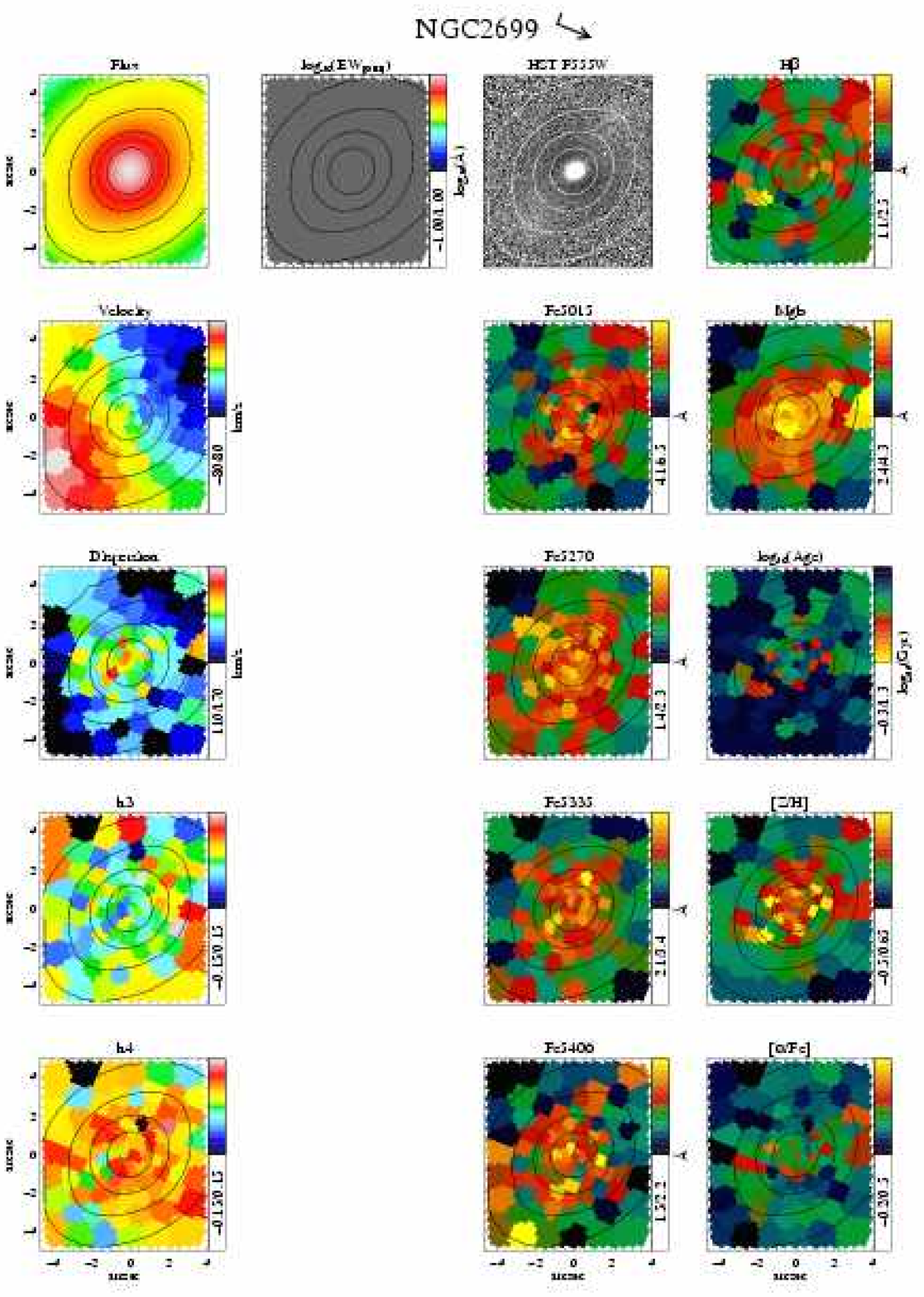}
 \end{center}
 \caption{{\it continued}}
 \label{fig:ngc2699}
\end{figure*}

\addtocounter{figure}{-1}

\begin{figure*}
 \begin{center}
  \includegraphics[width=16cm]{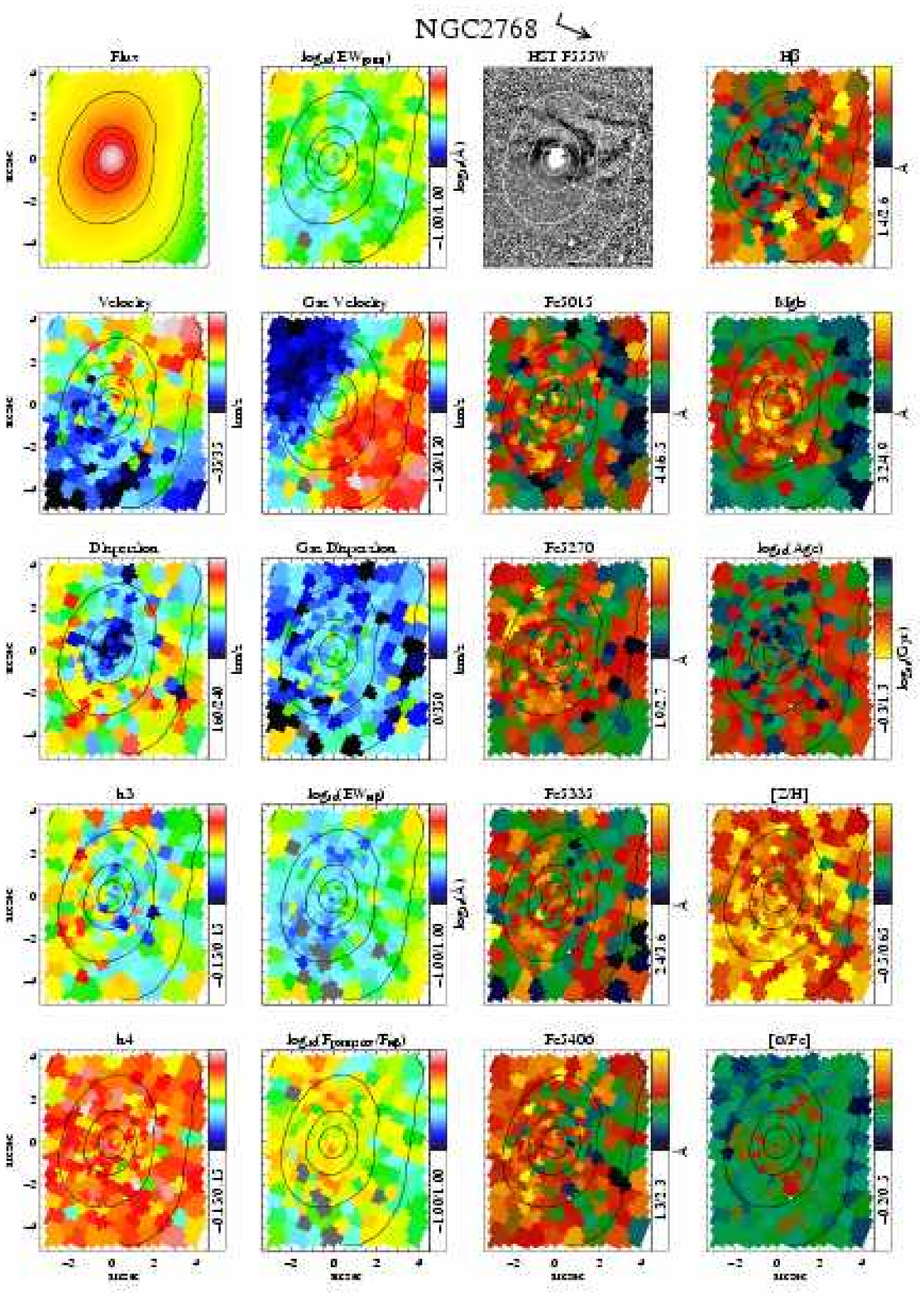}
 \end{center}
 \caption{{\it continued}}
 \label{fig:ngc2768}
\end{figure*}

\addtocounter{figure}{-1}

\begin{figure*}
 \begin{center}
  \includegraphics[width=16cm]{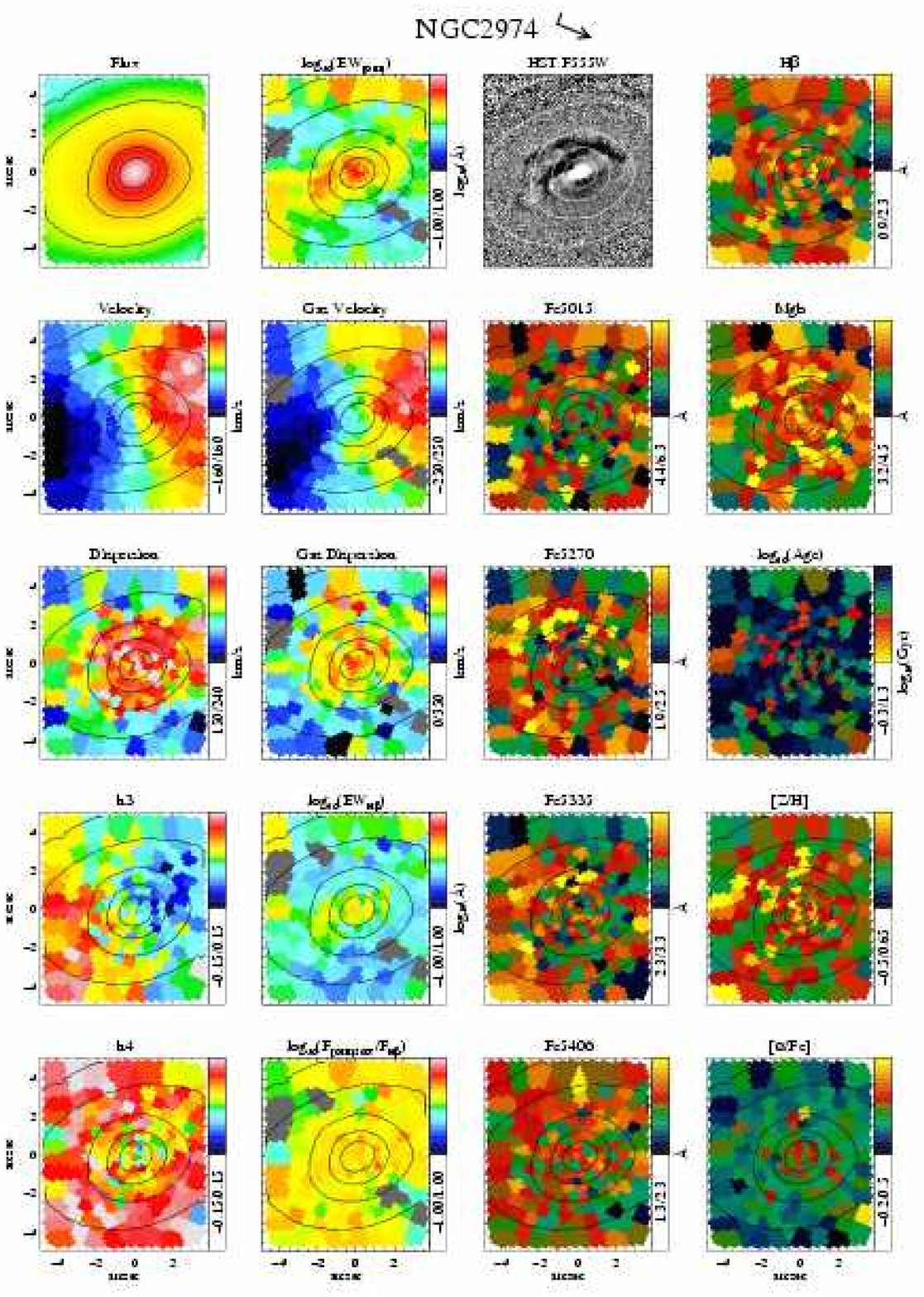}
 \end{center}
 \caption{{\it continued}}
 \label{fig:ngc2974}
\end{figure*}

\addtocounter{figure}{-1}

\begin{figure*}
 \begin{center}
  \includegraphics[width=16cm]{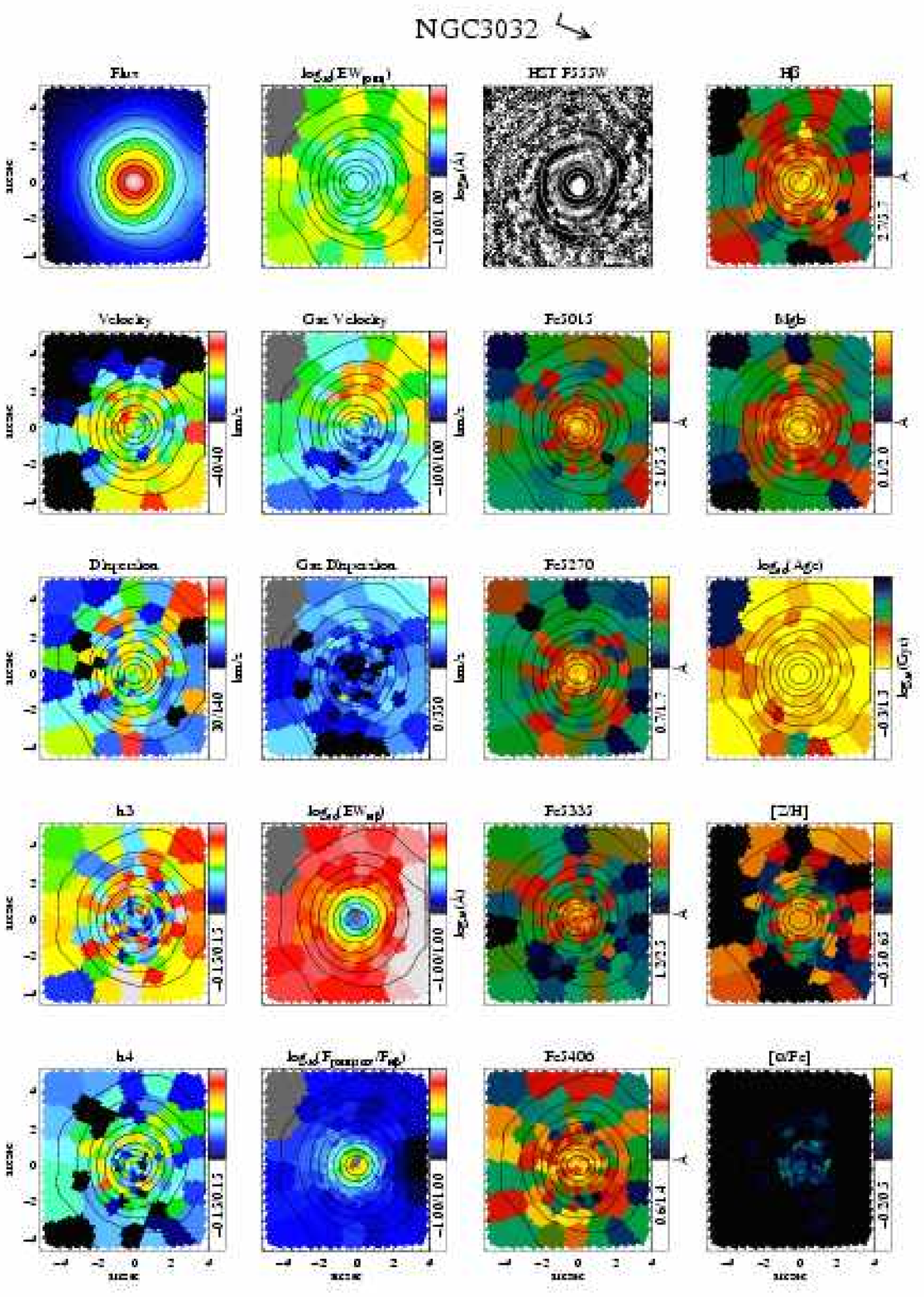}
 \end{center}
 \caption{{\it continued}}
 \label{fig:ngc3032}
\end{figure*}

\addtocounter{figure}{-1}

\begin{figure*}
 \begin{center}
  \includegraphics[width=16cm]{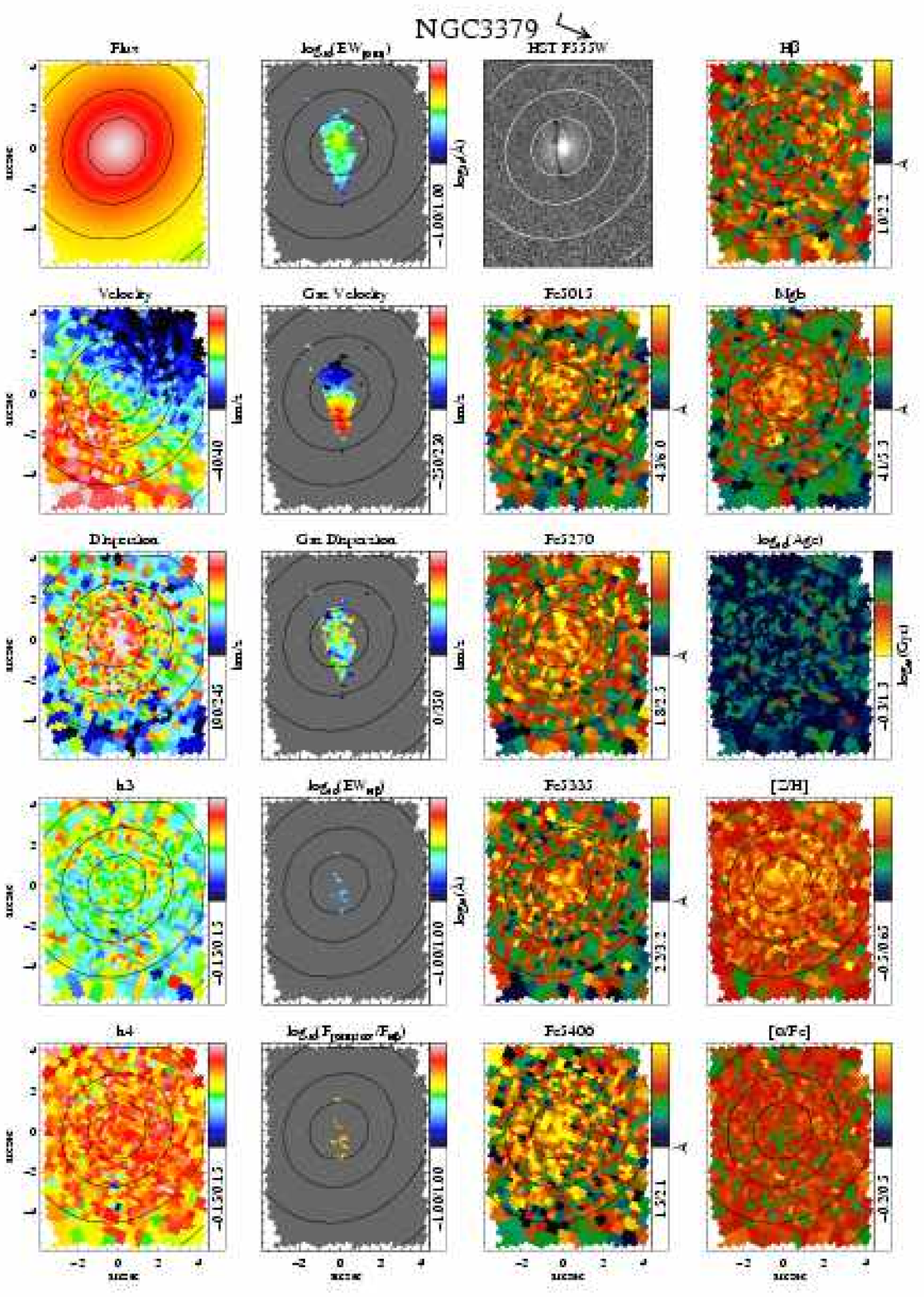}
 \end{center}
 \caption{{\it continued}}
 \label{fig:ngc3379}
\end{figure*}

\addtocounter{figure}{-1}

\begin{figure*}
 \begin{center}
  \includegraphics[width=16cm]{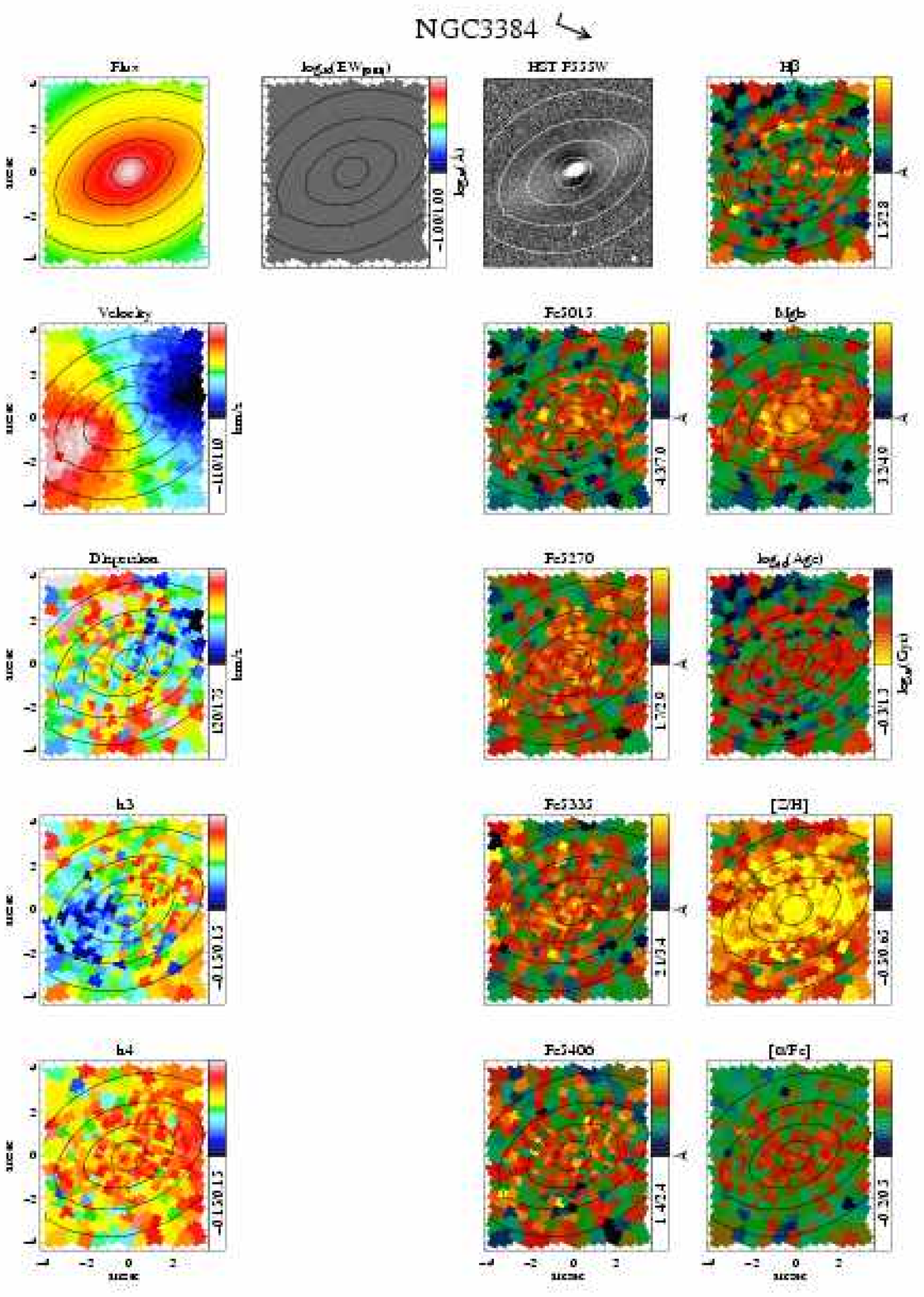}
 \end{center}
 \caption{{\it continued}}
 \label{fig:ngc3384}
\end{figure*}

\addtocounter{figure}{-1}

\begin{figure*}
 \begin{center}
  \includegraphics[width=16cm]{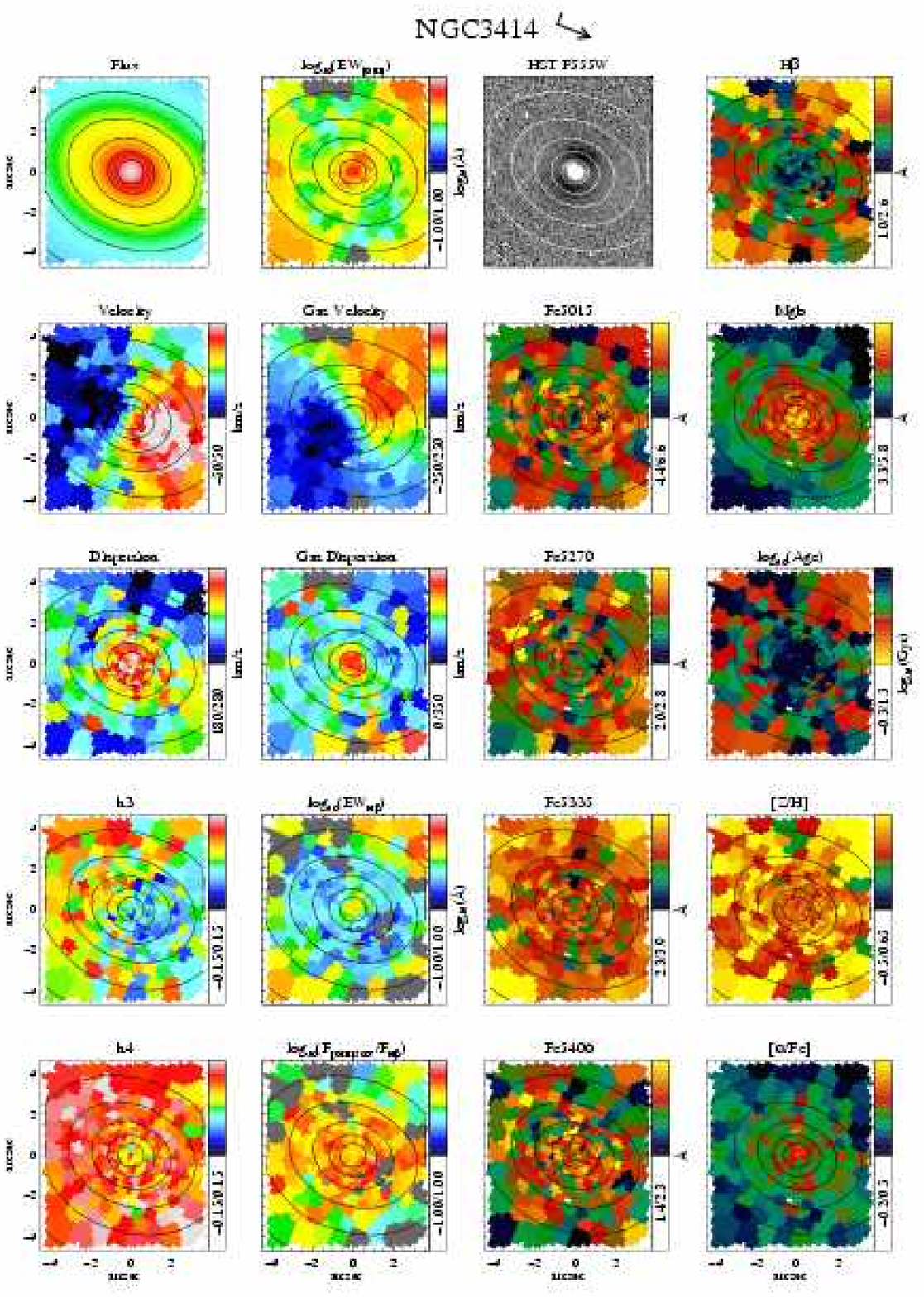}
 \end{center}
 \caption{{\it continued}}
 \label{fig:ngc3414}
\end{figure*}

\addtocounter{figure}{-1}

\begin{figure*}
 \begin{center}
  \includegraphics[width=16cm]{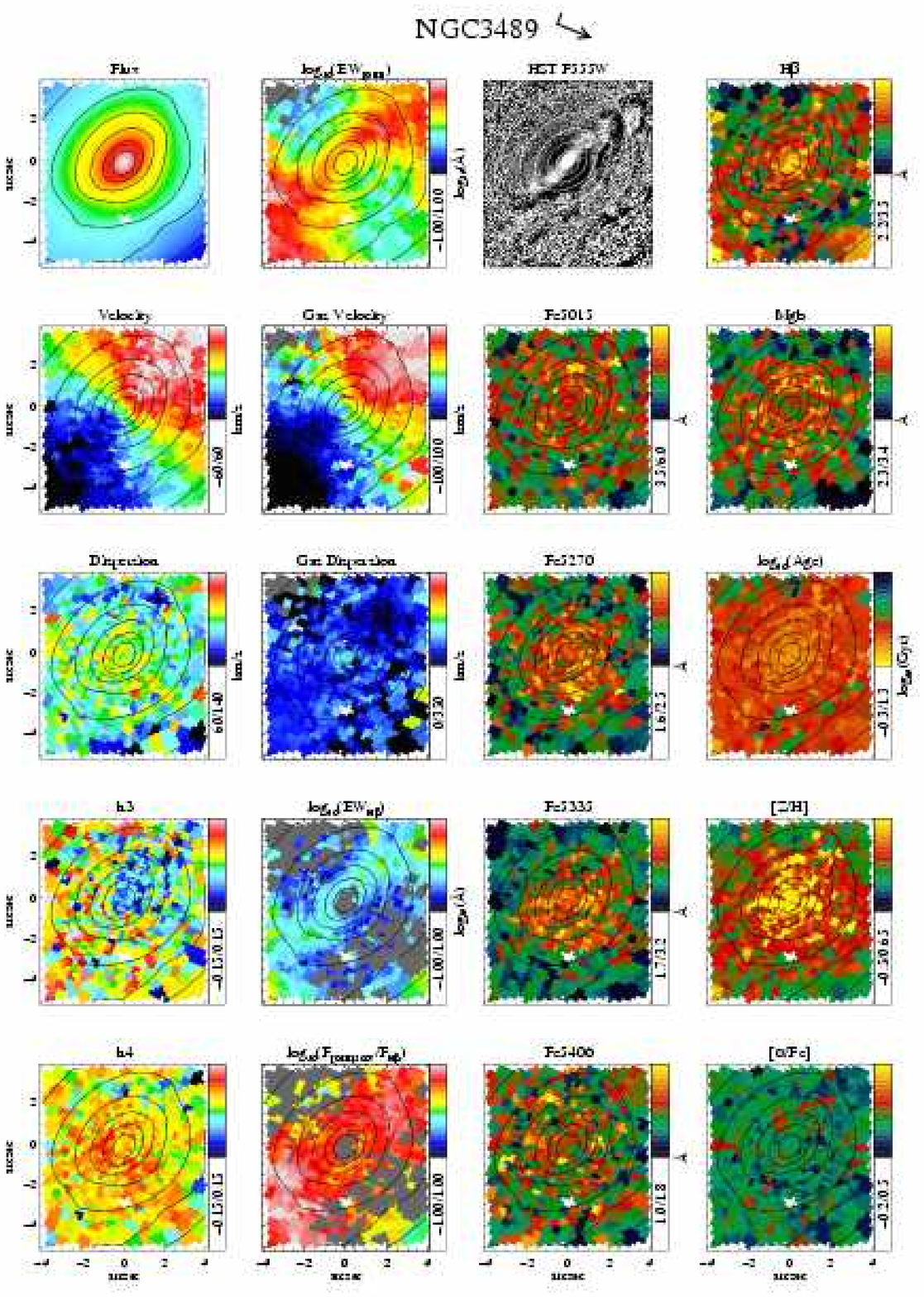}
 \end{center}
 \caption{{\it continued}}
 \label{fig:ngc3489}
\end{figure*}

\addtocounter{figure}{-1}

\begin{figure*}
 \begin{center}
  \includegraphics[width=16cm]{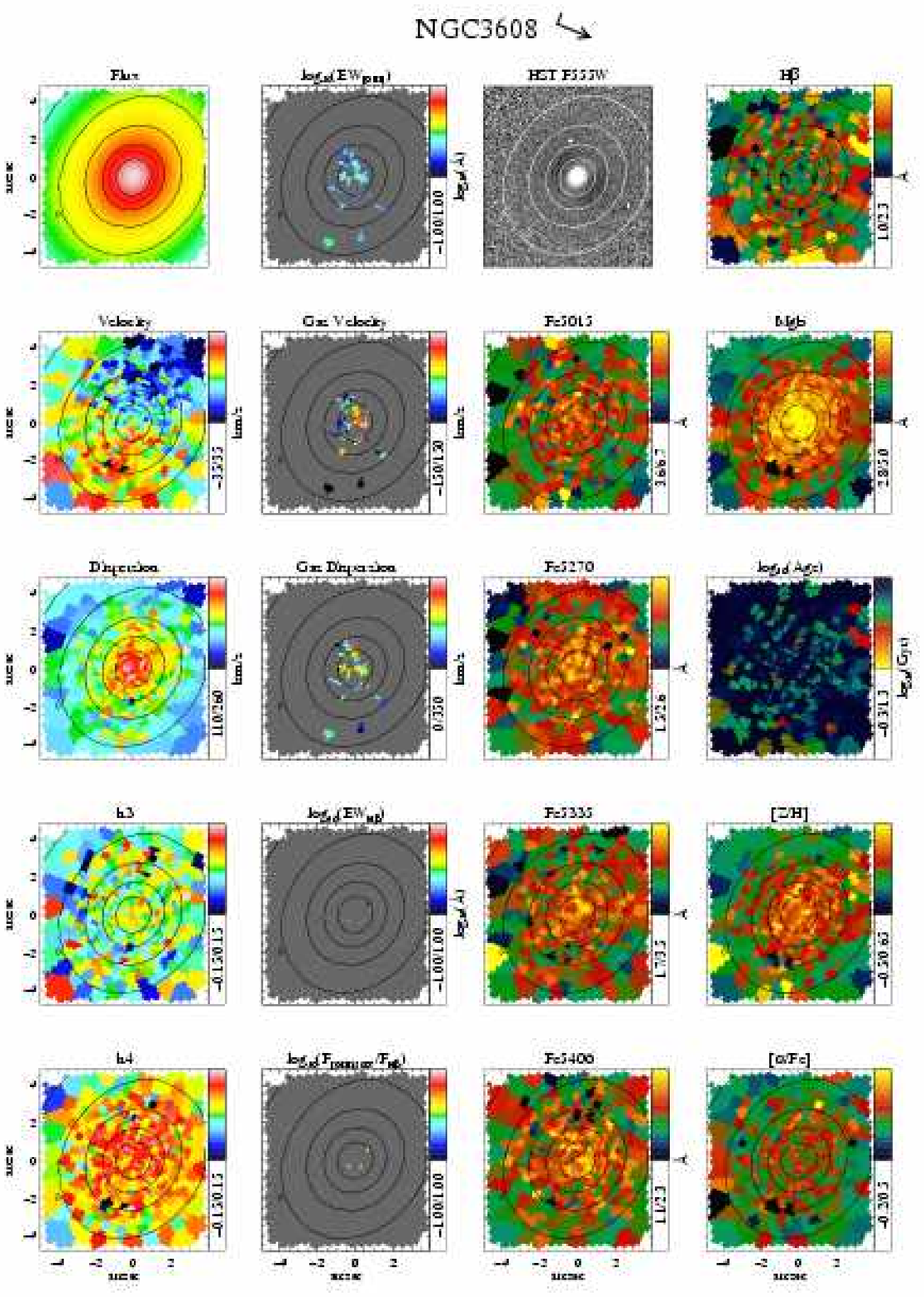}
 \end{center}
 \caption{{\it continued}}
 \label{fig:ngc3608}
\end{figure*}

\addtocounter{figure}{-1}

\begin{figure*}
 \begin{center}
  \includegraphics[width=16cm]{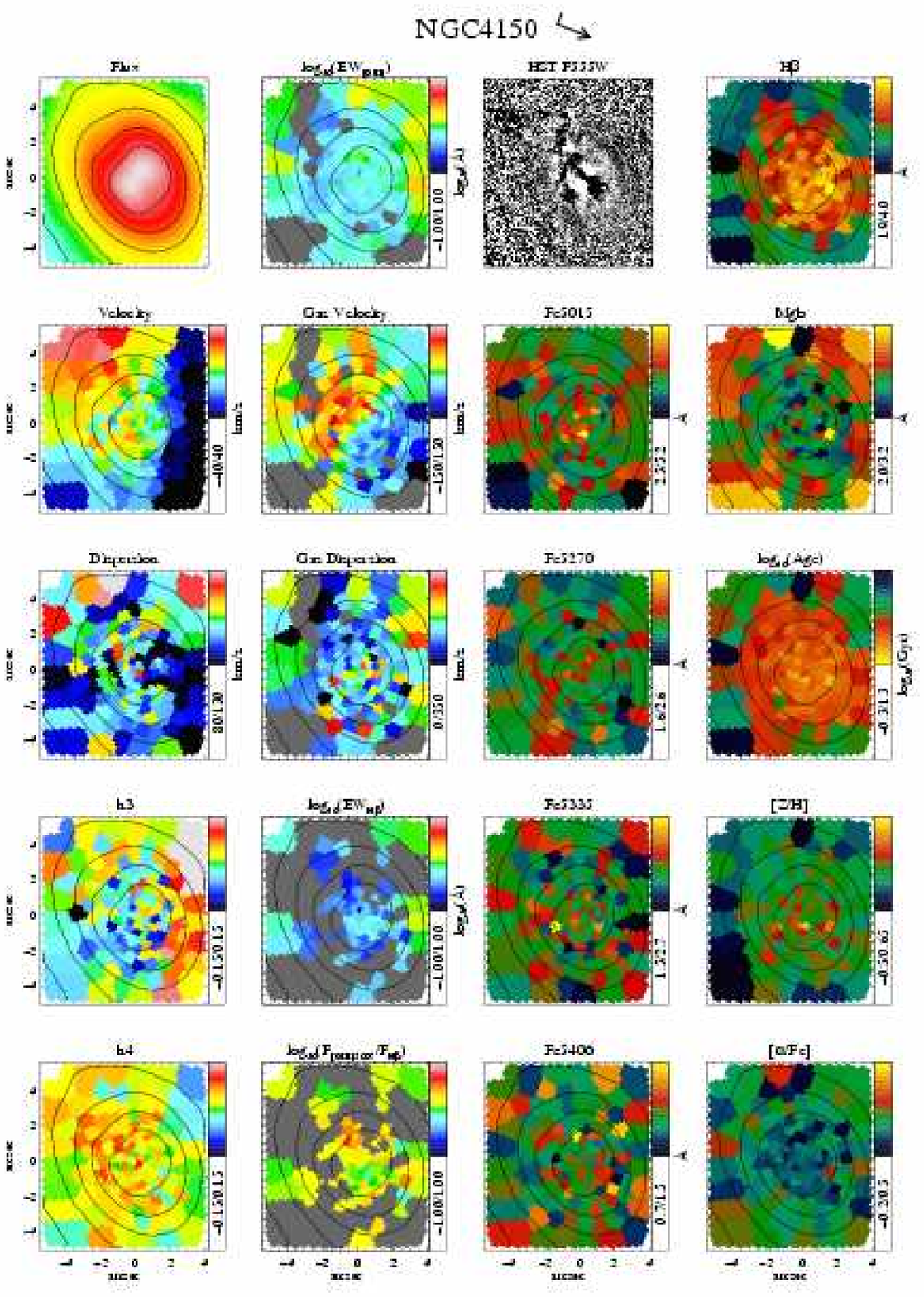}
 \end{center}
 \caption{{\it continued}}
 \label{fig:ngc4150}
\end{figure*}

\addtocounter{figure}{-1}

\begin{figure*}
 \begin{center}
  \includegraphics[width=16cm]{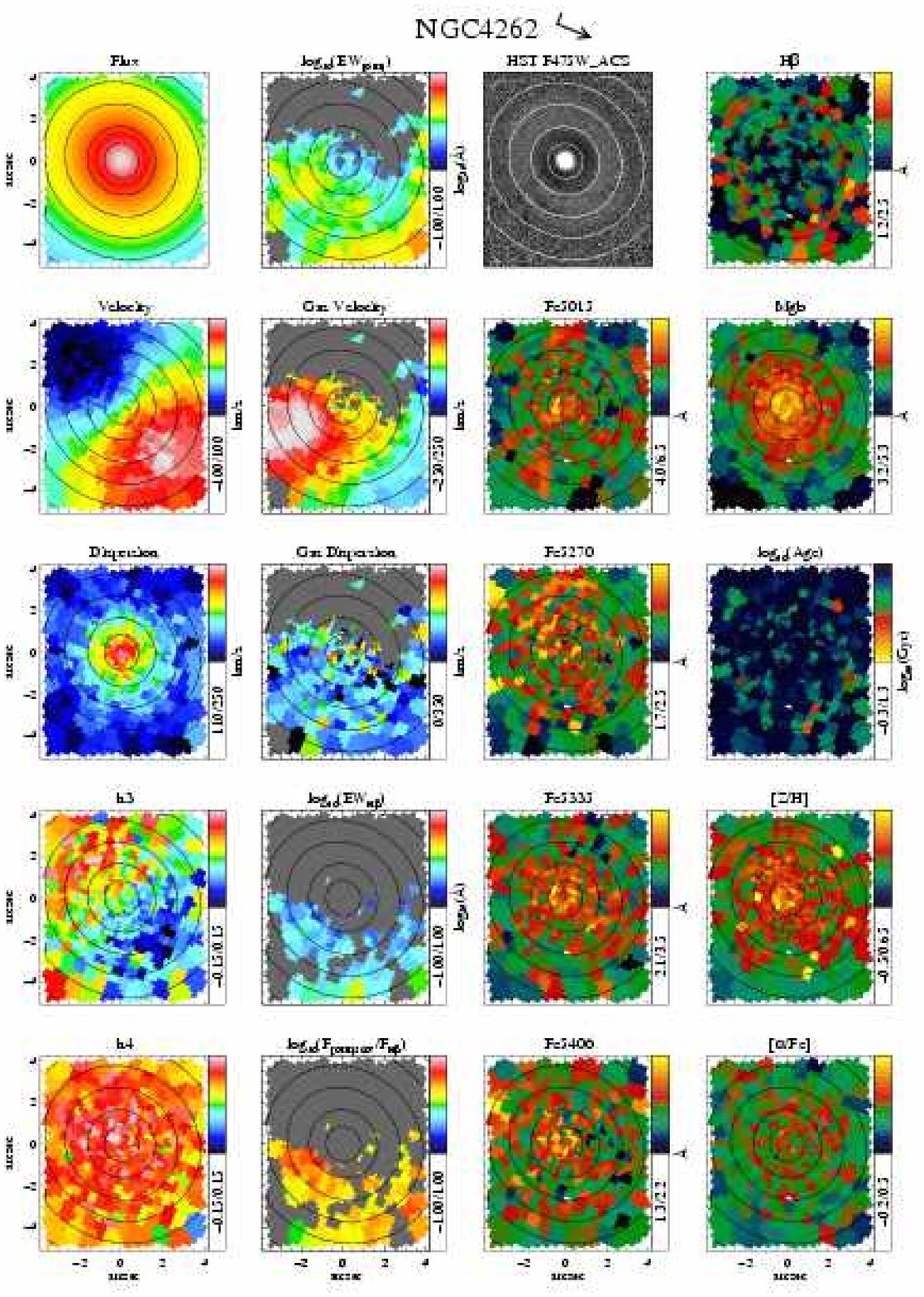}
 \end{center}
 \caption{{\it continued}}
 \label{fig:ngc4262}
\end{figure*}

\addtocounter{figure}{-1}

\begin{figure*}
 \begin{center}
  \includegraphics[width=16cm]{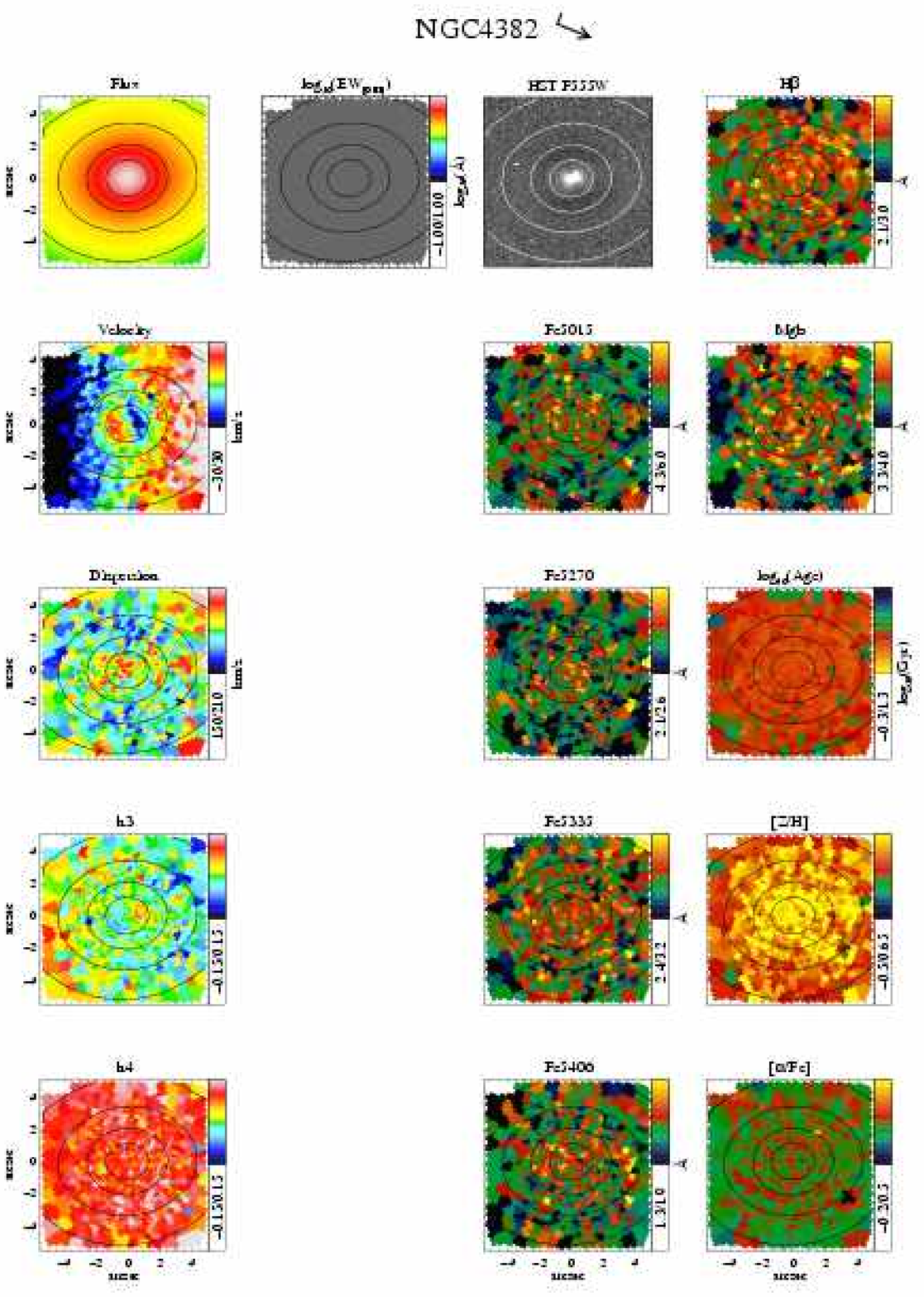}
 \end{center}
 \caption{{\it continued}}
 \label{fig:ngc4382}
\end{figure*}

\addtocounter{figure}{-1}

\begin{figure*}
 \begin{center}
  \includegraphics[width=16cm]{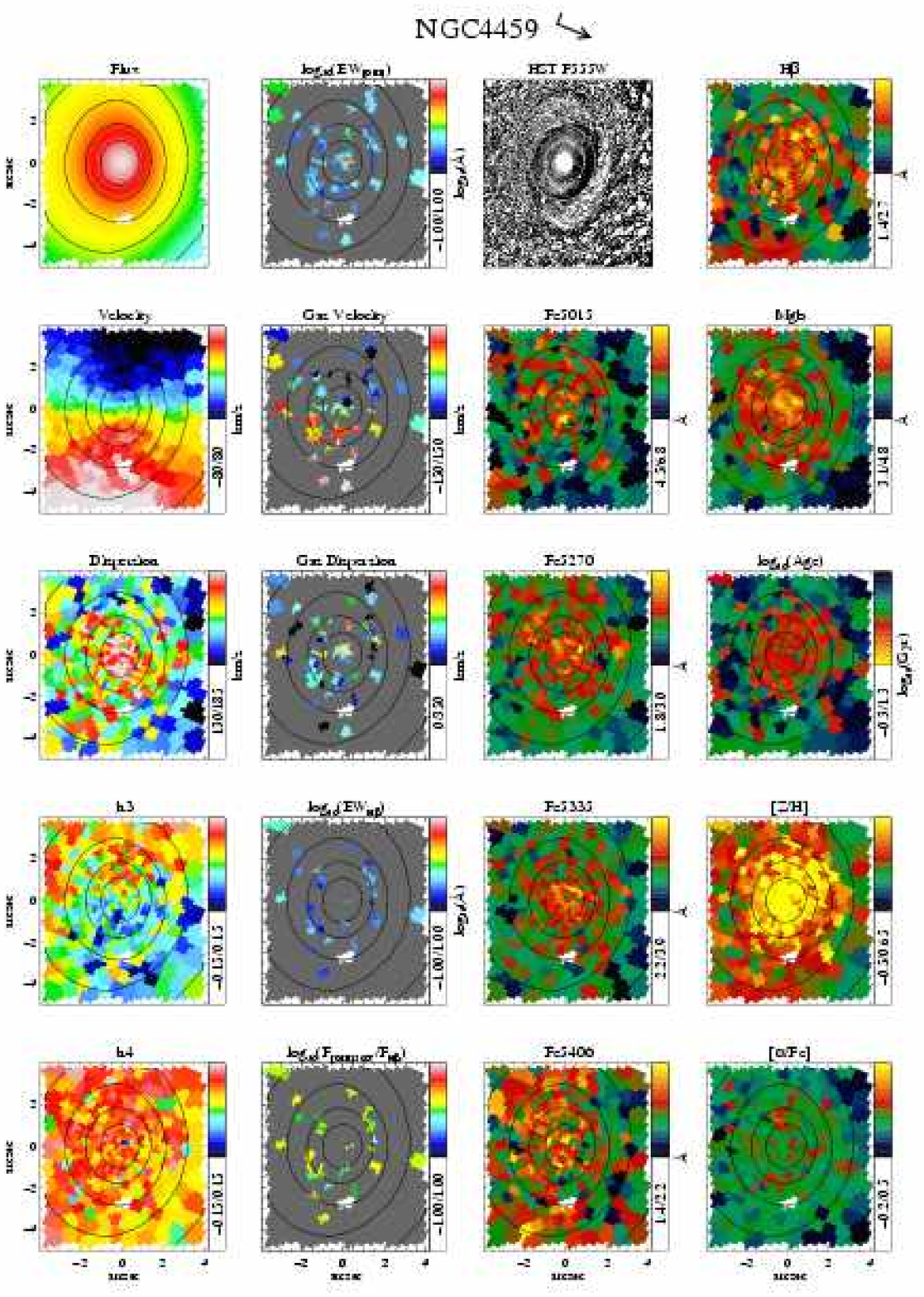}
 \end{center}
 \caption{{\it continued}}
 \label{fig:ngc4459}
\end{figure*}

\addtocounter{figure}{-1}

\begin{figure*}
 \begin{center}
  \includegraphics[width=16cm]{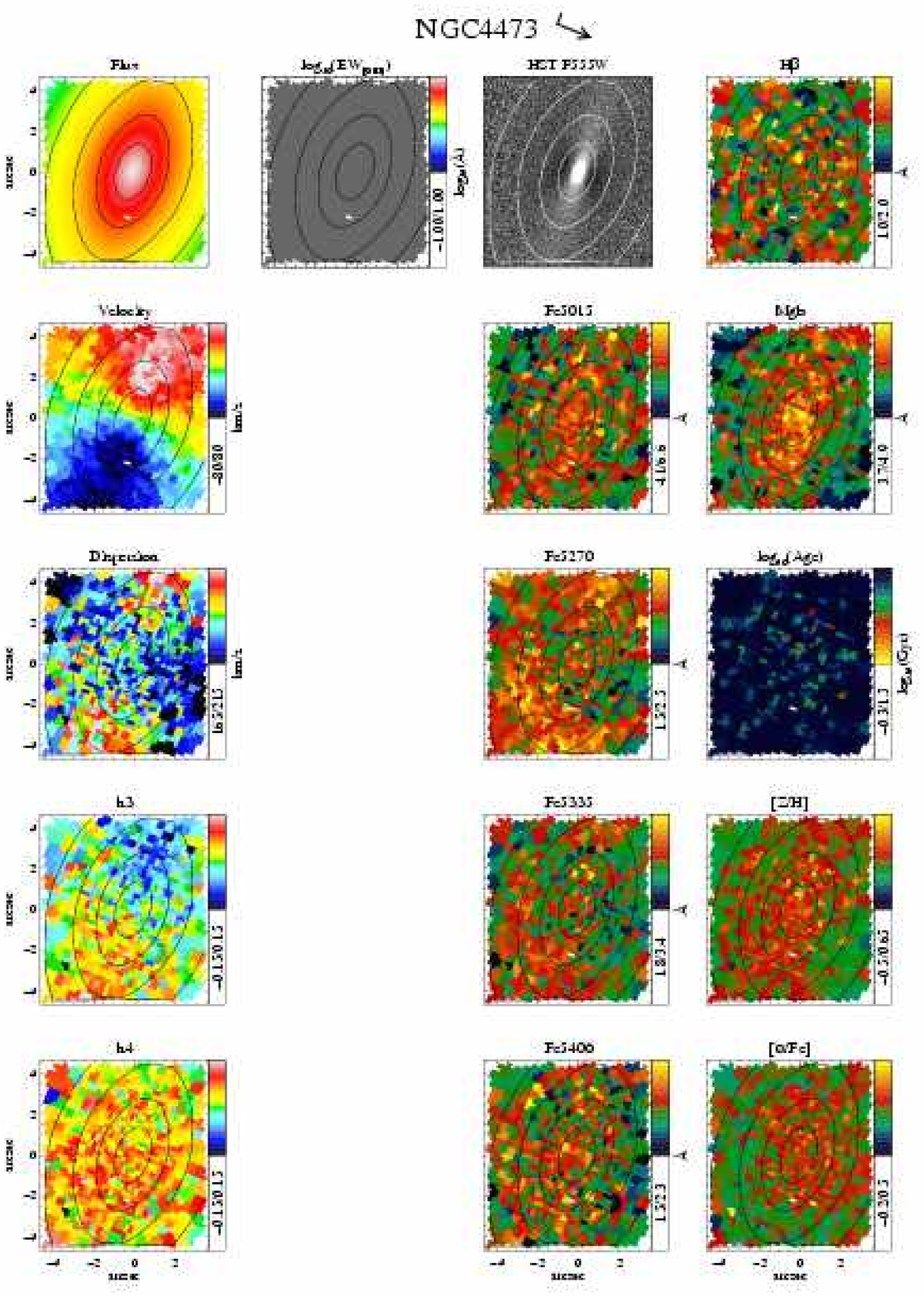}
 \end{center}
 \caption{{\it continued}}
 \label{fig:ngc4473}
\end{figure*}

\addtocounter{figure}{-1}

\begin{figure*}
 \begin{center}
  \includegraphics[width=16cm]{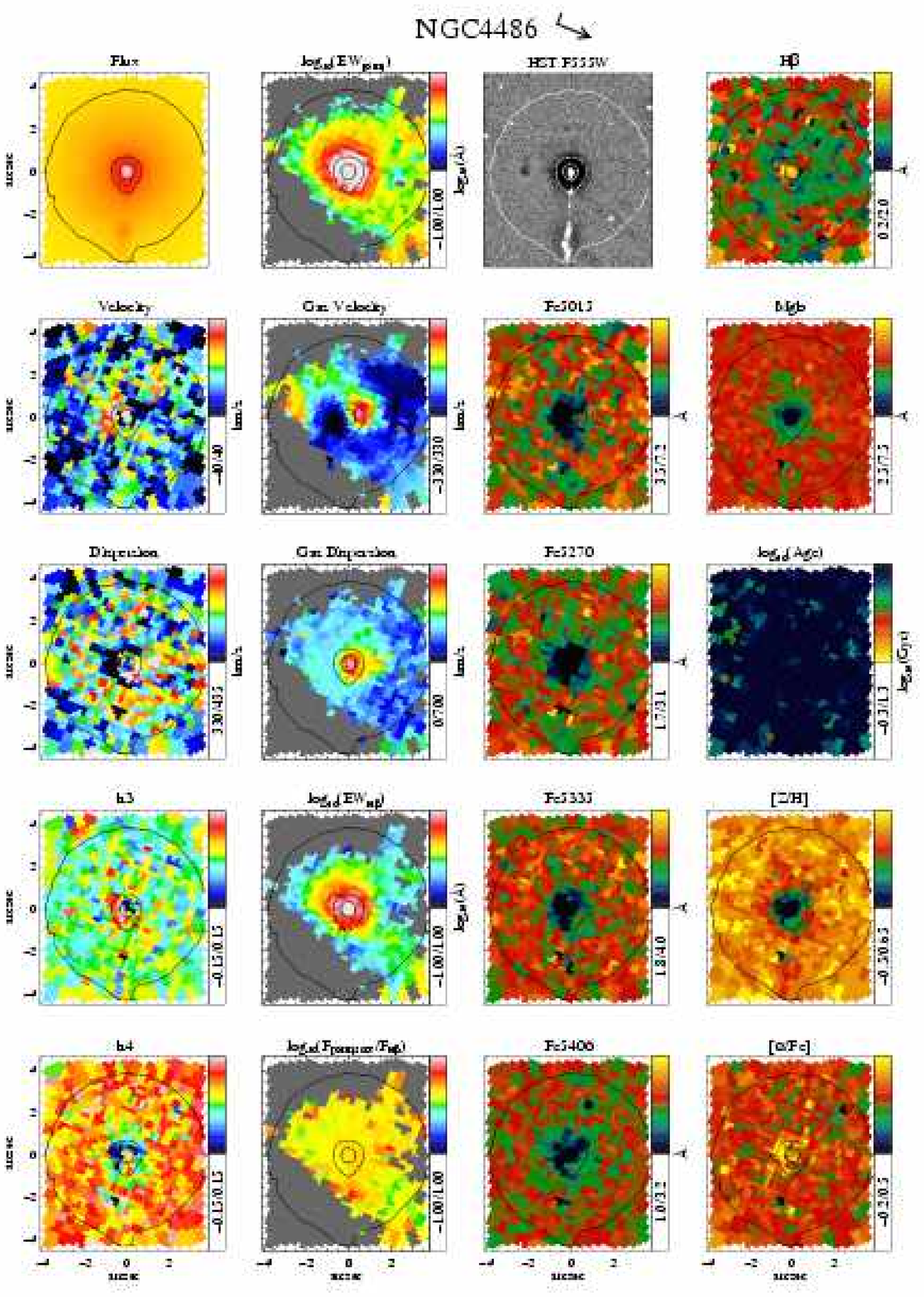}
 \end{center}
 \caption{{\it continued}}
 \label{fig:ngc4486}
\end{figure*}

\addtocounter{figure}{-1}

\begin{figure*}
 \begin{center}
  \includegraphics[width=16cm]{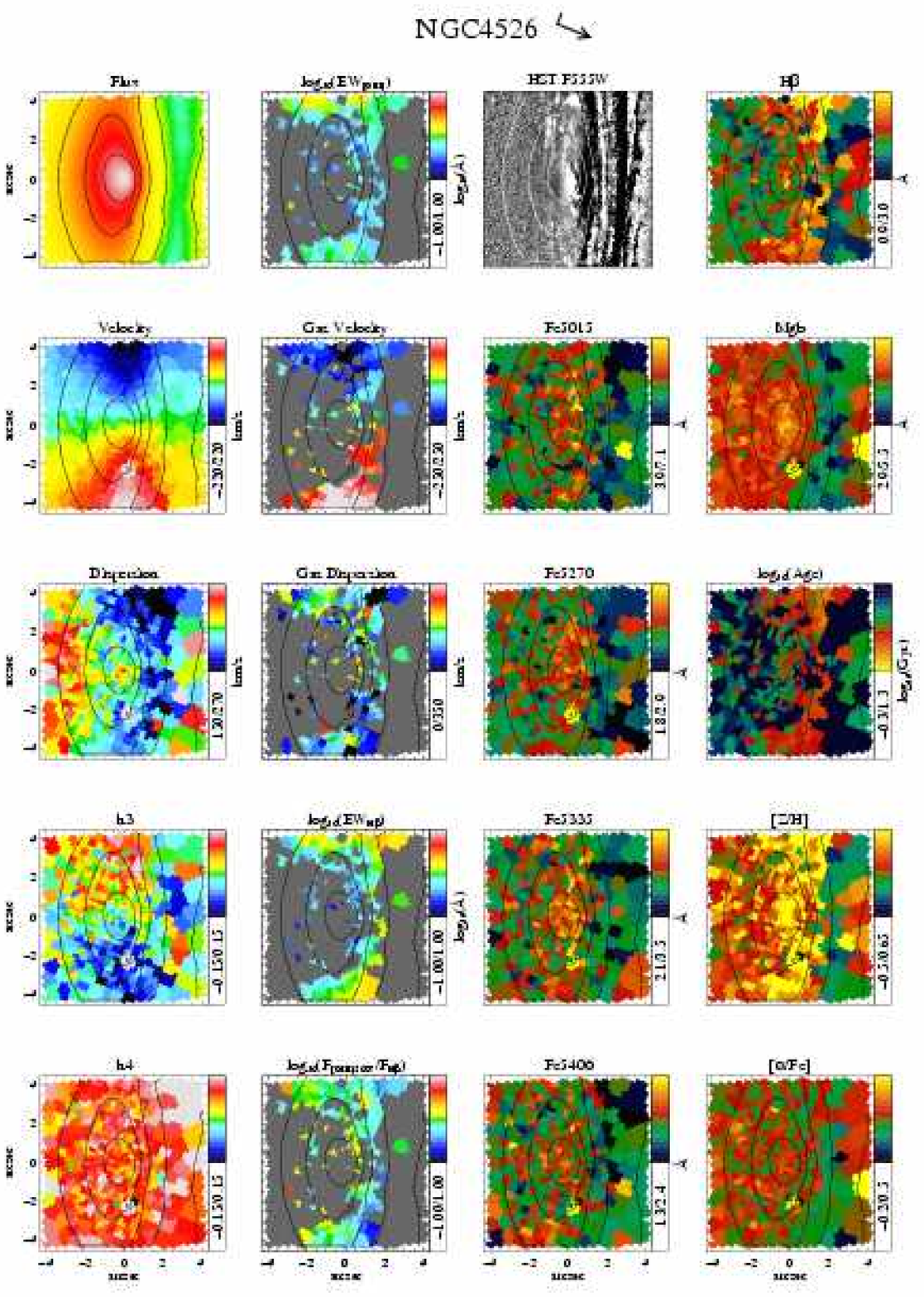}
 \end{center}
 \caption{{\it continued}}
 \label{fig:ngc4526}
\end{figure*}

\addtocounter{figure}{-1}

\begin{figure*}
 \begin{center}
  \includegraphics[width=16cm]{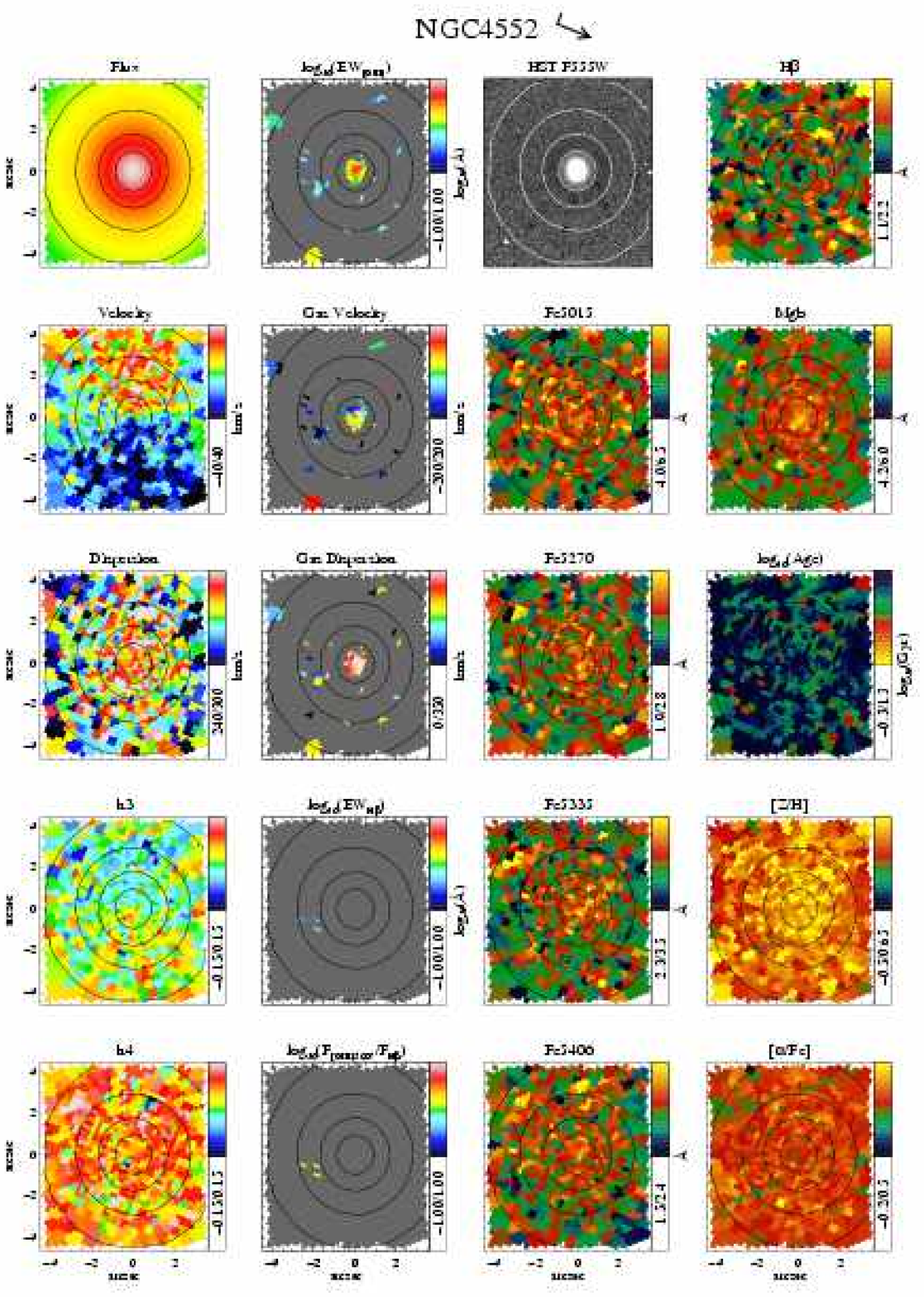}
 \end{center}
 \caption{{\it continued}}
 \label{fig:ngc4552}
\end{figure*}

\addtocounter{figure}{-1}

\begin{figure*}
 \begin{center}
  \includegraphics[width=16cm]{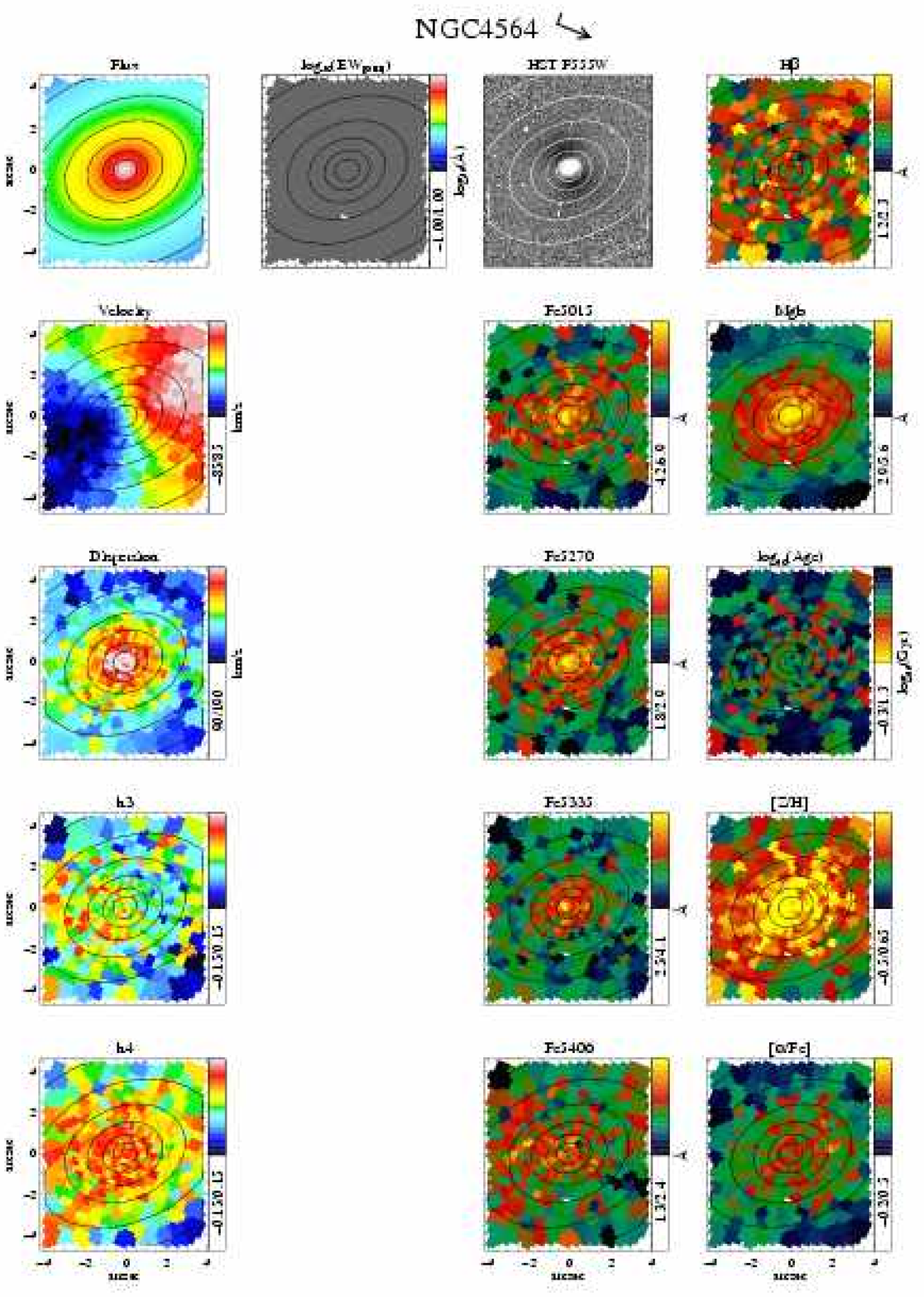}
 \end{center}
 \caption{{\it continued}}
 \label{fig:ngc4564}
\end{figure*}

\addtocounter{figure}{-1}

\begin{figure*}
 \begin{center}
  \includegraphics[width=16cm]{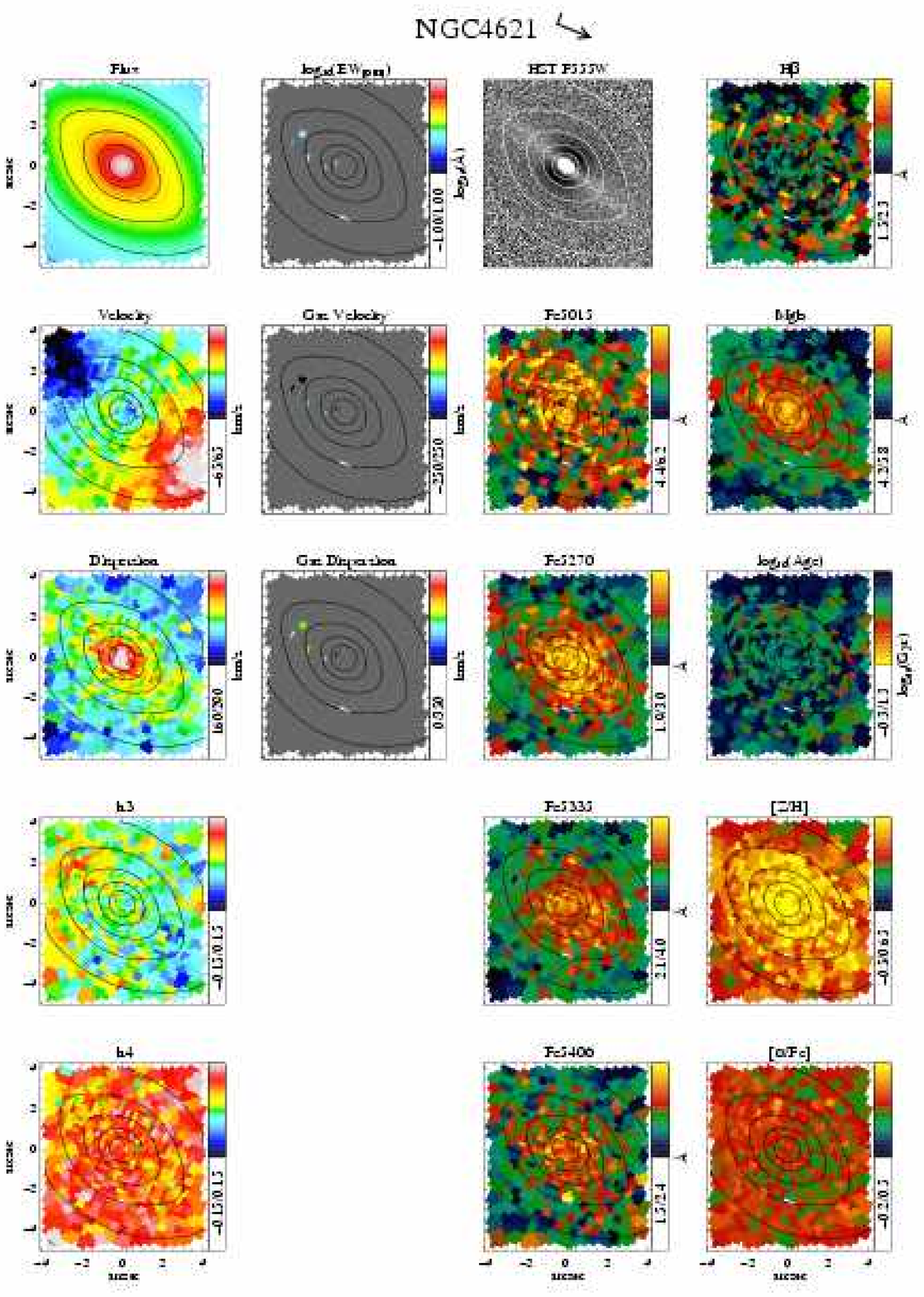}
 \end{center}
 \caption{{\it continued}}
 \label{fig:ngc4621}
\end{figure*}

\addtocounter{figure}{-1}

\begin{figure*}
 \begin{center}
  \includegraphics[width=16cm]{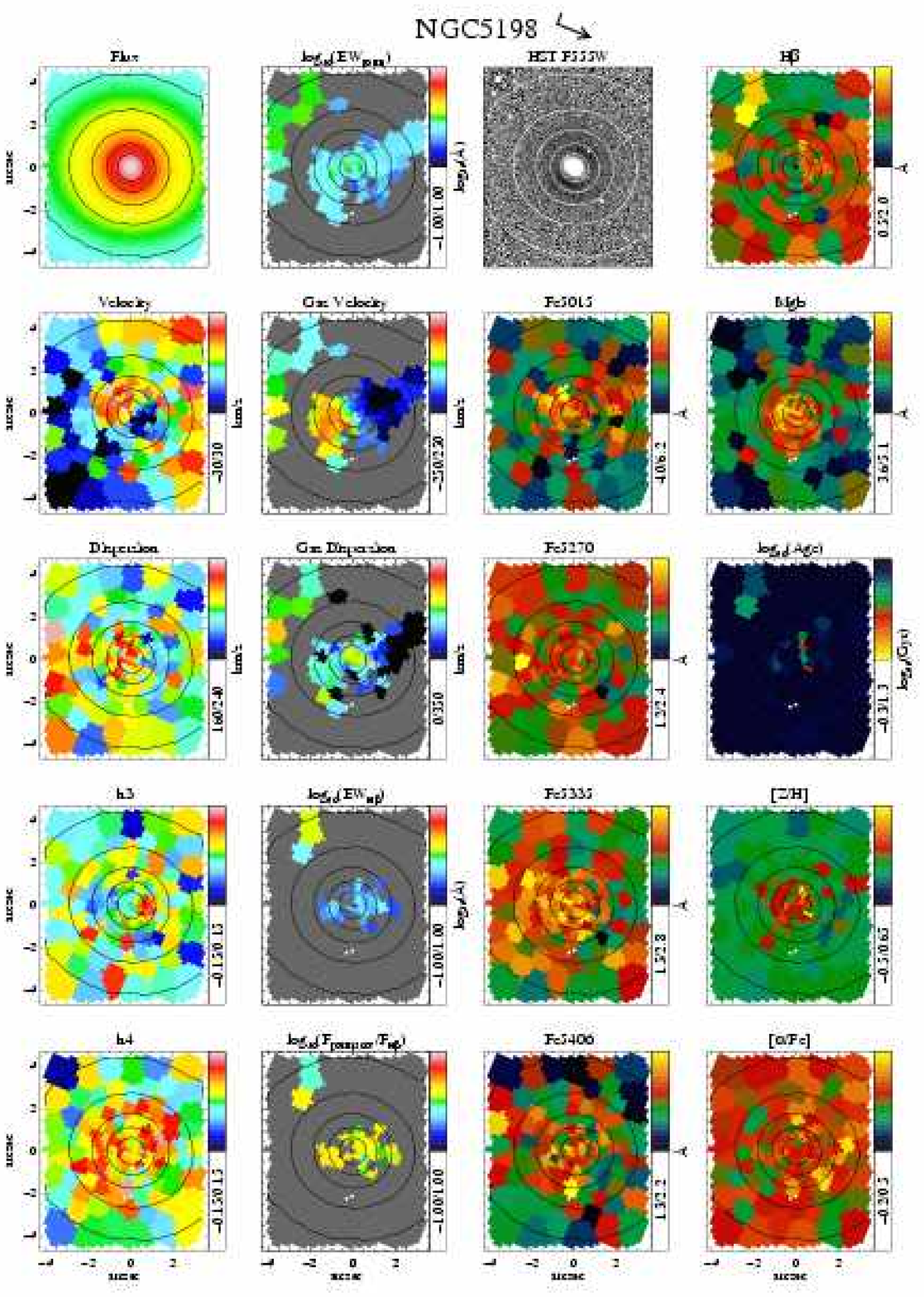}
 \end{center}
 \caption{{\it continued}}
 \label{fig:ngc5198}
\end{figure*}

\addtocounter{figure}{-1}

\begin{figure*}
 \begin{center}
  \includegraphics[width=16cm]{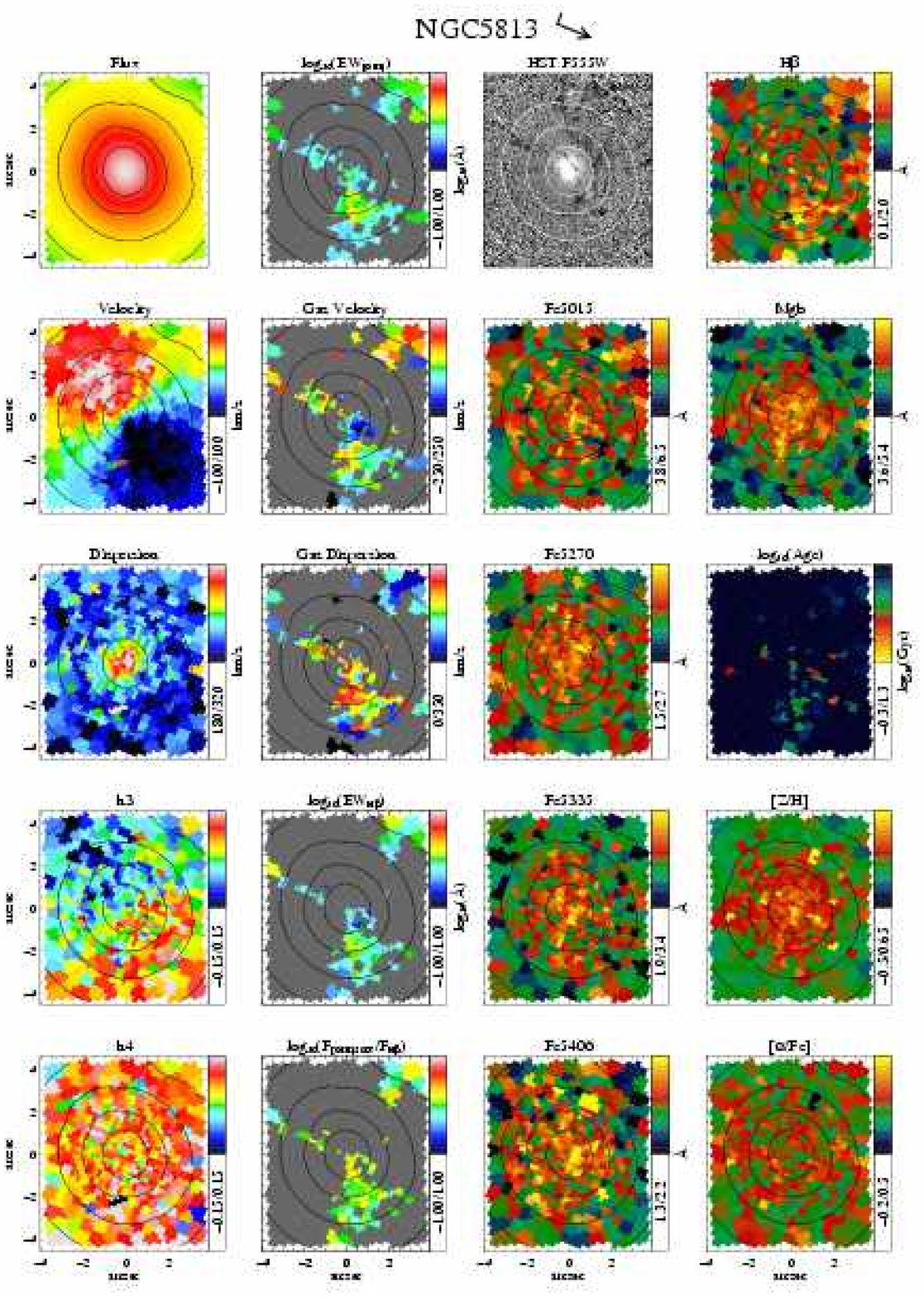}
 \end{center}
 \caption{{\it continued}}
 \label{fig:ngc5813}
\end{figure*}

\addtocounter{figure}{-1}

\begin{figure*}
 \begin{center}
  \includegraphics[width=16cm]{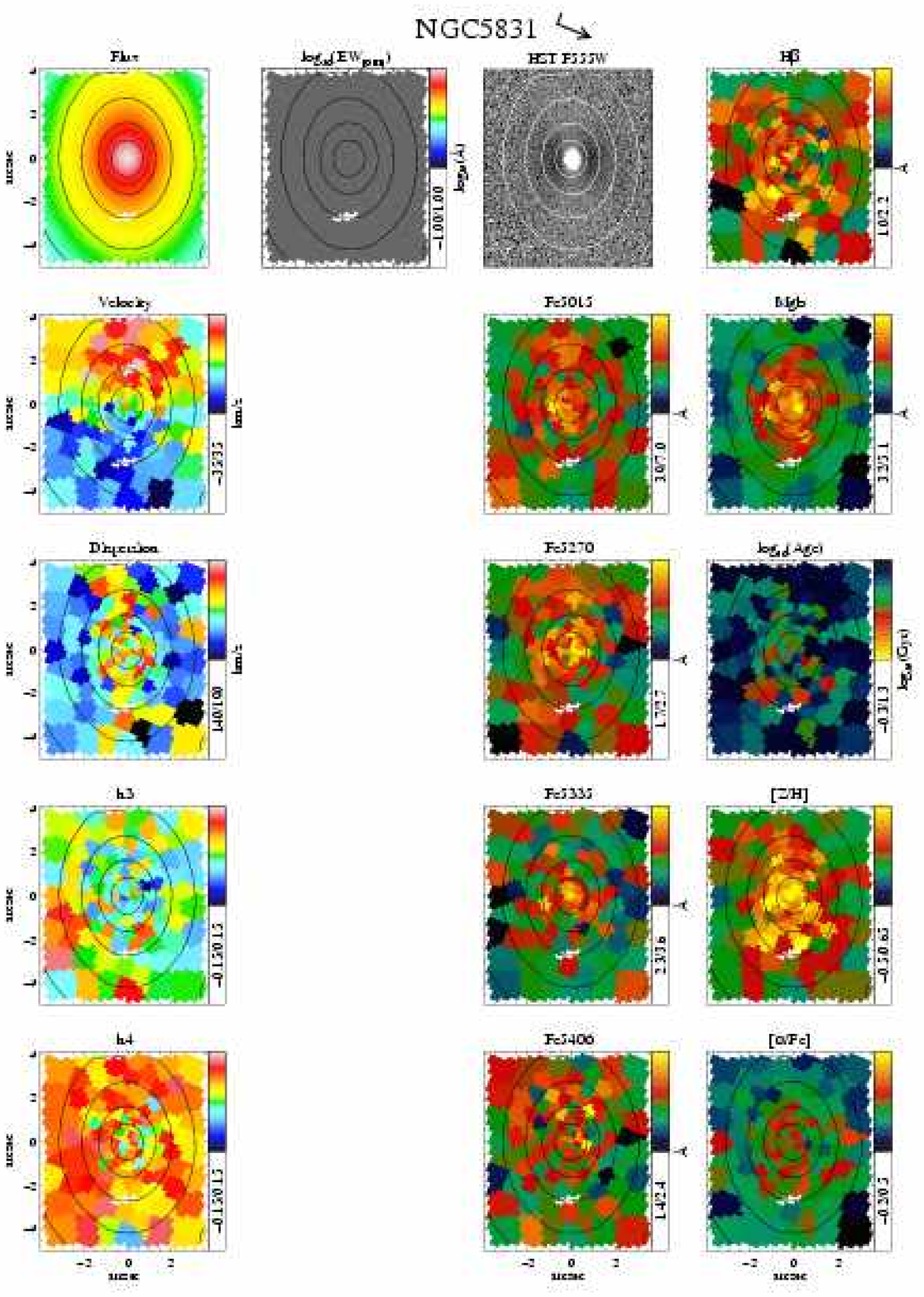}
 \end{center}
 \caption{{\it continued}}
 \label{fig:ngc5831}
\end{figure*}

\addtocounter{figure}{-1}

\begin{figure*}
 \begin{center}
  \includegraphics[width=16cm]{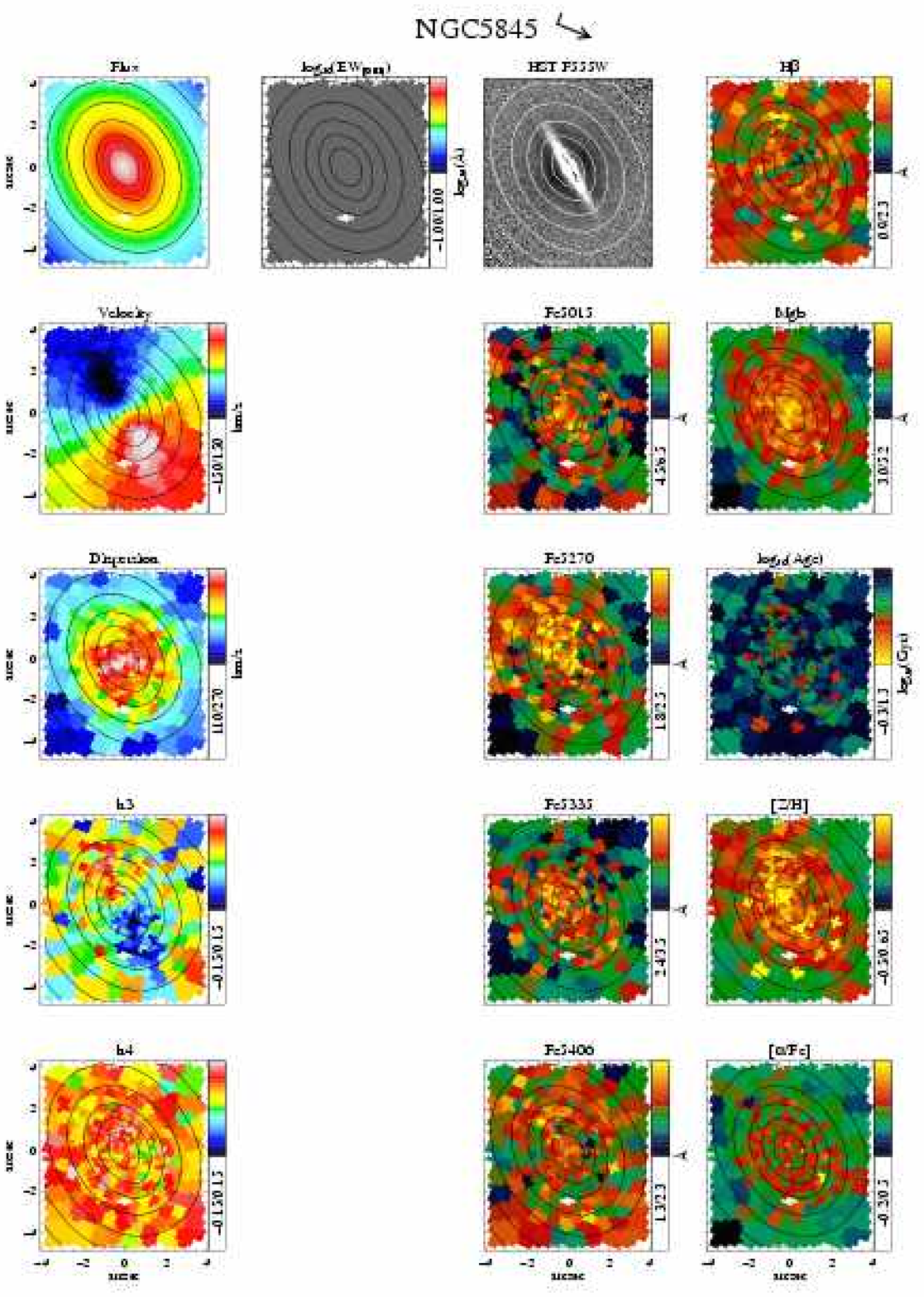}
 \end{center}
 \caption{{\it continued}}
 \label{fig:ngc5845}
\end{figure*}

\addtocounter{figure}{-1}

\begin{figure*}
 \begin{center}
  \includegraphics[width=16cm]{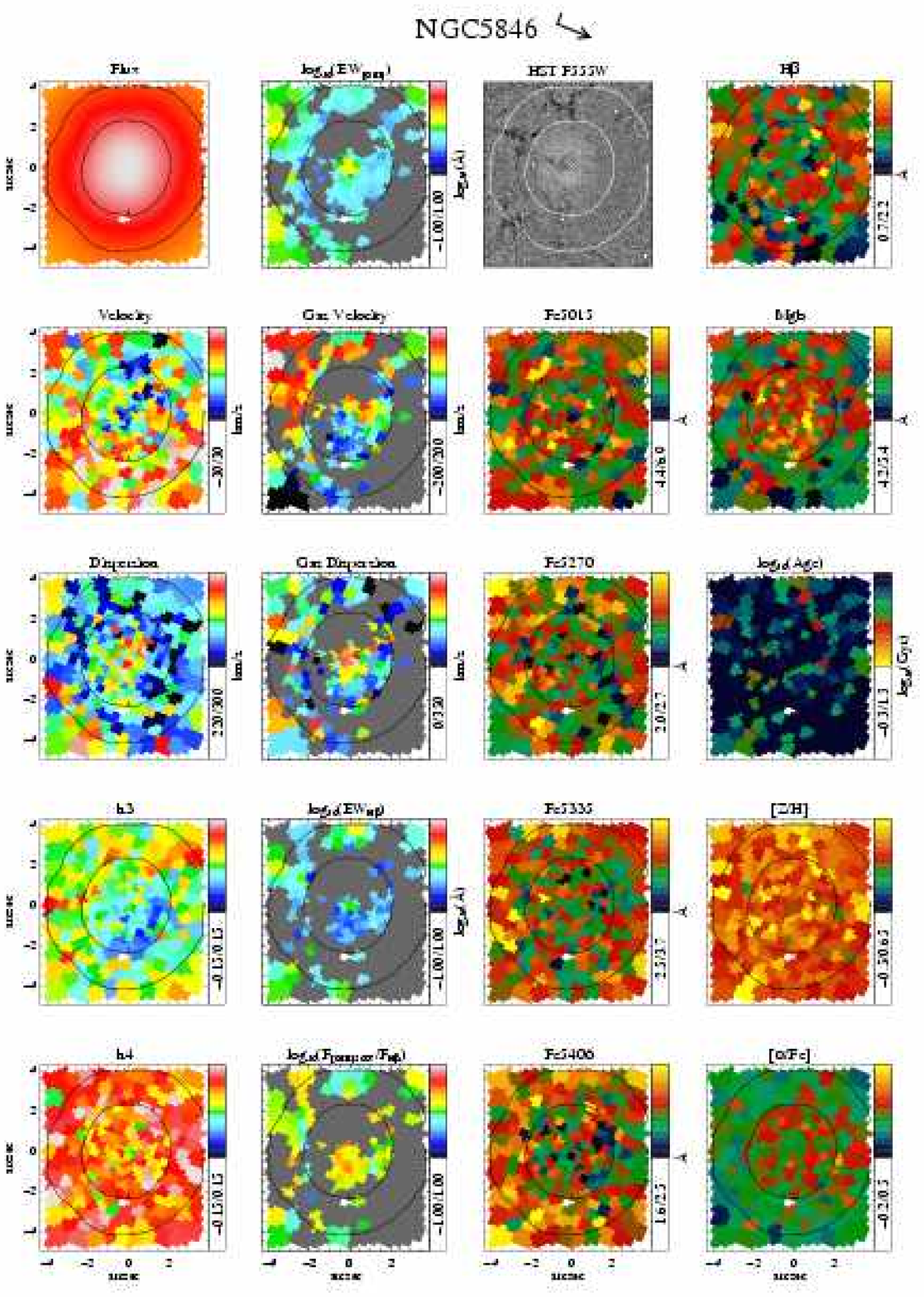}
 \end{center}
 \caption{{\it continued}}
 \label{fig:ngc5846}
\end{figure*}

\addtocounter{figure}{-1}

\begin{figure*}
 \begin{center}
  \includegraphics[width=16cm]{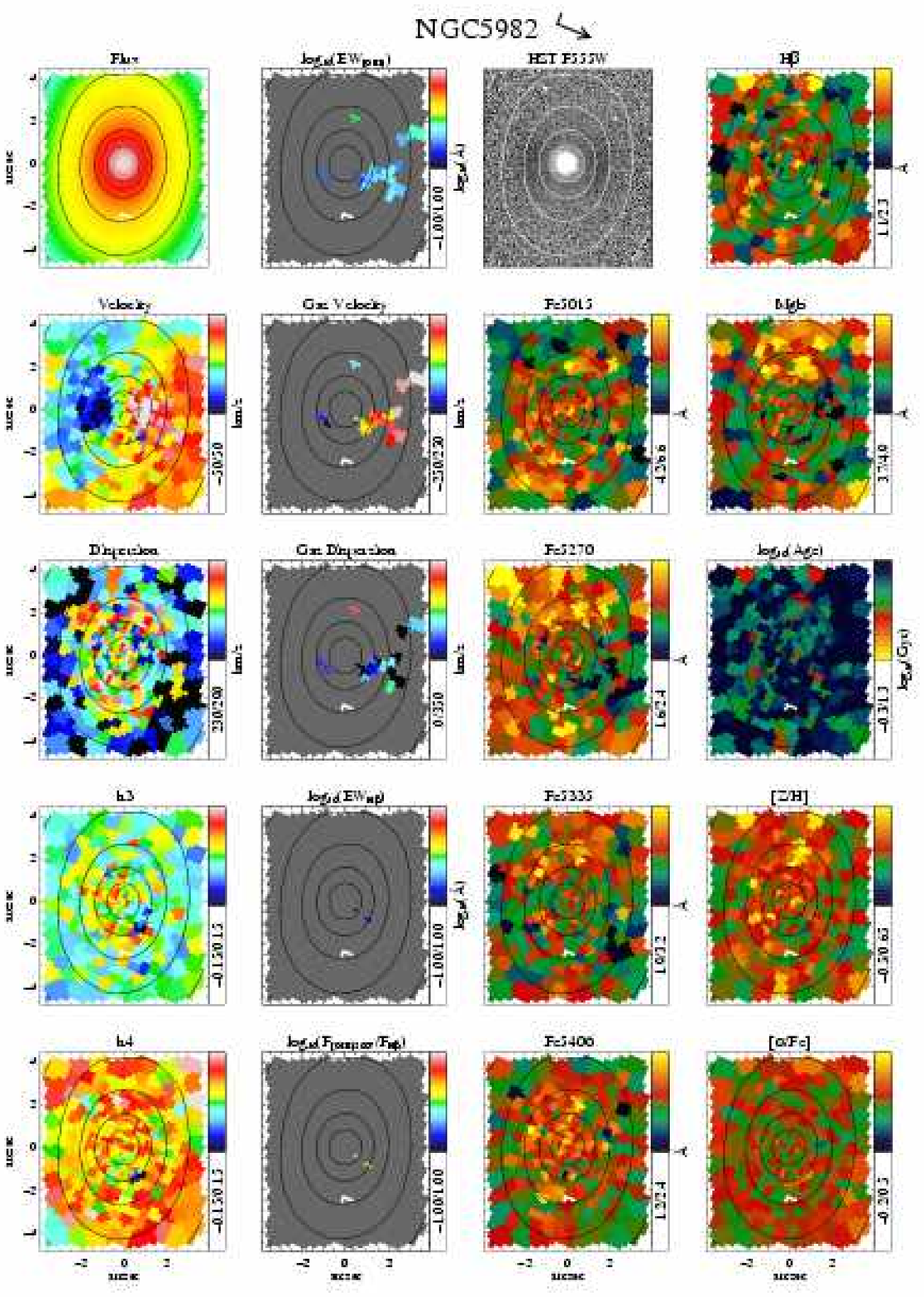}
 \end{center}
 \caption{{\it continued}}
 \label{fig:ngc5982}
\end{figure*}

\renewcommand{\thefigure}{\arabic{figure}}

\begin{table}
\begin{center}
\caption{Summary of central dust and ionised gas properties observed with \oasis.}
\begin{tabular}{lcccl}
\hline
NGC	&\oiii	&\hb	&\nii	&Dust			\\
(1)	&(2)	&(3)	&(4)	&(5)			\\
\hline
1023	&trace	&no	&no	&trace (filaments)	\\
2549	&yes	&trace	&no	&no			\\
2695	&no	&no	&no	&no image		\\
2699    &no	&no	&no	&no			\\
2768	&yes	&yes	&yes	&spiral			\\
2974	&yes	&yes	&yes	&spiral			\\
3032	&yes	&yes	&yes	&spiral			\\
3379	&yes	&yes	&no	&ring			\\
3384	&no	&no	&no	&no			\\
3414	&yes	&yes	&yes	&trace (filaments/spiral)\\
3489	&yes	&yes	&no	&spiral			\\
3608	&yes	&trace	&no	&no			\\
4150	&yes	&yes	&no	&spiral			\\
4262	&yes	&yes	&no	&trace (spiral)	\\
4382	&no	&no	&no	&no			\\
4459	&trace	&trace	&no	&spiral			\\
4473    &no	&no	&no	&no			\\
4486    &yes	&yes	&yes	&trace (filaments)	\\
4526    &yes	&yes	&no	&edge-on dust lanes/spiral\\
4552    &yes	&trace	&trace	&filaments		\\
4564    &no	&no	&no	&no			\\
4621    &trace	&no	&no	&no			\\
5198    &yes	&yes	&no	&trace (filaments)	\\
5813    &yes	&yes	&no	&filaments		\\
5831	&no	&no	&no	&no			\\
5845	&no	&no	&no	&central disk		\\
5846	&yes	&yes	&yes	&filaments		\\
5982	&yes	&trace	&no	&no			\\
\hline
\label{tab:gas}
\end{tabular}
\end{center}
Notes: (1)~NGC number. (2) Presence of \oiii\/ emission. (3) Presence of \hb\/ emission. (4) Presence of \nii\/ emission. (5) Presence of visible dust features in the unsharp-masked {\it HST} image, with some qualitative description of the dust morphology. The F555W filter was used wherever available. For all columns, the term `trace' implies that features were found close to the limit of detection.
\end{table}

\subsection{Ionised Gas}
\label{sec:results_gas}

In Paper V we have shown that many of the galaxies in our sample contain large quantities of ionised gas, spanning a range of distributions from rotating disks to irregular filaments. Here we discuss the incidence of ionised gas within the central regions of our sample galaxies observed with \oasis, and describe the various distributions and kinematics we find.\looseness-2

\subsubsection{Incidence of Ionised Gas}

The incidence of ionised gas in our sample is summarised in Table \ref{tab:gas}. We find that 14 of the 28 galaxies in our sample have clear evidence of both \hb\/ and \oiii\/ emission in their central regions, and a further four objects with only \oiii\/ clearly detected. Three galaxies exhibit faint \oiii\/ close to the detection limit (marked as `trace' in Table \ref{tab:gas}), with either undetected or similarly faint \hb. Seven galaxies show evidence of \nii\/ emission, which is generally confined to the central regions.

This incidence of gas is comparable to that found with \sauron, although there are a number of galaxies for which \sauron\/ shows stronger detection of emission, or shows clearer structures where the \oasis\/ data appear patchy or even show no emission at all. This is simply a reflection of the sensitivity limits of the two data sets, arising mainly from the generally lower $S/N$ of the \oasis\/ spectra compared to the \sauron\/ spectra, as discussed in Section \ref{sec:gas}. Appendix \ref{appendix:d} quantifies this further, showing for the apparently discrepant cases the different $S/N$ values of the \sauron\/ and \oasis\/ data sets as a function of circular radius, and how this translates into equivalent width limits for the two instruments. The \oiii\/ emission detected by \sauron\/ falls below the equivalent-width limit of our \oasis\/ data, thus accounting for the difference in detection.

\subsubsection{Ionised Gas Kinematics: Rotation}

The mean motion of the ionised gas is characterised by three main morphologies: regular circular rotation, consistent with a disk; warped or twisted rotation; and irregular motion. We discuss each of these in turn.

The following galaxies show gas kinematics consistent with disks: NGC\,3379, NGC\,3608, NGC\,4486 (centre), NGC\,4526, NGC\,4552, NGC\,5198, and NGC\,5813 (centre). Of these, NGC\,4486 is worth a special mention. Firstly, the PSF of these observations is the best of our sample (0\farcsec57). The effect of this excellent PSF is most obvious in the central velocity field of the ionised gas. The large-scale stream of gas extending to the centre from larger radii is clearly decoupled from a rapidly rotating central disk component. This disk of gas is well known from narrow band imaging studies, also from {\it HST} \citep[e.g.][]{ford94}. A closer examination of Fig.~\ref{fig:ngc4486} shows that the velocity dispersion of the gas in this galaxy is also very high in the central parts, suggesting the unresolved superposition of two or more velocity components. Van der Marel (\citeyear{vdmarel94}) also found a large central velocity dispersion of the \hg\/ emission line, although his value of 516~\kms\/ is significantly lower than our central value based on \oiii\/ ($> 700$~\kms). This difference can be attributed to the higher spatial resolution of our data (0\farcsec57 FWHM sampled with 0\farcsec27 lenses, compared with 0\farcsec79 FWHM sampled with a 1\arcsec\/ long-slit). Fig.~\ref{fig:ngc4486_central} presents the central spectrum of the \oasis\/ data, showing that the observed emission lines are mainly composed of two strong components, belonging to the approaching and receding sides of an unresolved disk, where the absolute velocity amplitude is over 1000~\kms. This is in agreement with spectroscopic measurements with {\it HST} \citep[e.g.][]{harms94}, and is thought to indicate emission from a gas disk orbiting in the Keplerian potential of a $3 \times 10^9$~\msun\/ black hole \citep{macchetto97}.

\begin{figure}
 \begin{center}
  \includegraphics[height=8cm, angle=90]{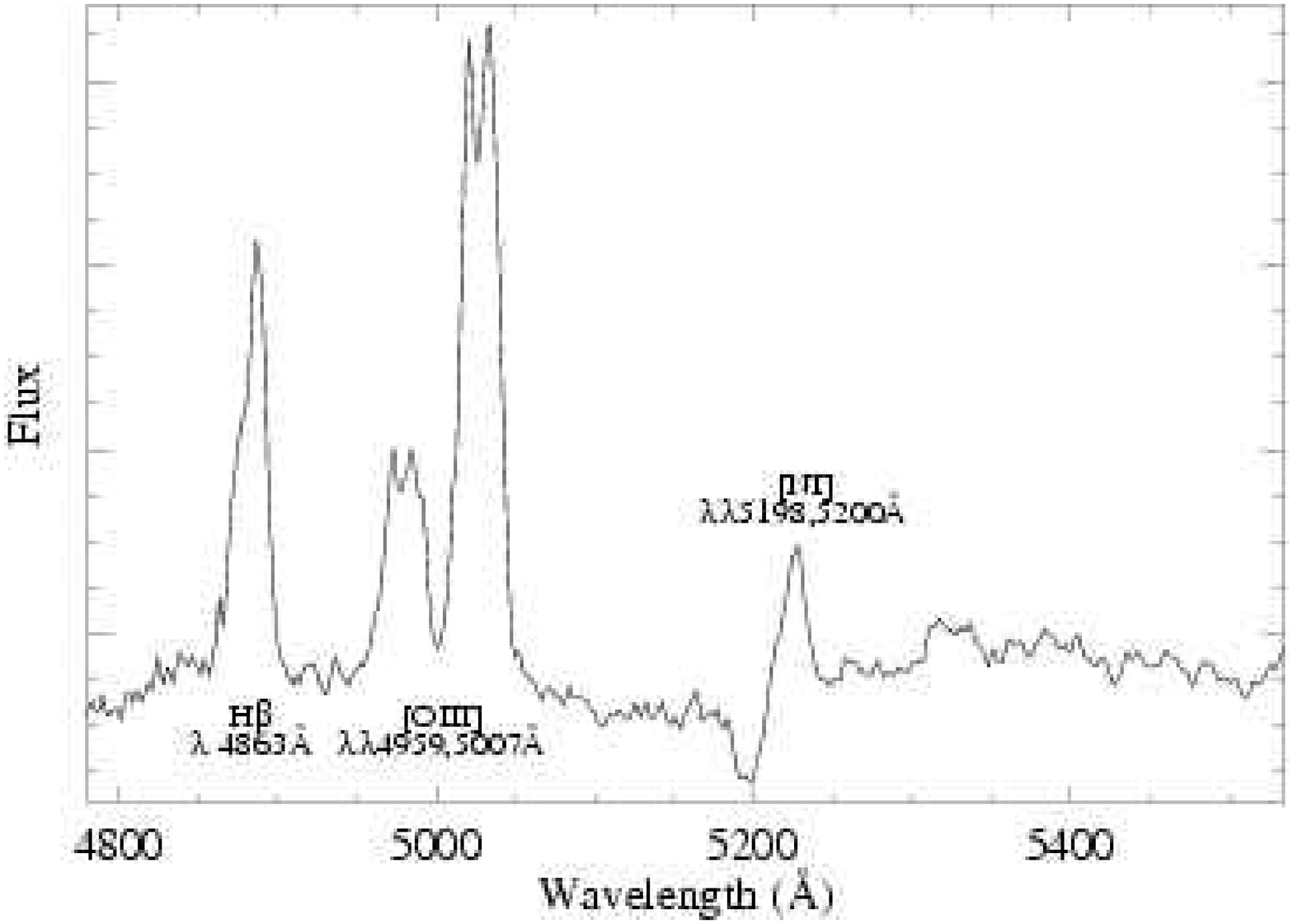}
 \end{center}
 \caption{Central spectrum of NGC\,4486, showing double-peaked emission lines. This is indicative of unresolved gas rotating rapidly ($\sim \pm 500$~\kms) in a nuclear disk surrounding the central black hole.}
 \label{fig:ngc4486_central}
\end{figure}

Another interesting case is the central disk in NGC\,3379. This disk appears to be rather misaligned with the stellar kinematics, although is very well aligned with the central dust ring visible in the unsharp-masked {\it HST} image \citep[see also][]{gebhardt00}. The nature of this central gas structure is explored in detail by \citet{shapiro06}, combining our \oasis\/ data with {\tt STIS} spectroscopy.

The gas disks in our sample are generally quite faint and difficult to detect, as well as being restricted to the very central regions. By contrast, several galaxies show much higher emission-line flux distributed across the whole \oasis\/ field. The velocity field morphology of these objects generally shows a twisted structure, where the axis of rotation smoothly changes with radius. Galaxies showing this behaviour include: NGC\,2768, NGC\,2974, NGC\,3032, NGC\,3414, NGC\,3489, NGC\,4150, NGC\,4486 (in the outer regions), and NGC\,5846. Several of these objects also show an elongated structure in the gas, which tends to be almost aligned with the rotation of the object, indicating either a close to edge-on disk (from the high observed flattening) or perhaps a bar structure.\looseness-2

Galaxies exhibiting clearly detectable emission, but which have gas that is irregularly distributed or does not show a clear sense of rotation are NGC\,2549, NGC\,4262 and NGC\,5982. In NGC\,2549 and NGC\,5982, the gas structure indicates the presence of streams of material. This explanation is supported by the \sauron\/ observations, which show these structures extending to larger radii. In the case of NGC\,4262, the physical explanation for this lopsided structure is less clear. The \sauron\/ data show ionised gas across most of the field of view, indicating a significant amount of gas, rather than a relatively small filament. The central regions, however, show a large asymmetric feature, which connects to the same feature in the \oasis\/ data. The unsharp-masked {\it HST} image also shows a peculiar ring-like feature, with the signature of faint dust lanes across it, co-spatial with the strongest emission features. This galaxy exhibits a strong stellar bar, the dynamical influence of which may help explain these peculiar features.\looseness-2

Three objects have detectable emission lines, but have a very `patchy' distribution, generally exhibiting emission close to our detection limit: NGC\,1023, NGC\,4459, NGC\,4621. These galaxies also show evidence for emission in the \sauron\/ data, albeit at lower spatial resolution. Due to the patchy distribution in our \oasis\/ data, the rotation structure in these objects is difficult to determine, although in the case of NGC\,4459, rotation follows the stars.\looseness-2

\subsubsection{Ionised Gas Kinematics: Dispersion}

Even with our detailed technique for separating the emission and absorption components of our spectra, measuring the velocity dispersion of the gas is much more sensitive to noise and template mismatch than the flux or velocity. Thus, in the case of weak emission, there is a large scatter in the determined dispersion values, with a tendency to measure broad lines (Paper V). A more difficult problem occurs with galaxies that have large stellar velocity dispersions, in that small discrepancies between our template library and the actual galaxy stellar population caused by differences in abundance ratios give rise to template mismatch effects, which can be compensated by including an artificially broad emission component in the fit. Where emission is strong enough, however, the width of the emission lines can be measured reliably without strong biases.

In general, the emission-line dispersion maps show values below that of the stars, but with a strong central peak, which can be as large or larger than the central stellar dispersion at the same position (e.g. NGC\,2974, NGC\,3414). For most objects, this central peak represents unresolved rotation around the galaxy nucleus. This effect is most clear in NGC\,4486, where the measured velocity dispersion becomes enormous ($> 700$~\kms). As mentioned above, this comes from fitting two lines, which are separated by around 1000~\kms, with a single Gaussian. This is an extreme case: no other examples of clearly separate lines are found, and we consider a single Gaussian to be a good representation of the emission lines in our data.

Where the gas rotation is regular and spatially well-resolved, the dispersion is generally low (e.g. NGC\,2768, NGC\,3032, NGC\,3414, NGC\,3489, NGC\,4459, NGC\,4526 and NGC\,5198). In this respect, NGC\,2974 is peculiar, in that the rotation field appears rather regular, but the dispersion is quite high within most of the \oasis\/ field. At larger radii, \citet{krajnovic05} were able to model the gas in this galaxy with near-circular motion under the assumptions of asymmetric drift, treating the gas as collisionless `clumps' moving under the influence of gravity. In the central regions, this treatment was still unable to reproduce the spatially-extended high velocity dispersion of the gas there. From TIGER integral-field spectroscopy, \citet{emsellem03} also found an increase in the central emission line widths, and account for this as non-circular motions giving rise to a complex line profile composed of multiple components,  indicating the presence of a bar structure. Fitting this with a single Gaussian gives rise to the high broadening we observe. This structure is visible in their \ha\/ spectral region emission lines (spectral resolution $\sim 3$~\AA), although we do not detect statistically significant departures from Gaussian profiles in the fainter blue emission lines in our lower-spectral resolution data.

\subsubsection{\oiii /\hb\/ Ratios}

The two strongest emission lines in the wavelength region used, namely \hb\/ and the \oiii\/ doublet, may originate from different kinds of gas clouds along our line-of-sight, excited by different ionisation mechanisms, and emitting different amounts of \hb\/ and \oiii\/ photons. The ratio of the fluxes of these two lines \oiii /\hb\/ holds information about the physical state of the gas, and in particular on its degree of ionisation. These two lines alone are not sufficient, however, to understand the r\^ole of the various ionising mechanisms, although very low values of \oiii /\hb\/ ($\lsim 0.33$) are generally observed only in the presence of star formation. Indeed, \hb\/ emission generally dominates over \oiii\/ in star-forming regions, unless the metallicity of the gas is sufficiently low. As in Paper~V, we consider it unlikely that the metallicity of the gas varies abruptly within a given galaxy, although it may vary between objects, should the gas have a different origin or enrichment history.

The following discussion is restricted to the 14 objects exhibiting clear evidence of both emission lines. Within this subset, there are a large variety of global ratio values and local structures. From the maps in Fig. \ref{fig:ngc1023}, the clearest connection between the \oiii /\hb\/ ratio and other galaxy properties is with the dust. Similarly to Paper V, we find that the galaxies with the lowest \oiii /\hb\/ ratio (NGC\,3032, NGC\,4459 and NGC\,4526, which all have \oiii /\hb\/ significantly less than unity) also show extensive dust in the form of a regular and settled disk. The other objects which show clear signs of dust extinction in their unsharp-masked {\it HST} image (NGC\,2768, NGC\,2974, NGC\,3379, NGC\,3489, NGC\,4150, NGC\,4262, NGC\,4486 and NGC\,5813), however, have \oiii /\hb\/ values around unity, and in some cases, significantly larger. In these latter objects, the dust is either less significant, or has an irregular (but apparently planar) distribution. This suggests that star-formation occurs more readily where the dust indicates cold, relaxed disk-like structures.\looseness-2

Although we do not have sufficient emission-line diagnostics to interpret our line ratios directly, in some cases there is additional evidence that very low \oiii /\hb\/ values indicate star-formation. In NGC\,4526, the regions of strong dust extinction closely correspond to regions of low \oiii /\hb\/ values, which in turn correspond to regions of young stellar populations. The metallicity of the stars in these regions also appears to decrease, indicating the gas is probably not very metal-rich, thus supporting our interpretation that star-formation is the dominant ionisation mechanism in those regions. The counter-example to this is NGC\,3489, which also shows dust and young stellar populations of moderate metallicity. This galaxy, however, exhibits the highest \oiii /\hb\/ in our sample, illustrating the limitations of interpreting \oiii /\hb\/ on its own. Clearly additional diagnostics, such as \ha\/ and \niii, would be valuable in discriminating between the true ionisation mechanisms in our galaxies, and establishing the role of dust and gas dynamics in the star-formation process.

\subsection{Line Strengths}
\label{sec:results_ls}

Figure \ref{fig:ngc1023} includes maps of the six Lick indices within our \oasis\/ spectral range. We measure four separate iron indices, of which Fe5015 is affected by \oiii\/ emission. It is reassuring that, even in galaxies with significant emission, the morphology of the different iron index maps is generally consistent (e.g. the central peak in NGC\,3032, or the dusty region in NGC\,4526). The two other measured indices, namely \hb\/ and \mgb, are often used as diagnostics of stellar population age and metallicity (respectively), usually in combination with additional iron features.

Applying the method described in Section \ref{sec:ssp}, we also present maps of luminosity-weighted stellar age, metallicity and abundance ratios using the SSP models of \citet{thomas03}. Fig.~\ref{fig:sigma_pops} presents the age, metallicity \zh\/ and abundance ratio \afe\/ derived from the 4\arcsec\/ aperture spectrum described in previous sections, each plotted as a function of the stellar velocity dispersion measured inside the same aperture. The values and errors (marginalised over the other two free parameters) of these parameters are given in Table \ref{tab:aperture_pops}, as is the fraction of the galaxies' effective radii covered by the 4\arcsec\/ aperture. Given our limited field of view and the large range in effective radius of our targets, no attempt was made to account for the different intrinsic aperture sizes. We perform a robust straight-line fit to the data, the parameters of which are:

\begin{equation}
\label{equ:age_sig}
\log t      = -3.95 + 2.04 \log \sigma \mathrm{,}
\end{equation}
\begin{equation}
\label{equ:z_sig}
[Z/H]       =  0.02 + 0.14 \log \sigma \mathrm{,}
\end{equation}
\begin{equation}
\label{equ:afe_sig}
[\alpha/Fe] = -0.45 + 0.26 \log \sigma \mathrm{.}
\end{equation}

As found by many authors \citep[e.g.][]{carollo93,trager2000,kuntschner01,proctor02,bernardi05,thomas05}, we find that the stellar population parameters correlate with velocity dispersion. This is in addition to the empirical trends of individual indices and combinations of indices with velocity dispersion, such as the well-known tight correlation with Mg absorption strength \citep[e.g.][and others]{bernardi98,colless99,worthey03}. We find rather a strong trend with age, such that lower-dispersion galaxies show younger luminosity-weighted ages. This is consistent with the `downsizing' scenario of galaxy evolution \citep{cowie96}, where the characteristic mass of star-forming galaxies decreases with time. Our small magnitude-limited sample is biased by some degree at lower velocity dispersions (masses) towards young objects, resulting in a steeper trend. Given the small observed range in mass-to-light ratio (M/L) with velocity dispersion \citep[e.g.][]{bernardi03}, however, this effect is small, and will not change the trend of decreasing age with dispersion. We do note, however, that we derive a significantly stronger gradient than \cite{thomas05}.

A correlation also exists with \afe, such that higher-dispersion galaxies have larger \afe\/ values. This trend has been inferred empirically from index ratios \citep[e.g.][]{kuntschner00}, and also measured directly (as here) using population models with variable abundance ratios \citep[e.g.][]{proctor02,thomas05}. This trend implies a variation in star-formation timescales, such that higher-dispersion (higher-mass) systems have experienced more rapid star formation than lower-dispersion galaxies. Linking to the trend with age, this rapid star-formation phase happened at early epochs, leading to the existence of massive star-forming systems at high redshift, perhaps explaining the existence of a population of bright sub-mm sources \citep[e.g.][]{smail02}. We find only a weak relation between decreasing metallicity and velocity dispersion, and there is a broad range of metallicity values for any given velocity dispersion. Other authors \citep[e.g.][]{thomas05} find a clearer relation, albeit in the same sense as ours.

The inhomogeneous aperture and small sample size strongly limit the generality of trends we find from these aperture measurements, although they are in broad agreement with other authors. In the following sections, we focus on the structure found in the two-dimensional maps of population parameters.

\begin{figure}
 \begin{center}
  \includegraphics[width=8cm]{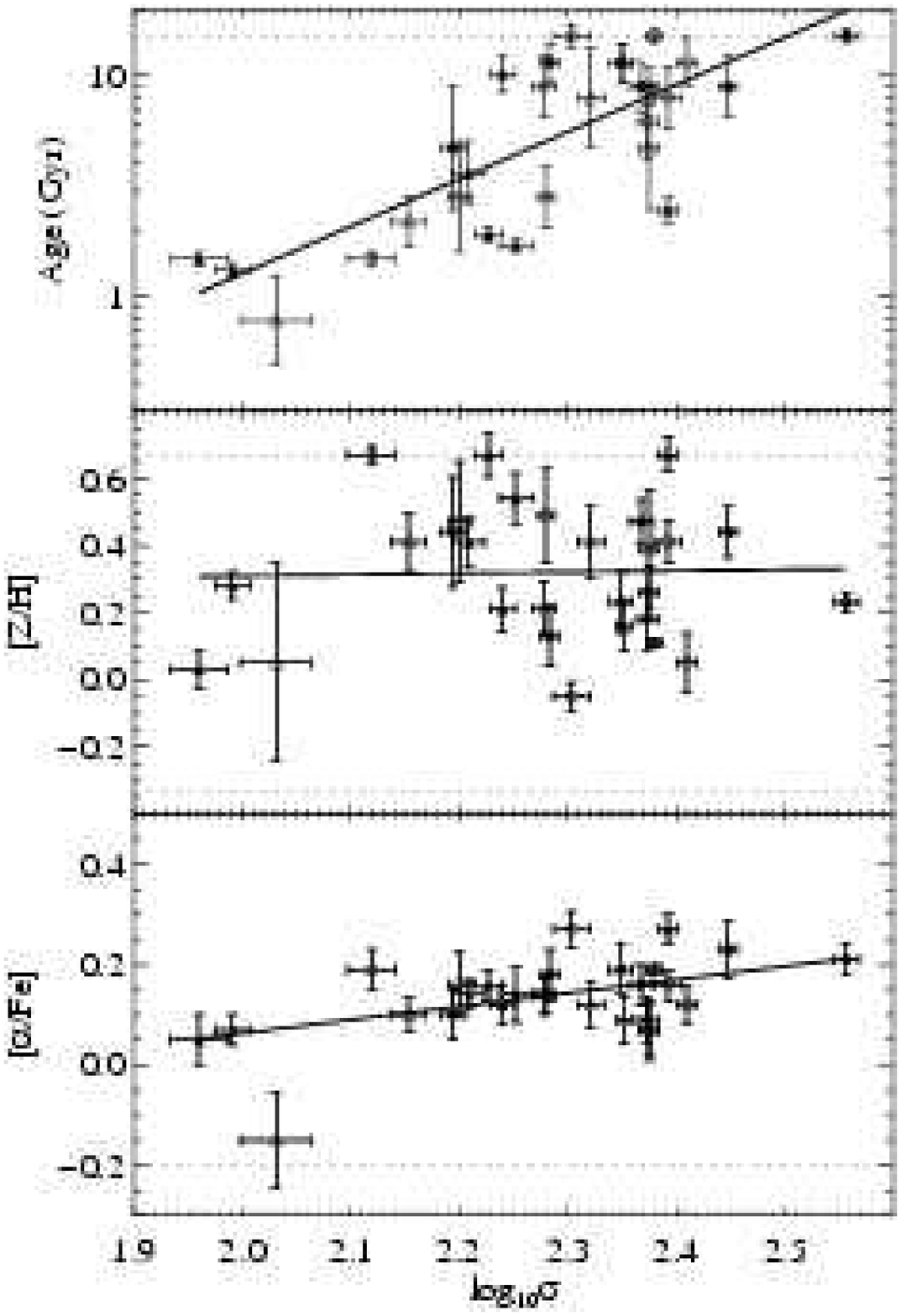}
 \end{center}
 \caption{Relations between stellar velocity dispersion $\sigma$ and age, metallicity and abundance of our galaxy sample, measured using the \citet{thomas03} stellar population models. Values and errors are given in Table~\ref{tab:aperture_pops}. The models are fitted to the six Lick indices in our spectral range, measured from the 4\arcsec\/ radius circular aperture integrated spectrum described in previous sections. Symbols indicate the RC3 classification: circles = ellipticals, triangles = lenticulars. Filled symbols indicate the cluster environment sample, open symbols indicate field galaxies. Overplotted are robust straight line fits to our data, with parameters given in eqs.(\ref{equ:age_sig}-\ref{equ:afe_sig}). Dotted horizontal lines indicate the limits of the SSP model range.}
 \label{fig:sigma_pops}
\end{figure}

\renewcommand\arraystretch{1.3}
\setlength{\tabcolsep}{1.5mm}
\begin{table}
 \begin{center}
  \caption{Central stellar population parameters, determined from six Lick indices measured on a single \oasis\/ spectrum integrated within a 4\arcsec\/ circular aperture.}
  \begin{tabular}{lccccc}
  \hline
NGC	&R$_{e}$	&$\sigma$	&Age	&[Z/Fe]	&\afe	\\[-4pt]
(1)	&(2)		&(3)		&(4)	&(5)	&(6)	\\
   \hline
1023   &  0.08  & 210 $\pm$  6 &  4.7$^{+3.6}_{-2.0}$ &  0.52 $\pm$  0.12 &  0.14 $\pm$  0.05 \\
2549   &  0.20  & 142 $\pm$  5 &  2.5$^{+1.2}_{-0.8}$ &  0.37 $\pm$  0.11 &  0.09 $\pm$  0.04 \\
2695   &  0.19  & 225 $\pm$  4 & 10.2$^{+3.9}_{-2.8}$ &  0.22 $\pm$  0.08 &  0.09 $\pm$  0.05 \\
2699   &  0.29  & 132 $\pm$  7 &  1.5$^{+0.2}_{-0.2}$ &  0.67 $\pm$  0.04 &  0.21 $\pm$  0.05 \\
2768   &  0.06  & 191 $\pm$  3 &  2.5$^{+0.7}_{-0.6}$ &  0.60 $\pm$  0.11 &  0.16 $\pm$  0.04 \\
2974   &  0.17  & 236 $\pm$  5 &  7.9$^{+3.7}_{-2.5}$ &  0.15 $\pm$  0.04 &  0.07 $\pm$  0.05 \\
3032   &  0.24  & 107 $\pm$  8 &  0.9$^{+0.5}_{-0.3}$ & -0.08 $\pm$  0.38 & -0.13 $\pm$  0.11 \\
3379   &  0.10  & 224 $\pm$  6 & 11.5$^{+3.3}_{-2.6}$ &  0.22 $\pm$  0.12 &  0.18 $\pm$  0.08 \\
3384   &  0.15  & 162 $\pm$  5 &  3.2$^{+1.5}_{-1.0}$ &  0.44 $\pm$  0.11 &  0.16 $\pm$  0.04 \\
3414   &  0.12  & 237 $\pm$  5 &  3.6$^{+3.7}_{-1.8}$ &  0.44 $\pm$  0.22 &  0.09 $\pm$  0.05 \\
3489   &  0.21  &  98 $\pm$  4 &  1.7$^{+0.2}_{-0.2}$ &  0.22 $\pm$  0.08 &  0.03 $\pm$  0.04 \\
3608   &  0.10  & 190 $\pm$  5 &  8.9$^{+4.1}_{-2.8}$ &  0.22 $\pm$  0.12 &  0.14 $\pm$  0.04 \\
4150   &  0.27  &  91 $\pm$  6 &  1.5$^{+0.3}_{-0.3}$ &  0.07 $\pm$  0.08 &  0.07 $\pm$  0.06 \\
4262   &  0.40  & 174 $\pm$  5 & 10.2$^{+2.9}_{-2.3}$ &  0.22 $\pm$  0.08 &  0.12 $\pm$  0.05 \\
4382   &  0.06  & 179 $\pm$  6 &  1.7$^{+0.3}_{-0.2}$ &  0.52 $\pm$  0.15 &  0.12 $\pm$  0.06 \\
4459   &  0.11  & 169 $\pm$  5 &  1.9$^{+0.1}_{-0.1}$ &  0.67 $\pm$  0.04 &  0.14 $\pm$  0.04 \\
4473   &  0.15  & 192 $\pm$  4 & 13.2$^{+4.0}_{-3.1}$ &  0.07 $\pm$  0.12 &  0.16 $\pm$  0.05 \\
4486   &  0.04  & 361 $\pm$  9 & 15.1$^{+1.1}_{-1.0}$ &  0.22 $\pm$  0.04 &  0.19 $\pm$  0.03 \\
4526   &  0.10  & 247 $\pm$  6 &  2.8$^{+0.2}_{-0.2}$ &  0.67 $\pm$  0.04 &  0.25 $\pm$  0.05 \\
4552   &  0.13  & 281 $\pm$  5 &  8.9$^{+3.4}_{-2.5}$ &  0.44 $\pm$  0.11 &  0.23 $\pm$  0.05 \\
4564   &  0.19  & 156 $\pm$  4 &  3.6$^{+3.3}_{-1.7}$ &  0.52 $\pm$  0.15 &  0.12 $\pm$  0.05 \\
4621   &  0.09  & 233 $\pm$  6 &  6.9$^{+3.9}_{-2.5}$ &  0.52 $\pm$  0.08 &  0.16 $\pm$  0.05 \\
5198   &  0.16  & 201 $\pm$  8 & 15.1$^{+2.0}_{-1.8}$ & -0.08 $\pm$  0.04 &  0.30 $\pm$  0.04 \\
5813   &  0.08  & 240 $\pm$  4 & 15.1$^{+1.1}_{-1.0}$ &  0.15 $\pm$  0.04 &  0.19 $\pm$  0.03 \\
5831   &  0.11  & 158 $\pm$  4 &  3.2$^{+3.4}_{-1.7}$ &  0.44 $\pm$  0.22 &  0.14 $\pm$  0.06 \\
5845   &  0.87  & 237 $\pm$  5 &  8.9$^{+2.6}_{-2.0}$ &  0.30 $\pm$  0.11 &  0.09 $\pm$  0.07 \\
5846   &  0.05  & 247 $\pm$  7 &  7.9$^{+5.2}_{-3.2}$ &  0.44 $\pm$  0.11 &  0.16 $\pm$  0.05 \\
5982   &  0.15  & 257 $\pm$  6 & 13.2$^{+2.8}_{-2.3}$ &  0.00 $\pm$  0.08 &  0.12 $\pm$  0.04 \\
   \hline
  \label{tab:aperture_pops}
  \end{tabular}
 \end{center}
Notes:
(1)~NGC number.
(2)~Fraction of effective radius covered by 4\arcsec\/ aperture.
(3)~Velocity dispersion measured within 4\arcsec\/ aperture, in \kms.
(4)~Luminosity-weighted age measured within 4\arcsec\/ aperture, in Gyr.
(5)~\zh\/ measured within 4\arcsec\/ aperture.
(6)~\afe\/ measured within 4\arcsec\/ aperture.
Errors are given as (one-dimensional) 1$\sigma$, marginalised over the other two parameters.
\end{table}

\subsubsection{Luminosity-weighted age}

Most objects in our sample show almost no significant variations in age within the \oasis\/ field of view. There are a number of notable exceptions to this, however, which show sharp changes in the spatial distribution of stellar ages. Three galaxies in our sample show a clear decrease in age towards the galaxy centre (NGC\,3032, NGC\,3489 and NGC\,4150). The most extreme case is NGC\,3032, which reaches a minimum age of 0.5~Gyr in the central \oasis\/ bin. These sharp drops in age derived from the models are driven by sharp increases in the \hb\/ line-strength clearly visible in the maps. Also, in both NGC\,3032 and NGC\,3489, the peak in \hb\/ absorption is coincident with a decrease in the \hb\/ emission equivalent width, indicating that the observed strong \hb\/ absorption is not an artifact of overestimated emission subtraction.\looseness-2

Another interesting case is NGC\,4526. The age map clearly shows a young population associated with the dusty disk embedded in the central 10\arcsec\/ of this galaxy. The foreground (northern) side of the disk shows the age decrease most clearly, although the structure can be traced also on the other side of the bulge. There is a sharp rise in stellar age at the northern edge of the field, which can be associated with the corresponding edge of the dust-lane in the unsharp-masked {\it HST} image. 

\subsubsection{Metallicity}

All galaxies in our sample show constant or increasing metallicity towards the galaxy centre, although the variations are generally small within the \oasis\/ field. In Paper VI, we found that flattened, fast-rotating galaxies show an \mgb\/ distribution which is flatter than the apparent surface brightness in the \sauron\/ data, suggesting the presence of a metallicity- or abundance ratio-enhanced disk embedded in these objects. We do not expect to find strong evidence of this effect in our \oasis\/ maps, since the disk component is not expected to dominate over the spheroidal component in the central regions. For some of the fast-rotating, flattened objects, however, we do find weak evidence of a metallicity distribution that is flatter than the surface brightness distribution (namely, NGC\,3384, NGC\,4473 and NGC\,5845). Of these, NGC\,5845 is particularly interesting. The stellar kinematics indicate the presence of an embedded disk (Section \ref{sec:res_stellar_kin_align}), which is clearly visible in the unsharp-masked {\it HST} image. In the map of metallicity, there is evidence for a weak enhancement along the major axis within the central 2\arcsec, which may be associated with this thin disk component. 

In the three galaxies with young nuclei mentioned above, the young population is associated with an increase in the apparent metallicity, implying that the young stars (which are dominating the light in these regions) are more metal rich than those at larger galactic radii. By contrast, within the young disk of stars associated with the dust structure in NGC\,4526, the metallicity shows a distinct decrease compared to the metal-rich bulge region. The absolute metallicity of the newly formed stars in all cases is, however, similar, and generally super-solar.

\subsubsection{Abundance ratio}

With the exception of two galaxies, the maps of abundance ratio either show a weak increase towards the centre, or are rather flat, showing little significant structure within our field of view. NGC\,4150 shows a weak {\em decrease} in \afe\/ towards the centre. In this case, the decrease is also coincident with younger stars and a decoupled kinematic component. This is also true for the disk of young stars in NGC4526, where the abundance decreases within the dusty region. With the limitations of interpreting integrated populations with SSP models in mind (see \ref{sec:ssp}), given the similar spatial structure in the maps of age and abundance for these two galaxies, it seems that the decrease in apparent abundance is due to the presence of the young stars.

NGC\,3032 shows a particularly low \afe\/ in the outer parts of our field, giving the only sub-solar abundance ratio in our sample. In the central regions, however, the abundance becomes near-solar coincident with the strong decrease in age. This relative enhancement of $\alpha$-elements with respect to iron suggests that the young stellar population is strongly enriched with material expelled from short-lived massive stars via type II supernovae. This implies that either the star-formation timescale within the central component is markedly shorter than in the main body of the galaxy, or that the initial mass function (IMF) is skewed towards higher-mass star in this region. A combination of these effects is also a possible explanation \citep[e.g.][]{matteucci94}.

Several authors have reported that KDCs show enhanced metallicity and/or abundance, inferred from changes in line-strength gradients coincident with the kinematic structure \citep[e.g.][]{bender92,carollo94b,morelli04}. We note that two of the objects mentioned above, NGC\,4150 and NGC\,3032, contain stellar KDCs, but show contrasting changes in abundance ratio within the KDC. These KDCs also show distinctly young ages (see following Section), unlike the objects studied previously, and so may not be comparable. Full consideration of this phenomenon for our sample requires also the use of \sauron, since most of the KDCs are larger than the \oasis\/ field. We therefore leave further discussion of this issue to future papers of this series.

\section{Discussion}
\label{sec:discussion}

\subsection{Stellar kinematics of young galaxy centers}

From Paper VI, \sauron\/ observations show that galaxies exhibiting the strongest global \hb\/ absorption (measured within one effective radius) also tend to show an increased \hb\/ line strength towards the central regions. Three of the four galaxies with global \hb\/ absorption around 3~\AA\/ or stronger are in our \oasis\/ sample, namely NGC\,3032, NGC\,4150 and NGC\,3489. As discussed above, these objects show a sharp decrease in their central luminosity-weighted age in our \oasis\/ field, indicating very young stellar components that are in some cases not spatially well-resolved, even at the spatial resolution of our data. Although these galaxies also show young or intermediate aged stars across their main body, their central regions ($\lsim$ few 100~pc) are significantly younger. These young centers are responsible for driving the spread in age covered by the \sauron\/ sample galaxies to include relatively low ages ($< 2$~Gyr).

From {\it HST} imaging studies, a number of early-type galaxies are found to contain point-like nuclei, unresolved even at {\it HST} resolutions \citep[$\sim$5~pc; e.g.][]{carollo97a,rest01,laine03,lauer05,cote06}. These nuclei are generally blue in broad-band colours, and have been attributed to both non-thermal sources and nuclear star-clusters. We differentiate these nuclei from the young centers found in our data, as we do not have sufficient spatial resolution to directly connect the \hb-enhanced structures to such small components. Also, the nuclei found with {\it HST} exist in early-type galaxies of a range of masses, whereas the young centers found in our data are confined to the low-mass objects. A second important difference is that the central structures found with our data are found to be young spectroscopically, thus largely avoiding the problems of the age-metallicity degeneracy and dust absorption that often hamper the interpretation of broadband colour gradients. We do not exclude, however, that the young central structures we observe may correspond to nuclei at 5~pc scales.

As well as the unresolved blue nuclei found with {\it HST}, more extensive blue components are often seen in so-called `E+A' galaxies: objects which are morphologically similar to early-type galaxies, but exhibit strong Balmer absorption typical of A-stars \cite[e.g.][]{dressler83,zabludoff96}. Moreover, compact blue components are found in these systems, unresolved at scales of $\sim 100$~pc \citep{yang04}. These are thought to be star clusters formed during a recent star-burst. Their locations within the main galaxy can vary, however, and they are not always confined only to the central parts. E+A galaxies are a somewhat mixed class of objects, whose members often show disturbed morphologies at large scales, suggesting a recent major merger. The morphologies of our young-center galaxies do not show obvious signs of a recent major merging event (e.g. tails, shells). However, the compact nature of our young components in our galaxies suggests that they may have experienced a similar mode of localised star-formation.

Clusters of recently formed stars are often found in galaxy centers of later type than E/S0 \citep[e.g. spirals and late-type disks:][and references therein; early-type bulges: \citealt{maoz96}; M31: \citealt{lauer98,brown98}; Seyfert galaxies: \citealt{heckman97,delgado98}]{carollo97b,carollo98,carollo99,carollo02}. The best candidate in our sample for such a cluster is the youngest object, NGC\,3032, an SAB0 galaxy that shows a bright central peak and possible circum-nuclear ring structure in the unsharp-masked {\it HST} image (see Fig. \ref{fig:ngc3032}), which could be forming stars. For the next-youngest objects, NGC\,3489 (SAB0) and NGC\,4150 (S0), this explanation is less compelling for their young centers, given the lack of bright central sources or star-forming rings, but this may be due to projection and dust. In any case, both objects exhibit recent circum-nuclear star-formation, and are classified as `transition' objects by \cite{ho97}, having emission-line properties between that those of H{\small II} nuclei and LINERS \citep{ho93}.

It has long been suggested that some early-type galaxies show young apparent ages due to a small fraction (by mass) of young stars within the galaxy \citep{worthey94,trager2000}. This `frosting' of young stars has a low mass-to-light ratio; thus while the young stars do not contribute much in terms of mass, their contribution to the integrated light is significant. However, it was not clear where these stars were located in the galaxy, or what triggered their recent formation. In Paper V we have shown that many of our sample galaxies contain large quantities of (ionised) gas that may exist in kiloparsec scale structures, suggesting that there are extended reservoirs of star-forming material available in these galaxies. It would seem, however, that only in the central regions, deep in the galaxy potential, are the conditions suitable to trigger efficient conversion of this material into stars.

Investigating the stellar kinematics of the three youngest galaxies in our sample, the two with the strongest \hb\/ absorption (NGC\,3032, NGC\,4150) harbour small KDCs, which are essentially aligned but counter-rotating with respect to the main body of the galaxy ($\Delta$\kinpa\/$\sim 180$\degree\/ and $\sim 190$\degree\/ respectively). The other galaxy (NGC\,3489) does not show evidence for counter-rotation, but does exhibit a mild twist in its velocity field ($\Delta$\kinpa$\sim 10$\degree) within a 2\arcsec\/ radius, as discussed in Section \ref{sec:results_kin}. Furthermore, the gas in this galaxy shows an elongated structure with enhancements at either end, and open spiral features in the dust extinction (Fig.~\ref{fig:ngc3489}), suggesting the possible presence of a central bar \cite[cf. NGC\,2974, ][]{emsellem03}.

\begin{figure*}
 \begin{center}
  \includegraphics[width=5.8cm, angle=0]{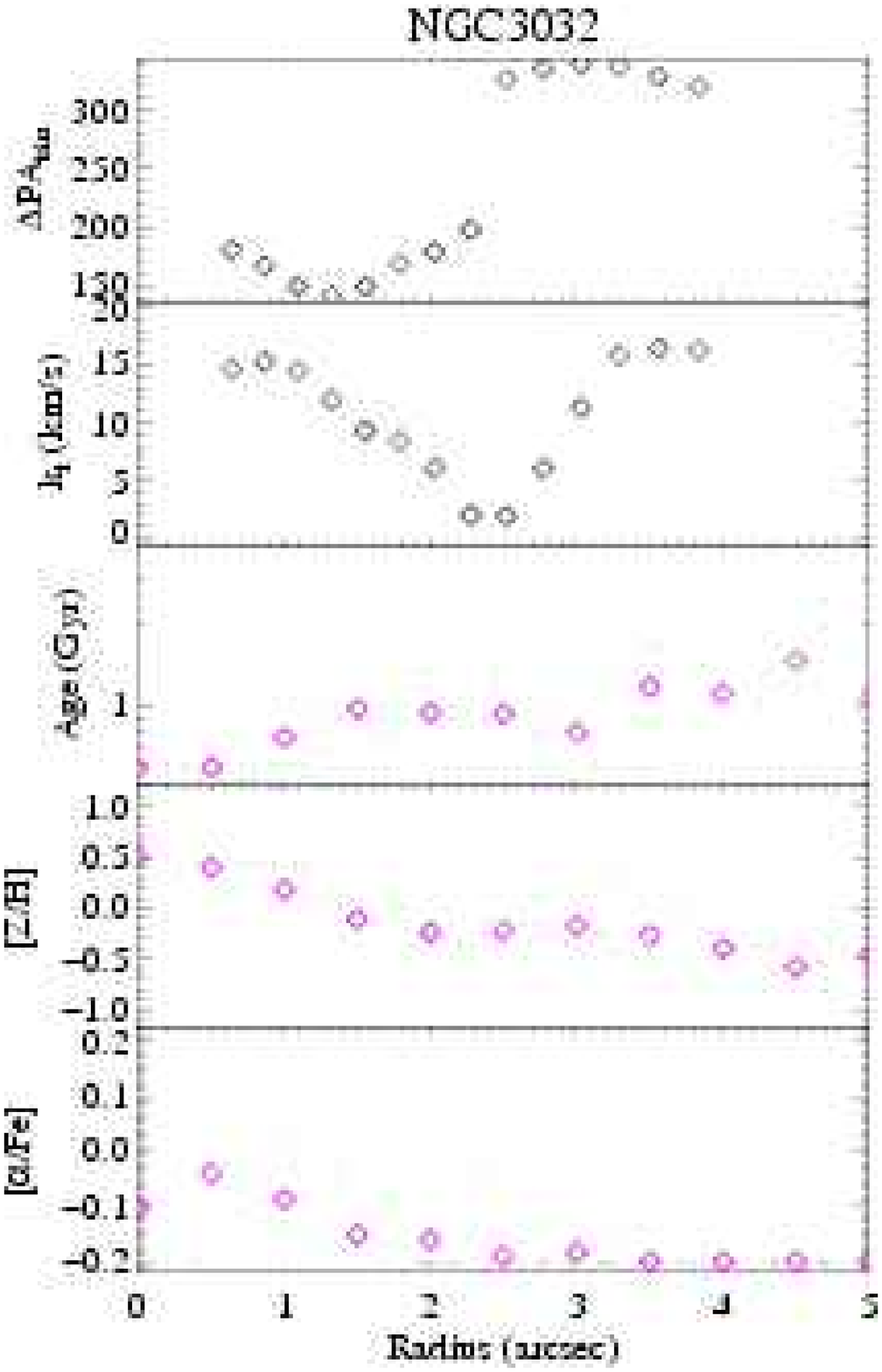}
  \includegraphics[width=5.8cm, angle=0]{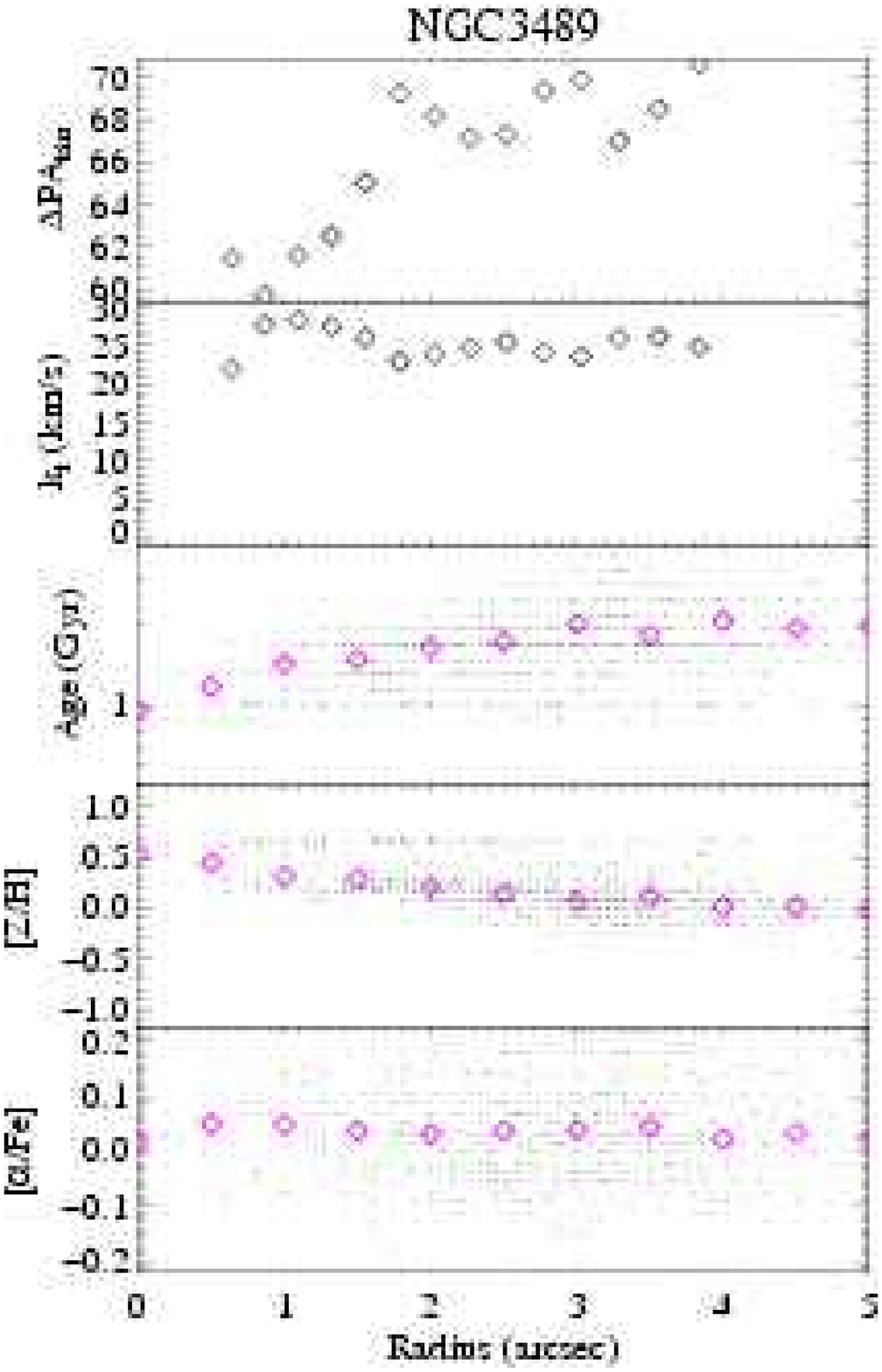}
  \includegraphics[width=5.8cm, angle=0]{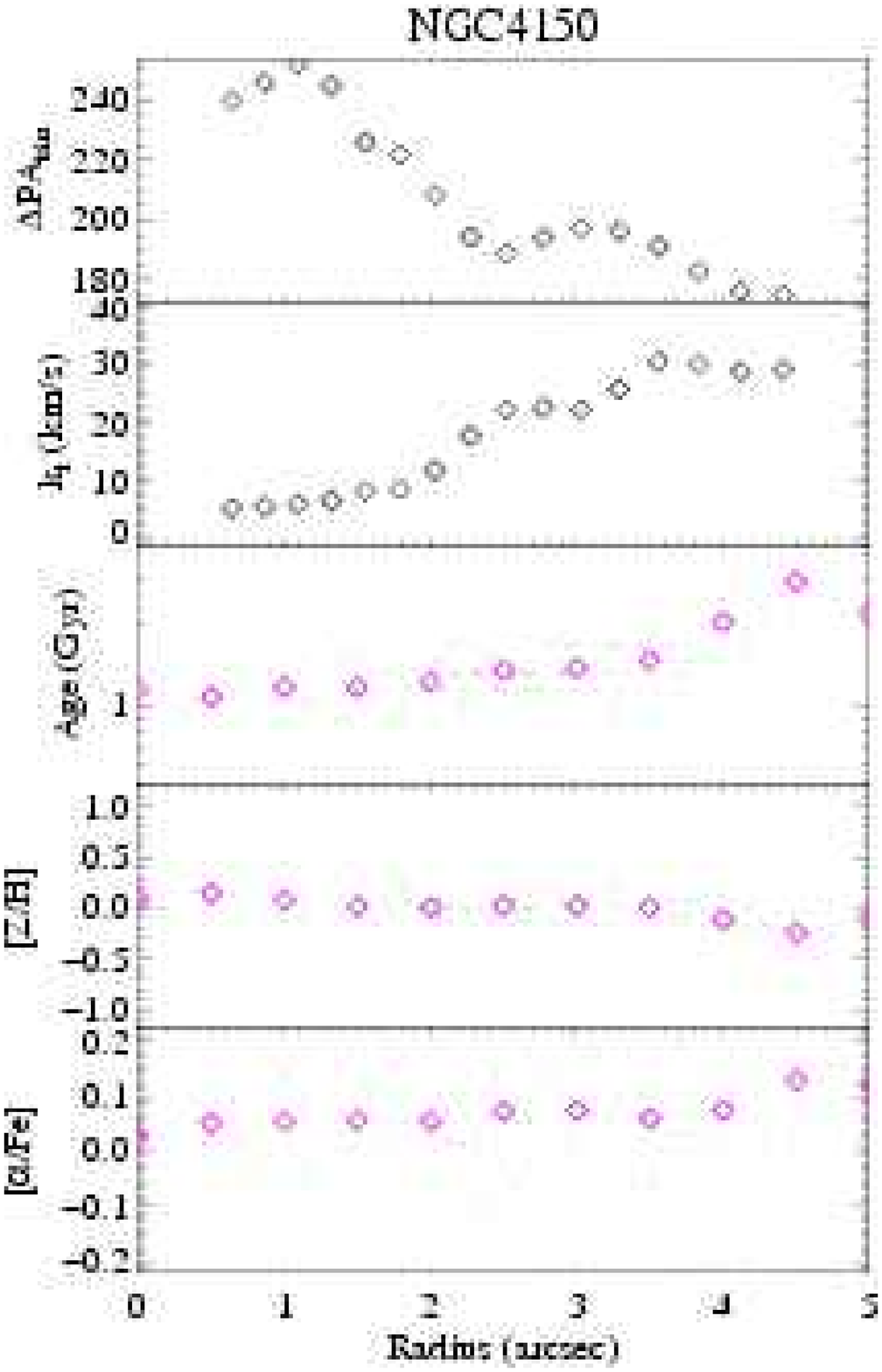}
 \end{center}
 \caption{Radial profiles of relative kinematic position angle ($\Delta$PA$_{\mathrm{kin}}$); maximum rotation velocity, as given by the $k_1$ term of the kinemetric analysis; and stellar population parameters: age, metallicity (Z) and abundance ratio (\afe). Small dots indicate the individual stellar population parameters measured from our index maps. Red diamonds indicate the biweight mean value of these individual measurements within a circular annulus, plotted against the geometric radius of this annulus. The galaxy name is indicated in the title of each panel.}
 \label{fig:kdc_kinemetry}
\end{figure*}

In all three galaxies, the region showing structure in the velocity field is coincident with the young stellar populations, implying that the two phenomena are directly linked. Fig.~\ref{fig:kdc_kinemetry} illustrates this, showing the radial profiles of both the kinemetric analysis (see Section \ref{sec:kinemetry}) and the stellar population parameters. The clearest association between the stellar kinematic structure and the population parameters is for NGC\,3032. The 180\degree\/ jump of the kinematic PA at a radius of $\sim 2$\arcsec\/ and the corresponding minimum in rotation between the outer body and the central KDC coincide closely with a downturn in the age, and upturns in the metallicity and abundance ratio. NGC\,4150 shows less dramatic variations in the kinematics, but is strongly affected by poor seeing. However, \kinpa\/ changes significantly within the central few arcseconds, and the mean rotation is almost zero. This region corresponds to the KDC, and also shows a decrease in age compared to the outer regions. There is also weak evidence of an increase in metallicity and a decrease in \afe. NGC\,3489 does not exhibit a counter-rotating component. Fig.~\ref{fig:kdc_kinemetry} shows, however, that \kinpa\/ changes by $\sim 10$\degree\/ within the region where the age decreases from 2 to 1~Gyr. The metallicity  also increases more steeply within this region. In all three cases, a change in the stellar populations, and most notably a decrease in the luminosity-weighted age, is accompanied by a change in the stellar kinematics, suggesting a link between the two.

\subsection{Stellar ages of KDCs}

In the above discussion, we highlighted that two of the three youngest galaxies in our sample have a KDC associated with the young stars. Here we extend this discussion to examine the stellar populations of all KDCs that have been found so far in the \sauron\/ survey, considering the combined information from both the \sauron\/ and \oasis\/ data wherever possible.\looseness-2

With integral-field spectroscopy, we can obtain a complete two-dimensional map of the KDC, allowing us to measure accurately the orientation and spatial extent of the structure, as well as determining the stellar populations. The \sauron\/ representative sample contains numerous examples of KDCs (Paper III), many of which are also in our present subsample. To these, we add the three KDCs which are visible in our \oasis\/ data, though not resolved with \sauron: NGC\,3032, NGC\,4382, and NGC\,4621.

Table \ref{tab:kdcs} lists the galaxies from the full \sauron\/ sample that we consider to harbour a KDC. These objects are selected as they exhibit subcomponents that either rotate around a different axis to the main galaxy, or that rotate around the same axis, but with an opposite sense of rotation. With this definition, we do not consider dynamically decoupled structures which co-rotate with the main galaxy, such as nuclear disks, even though there are several such structures in our sample. Misaligned and counter-rotating components offer the advantage, however, of being clearly distinguished from the rest of the galaxy, and may indeed have a distinct formation process from the co-rotating case.

Considering both the \oasis\/ and \sauron\/ data, there is a large range of apparent sizes for the KDCs in the \sauron\/ sample. We estimate the size of the KDC by using the isolating region in the velocity map which surrounds the KDC, where the superposition of the two components gives a local minimum in the mean velocity. Although approximate, this gives a straightforward estimate of the projected extent of the central structure. We also include three galaxies which show evidence of a KDC in the \sauron\/ velocity fields of Paper III, but for which no \oasis\/ observations were obtained (NGC\,4458, NGC\,7332 and NGC\,7457). In addition, we include the well-known KDC in NGC\,4365, which was observed with \sauron, though not as part of the main survey \citep{davies01}. We re-measure the KDC diameter for this galaxy using our method, and take the age and \hb\/ line-strength from \citet{davies01}. For each KDC, the data with the smallest PSF and with a large enough field of view to encompass the isolating ZVC was used to estimate the diameter. The apparent diameters are converted to intrinsic sizes using the surface brightness fluctuation (SBF) distance measurements of \citet{tonry01}. Distances, and apparent and intrinsic KDC diameters are listed in Table \ref{tab:kdcs}.

Table \ref{tab:kdcs} also lists the central value of the \hb\/ absorption-line index measured from a spectrum integrated within a 1\arcsec\/ circular aperture centred on the galaxy, using \oasis\/ data where possible. This index was combined with \mgb\/ and available iron indices within the same aperture to estimate the luminosity-weighted age, as in Section \ref{sec:results_ls}. For \sauron\/ age estimates, only the Fe5015 and Fe5270 iron indices are available.

\begin{table}
\setlength{\tabcolsep}{0.5mm}
 \begin{center}
  \caption{List of galaxies with KDCs from \sauron\/ and \oasis\/ stellar
kinematics.}
  \begin{tabular}{lccccc }
\hline
NAME    & Dist. & Diameter  & Diameter		& Age   & Rotation \\
(NGC)   & (Mpc) & (\arcsec) & (pc)        	&(Gyr)  &          \\
\hline
3032  &  22.0 (3.0) &   3.5  &    372 (73)    &   0.50 (0.1) & F \\
3414  &  25.2 (4.1) &  23.0  &   2813 (465)   &  18.00 (2.3) & S \\
3608  &  22.9 (1.5) &  21.0  &   2332 (164)   &  17.00 (1.0) & S \\
4150  &  13.7 (1.6) &   4.0  &    266 (45)    &   1.00 (0.1) & F \\
4382  &  18.5 (1.2) &   3.5  &    313 (49)    &   3.00 (0.3) & F \\
4458  &  17.2 (1.0) &  10.0  &    834 (63)    &  17.00 (1.5) & S \\
4621  &  18.3 (1.8) &   2.0  &    177 (47)    &  18.00 (2.1) & F \\
5198  &  36.3 (3.5) &   6.0  &   1056 (134)   &  15.00 (2.5) & S \\
5813  &  32.2 (2.8) &  20.0  &   3123 (280)   &  17.00 (2.5) & S \\
5831  &  27.2 (2.2) &  18.0  &   2370 (203)   &   9.00 (1.2) & S \\
5982  &  41.9 (4.0) &  10.0  &   2030 (220)   &   8.00 (1.5) & S \\
7332  &  23.0 (2.2) &   4.0  &    446 (70)    &   3.50 (0.2) & F \\
7457  &  13.2 (1.3) &   4.0  &    256 (41)    &   2.00 (0.2) & F \\[4pt]
4365  &  20.4 (1.7) &  15.0  &   1484 (130)   &  14.00 (2.5) & S  \\
\hline
\label{tab:kdcs}
\end{tabular}
\end{center}
Notes: Each column gives (from left to right): galaxy name; distance from surface brightness fluctuations of \citet{tonry01}; estimated apparent diameter from \sauron\/ or \oasis\/ velocity field; intrinsic diameter in pc; luminosity-weighted age estimate from line-strengths measured within 1\arcsec\/ using the population models of \citet{thomas03}; galaxy classification according to the measured value of specific angular momentum within one effective radius (Paper IX, in preparation). Errors for each parameter are given in parentheses. Errors in distance come from \cite{tonry01}, and the error in diameter combines the distance error with an estimated accuracy of 0\farcsec5 for measuring the KDC diameter from the velocity map.\looseness-2
\end{table}

The values given in Table \ref{tab:kdcs} cover a very broad range of parameter space, spanning more than an order of magnitude in both size and luminosity-weighted age. Fig.~\ref{fig:kdc_size_age} illustrates this, showing that the large ($\ge 1$~kpc) KDCs have predominantly old ($\ge 10$~Gyr) populations. The group of small ($< 1$~kpc) KDCs, by contrast, exhibit populations spanning the full range of ages, from $< 1$~Gyr to 15~Gyr, with five of the six small KDCs younger than 5~Gyr. The colours used in this figure indicate the class of rotation (also given in Table \ref{tab:kdcs}), being either a fast-rotating object, mainly supported by rotation (blue symbols); or a slow-rotating object, mainly supported by dynamical pressure (red symbols). There are clearly no examples of slow-rotating objects with young, compact KDCs.

In addition, the small, young KDCs observed with \oasis\/ are all very close (i.e. within 10-20\degree) to being counter-rotating (i.e. $\Delta$\kinpa$\sim 180$\degree), as far as our data can reveal. In the cases of NGC\,4150 and NGC\,4621, our measurements of \kinpa\/ for the KDC are strongly affected by seeing effects. For NGC\,3032 and NGC\,4382, however, the measurements are robust. The other two young KDCs (NGC\,7332 and NGC\,7457) are only observed with \sauron, and limited spatial information prevents a detailed measurement of the KDC \kinpa. By subtracting a model of the large-scale rotation, \citet{falcon04} estimate the \kinpa\/ of the KDC in NGC\,7332 as 155\degree\/ (25\degree\/ from the photometric major-axis), consistent with our general finding of near-aligned counter-rotation. The small misalignment of these KDCs raises the question of the existence of co-rotating counterparts. Such components have a more subtle effect on the velocity field, having similar rotation amplitude to the background large-scale rotation. For this reason, we limit the current analysis to the clearly visible KDCs, and note that the majority of young nuclei in our sample are in fact KDCs of this kind.\looseness-2

\begin{figure}
 \begin{center}
  \includegraphics[width=8cm]{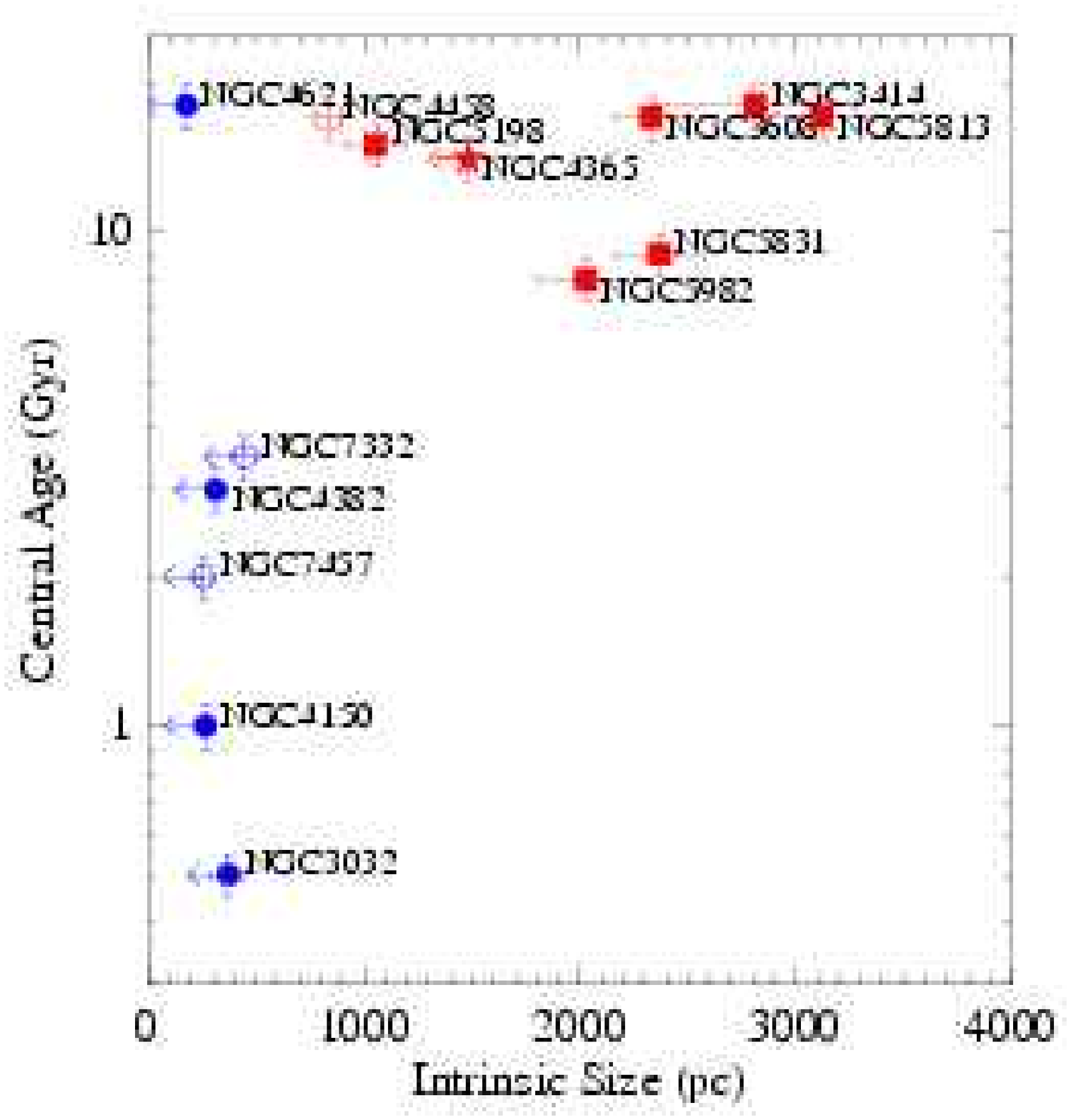}
 \end{center}
 \caption{Intrinsic size of KDCs versus their central luminosity-weighted age. Closed symbols show objects observed by both \sauron\/ and \oasis. Open symbols show objects only observed with \sauron. Red points indicate slow-rotators and blue points indicate fast rotators, as defined in Paper IX (in preparation). Arrows indicate where the measured diameter of the KDC is less than four times the seeing FWHM, implying that the KDC is not well-resolved and may be smaller than measured here.}
 \label{fig:kdc_size_age}
\end{figure}

An indicator of how significant the KDC is in terms of the global galaxy properties is the fraction of the galaxy's mass which is contained in the KDC. Accurately estimating this quantity is difficult. Perhaps the most accurate but complex way is via dynamical modelling, where the KDC can be separated in phase-space directly \citep[e.g.][]{cappellari02}. Such work is beyond the scope of this paper, but will be addressed in a future paper of this series. A very simple way to quantify the significance of the KDC is to consider its size with respect to the effective radius \re. The young KDCs generally extend to only $\lsim 0.1$~\re, whereas the older KDCs are generally 0.2 - 0.3~\re\/ in size. Moreover, the young KDCs have a low mass-to-light ratio, thus contributing significantly to the light, but relatively little in mass.\looseness-2

\subsection{Age distribution of compact KDCs}

Fig. \ref{fig:kdc_size_age} shows that the compact (i.e. smaller than 1~kpc diameter) counter-rotating KDCs occur only in the fast-rotating galaxies of our sample. Figure \ref{fig:kdc_age_hist} shows that the distribution of fast-rotating galaxies with counter-rotation in our sample is skewed towards objects with younger global luminosity-weighted mean ages. Five of the seven objects younger than $< 5$~Gyr show a counter-rotating core (and one of these has still to be observed with high spatial resolution, and so may yet reveal a KDC). Only one of the remaining thirteen objects older than 5~Gyr shows a detectable KDC.

%
%
\begin{figure}
\centering
\includegraphics[width=8cm]{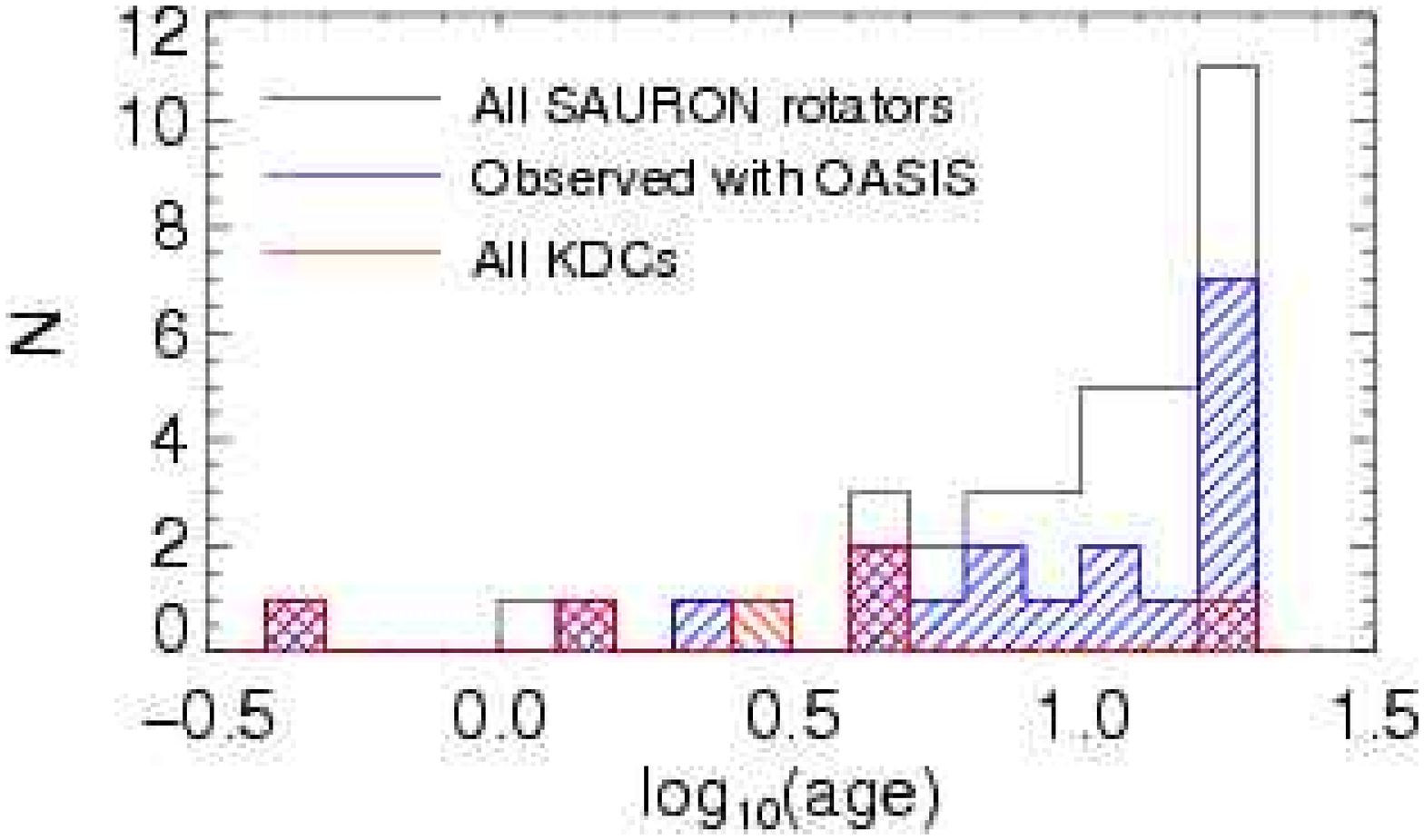}
\caption{Histogram of ages for the fast-rotating galaxies from the {\tt SAURON} survey. From the complete \sauron\/ sample (black line), we indicate those observed so far with {\tt OASIS} (blue), and those with a counter-rotating KDC (red). Ages are derived from the \sauron\/ lines-strengths of Paper VI.}
\label{fig:kdc_age_hist}
\end{figure}

The higher occurrence of counter-rotation in young galaxies may simply be the effect of luminosity weighting. These small components contribute relatively little in mass to the galaxy, but are still visible above the background of the main body because their young age (and therefore low M/L) makes them very bright. We explore this with the following `toy' model. Using the measured central line-strengths of NGC\,4150, we constrain the amount of mass of young stars within the central aperture that can be added to a background `base' population (assumed to be that of the outer parts of the galaxy). Our \sauron\/ data indicate an age of $\sim 5$~Gyr for the outer body of this galaxy. The central \hb\/ value is $\sim 3.6$~\AA, allowing a mass fraction of around 6\% of a 0.5~Gyr population to be added. Note that this is only the contribution within the 1\arcsec\/ line-of-sight, not to the entire system.

We simulate a two-component velocity field using a kinemetry Fourier expansion with ad hoc coefficient profiles, consisting of a large-scale rotating component, with a compact counter-rotating component. Both components have the same velocity dispersion. We assign the two velocity distribution fields to SSP model spectra of \cite{bruzual03}, assigning the young population to the compact counter-rotating component, and combine the spectra using the appropriate mass-weighting. Noise is added to the simulated data cube, which is then spatially binned, and the kinematics are extracted using pPXF, consistent with how we treat our observational data. The result is a realistic-looking KDC velocity field (Fig. \ref{fig:kdc_model}). This test demonstrates that our analysis technique (which uses a different library of SSP models as kinematic templates) can reliably recover the velocity distribution, showing that the observed KDCs are not spurious artifacts of our kinematics analysis due to the presence of strong population gradients. 

%
%
\begin{figure*}
\centering
\includegraphics[width=16cm]{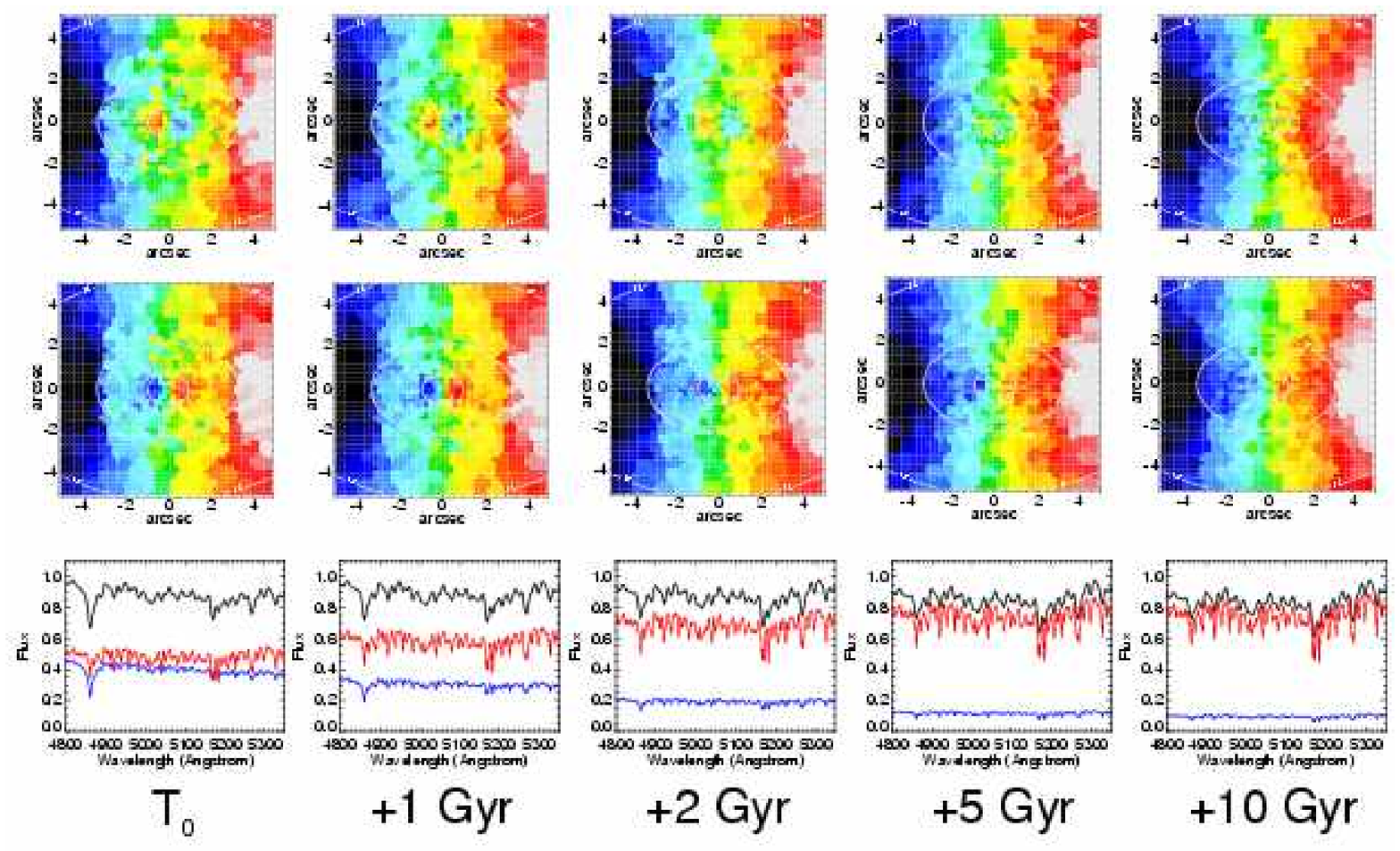}
\caption{Model velocity fields for an evolving counter-rotating (top row) and co-rotating (middle row) young subcomponent. The spectrum of the young decoupled component (blue) and older main component (red) are shown on the bottom row, indicating there luminosity-weighted contribution to the observed spectrum (black).}
\label{fig:kdc_model}
\end{figure*}

We then hold the mass fraction fixed, and `evolve' the populations in step by combining older SSP models with the same age difference. As the KDC population ages, its mass-to-light ratio increases, resulting in a dimming of the KDC stars. The effect is to `fade' the KDC into the background rotation field, as the luminosity-weighted contribution becomes less significant. After 5~Gyr, the KDC is barely visible. This helps explain the apparent lack of intermediate and old aged small KDCs.

Assuming that young co- and counter-rotating components trace the angular momentum vector of the `seed' material which has recently entered the system from outside, both types of component are equally likely to form. It is therefore somewhat surprising that five of the seven youngest galaxies in our sample show evidence of counter-rotating components. These are, however, small-number statistics, and considering errors in the age determination and what maximum age limit one considers, the significance that counter-rotating components tend to be young can be diminished. It is tempting, however, to conclude that counter-rotation is somehow connected to enhanced star-formation. \cite{kapferer05} explore this through combined N-body/hydrodynamic simulations, and find that counter-rotating encounters show a slight tendency to produce higher star-formation rates than their co-rotating counterparts, although there are a large number of exceptions to this.

We have focused our discussion on the counter-rotating case, mainly because these components leave an obvious signature in the observed velocity field. The middle row of Figure \ref{fig:kdc_model} uses the same kinematic components and populations as the top row, but in this case the subcomponent is co-rotating. The impact of the co-rotating component on the total observed velocity field is clearly more subtle, visible only as an additional `pinched' central isovelocity contour. The co-rotating component becomes difficult to separate from the background field after only $\sim 2$~Gyr. For this reason, the intrinsic distribution of co- and counter-rotating central young components could actually be rather similar, but since the co-rotating cases can only be clearly identified at very young ages, our sample of young galaxies is currently too small to determine this reliably.
}

\subsection{Interpretation}

The existence of KDCs has been known since the mid 1980's from long-slit studies \citep[e.g.][]{franx88, bender88}, and integral-field studies have reinforced this \citep[e.g.][Paper III]{davies01}, allowing their full kinematic structure to be mapped. Since the discovery of these structures, they have often been used as evidence that collisionless galaxy mergers have played a significant role in forming early-type galaxies, where the tightly-bound core of one galaxy becomes embedded in the centre of the remnant \citep[e.g.][]{kormendy84, bockelmann2000}. This scenario can naturally account for the often misaligned orientation of the KDC, as well as the dynamically hot nature of the remnant galaxy. 

An alternative explanation for the formation of KDCs is that they `grow' via star-formation inside the host galaxy. To account for misalignment of the KDC, the star-forming material is assumed to be accreted from outside the host galaxy \citep[e.g.][]{hernquist91,weil93}.\looseness-2

In both of these scenarios, the KDC should show distinct stellar populations from the host galaxy. A KDC `grown' inside the main galaxy would be composed of younger stars than its host. A KDC formed as the denser, more tightly-bound core of one of the progenitor galaxies of a major merger should also exhibit distinct populations due to the different star-formation histories of the two galaxies (which gives rise to the necessary density difference). In general, however, little evidence has been found to suggest that KDCs have particularly distinct star-formation histories from their host galaxy \citep[e.g.][but see \citealt{bender92,brown00}]{forbes96,carollo97a,carollo97b,davies01,brodie05}. This apparent lack in contrast between the stellar populations of the KDC and its host galaxy suggest that either the assembly of these objects is rather carefully arranged, or that they formed at early epochs, and the observable differences between the populations have been diluted by time, masking their distinct origins.

From our high spatial resolution observations, we find a number of compact KDCs which {\em do} show clear evidence of distinct, young stellar populations. Extending this connection between the stellar kinematics and populations for all KDCs in the \sauron\/ survey using both \oasis\/ and \sauron\/ data, we find two types of KDC: large, kiloparsec-scale KDCs which have old stellar populations, and are found in slowly-rotating systems; and very compact KDCs on 100~pc scales, which are generally very young, are found in aligned, fast-rotating objects, and are close to counter-rotating around the same axis as the outer parts..

This intuitively fits with the picture that fast-rotating galaxies (including possible sub-components) tend to form through dissipative processes, involving a significant amount of gas. There is a sizeable supply of star-forming material for these objects, provided it can reach a high enough density to induce star-formation. Such material will tend to fall to the centre of these objects, losing angular momentum via hydrodynamical processes. Reaching the critical density for star-formation is most likely at the centre of the potential, where the material is collected and compressed. This naturally results in compact central star formation. How efficiently gas can reach this region may be key to determining the degree of star-formation that can occur. In this respect, dynamical perturbations allow more efficient transfer of angular momentum, resulting in a more rapid build-up of material at the galaxy centre and consequent star-formation. In the presence of such a perturbation (e.g. a bar), the orientation of the angular momentum of the gas may be significantly altered, which may explain the presence of the stellar KDCs having formed from this material. The fact that the KDCs are rather well-aligned with the host galaxy supports the interpretation that dissipational processes play an important role in the formation of these components.

Slowly-rotating galaxies, on the other hand, are thought to form through predominantly dissipationless processes \citep{binney78,defis83,bbf92}, resulting in near-spheroidal systems, often with some degree of triaxiality. Since these objects have little star-forming material, they are less likely to contain young stellar populations. In addition, in a dynamically hot collisionless system, kinematic subcomponents (either acquired via interactions, or `grown' from accreted material at early epochs) can be dynamically stable over the lifetime of the object, provided they are massive enough to survive subsequent interactions. Given that there are no noticeable differences in the stellar age of the large KDCs and their host galaxies, it is likely that these components were formed early in these galaxies' lifetimes. Clues as to the formation scenario of these subcomponents may come from investigating their dynamical structure, using dynamical models.

\section{Conclusions}
\label{sec:conclusions}

We have presented high spatial resolution integral field spectroscopy, using the \oasis\/ spectrograph, of a subset of 28 elliptical and lenticular galaxy centres selected from the \sauron\/ survey. From these observations, we derive the distribution and kinematics of stars and gas, as well as stellar absorption-line indices independently calibrated to the Lick system. This data set has a factor four higher spatial sampling and almost a factor two increase in median spatial resolution to that of \sauron. We compare the results of our analysis to that of \sauron\/ spectra within equivalent integrated aperture spectra. The measured parameters, including velocity dispersion, emission line fluxes and stellar line-strength indices, are all consistent within the uncertainties, verifying our analysis.

Within the central regions probed by \oasis, there is a rich variety of structure in the observed parameters, comparable to that observed on larger scales with \sauron. In particular, we find a range of stellar dynamical substructure within the central kiloparsec of these objects, including central disks and counter-rotating cores, which we quantify using the technique of kinemetry. The ionised gas is similarly varied, being often dynamically decoupled from the stellar component, and varying flux ratios indicating a range of ionisation mechanisms, even within the same galaxy. We analyse the stellar populations by comparing the observed line-strength maps to model predictions, giving maps of luminosity-weighted stellar age, metallicity and abundance ratio. These maps reveal that the galaxies with the youngest global populations in our sample also have even younger centrally concentrated populations. These young nuclei may be explained by young nuclear star clusters or other recent circum-nuclear star formation as in H {\small II} nuclei.\looseness-2

The young nuclei show a propensity for harbouring compact kinematically decoupled stellar components, which can be associated directly with the young stars. Considering all galaxies in the \sauron\/ survey that show a clear kinematically decoupled stellar component, we explore the relationship between central stellar population and the size of the component. We find that, although the `classical' decoupled components show rather old stellar populations, there also exists a population of compact ($\lsim 100$~pc) decoupled components which show a range of ages, from 15~Gyr to $\lsim 0.5$~Gyr. Given the large difference in intrinsic size, we conclude that the two types of decoupled component are not directly related in the evolutionary sense, and that the compact variety are not necessarily an important feature {\em by mass}, although the young stars within these decoupled components are largely responsible for the significant spread in integrated luminosity-weighted ages observed for these galaxies.

Our data suggest a link between counter-rotation and enhanced star-formation. By consideration of a simple model, we conclude that older KDCs of similar mass fractions are rather difficult to detect, and so there may be a sizeable number of low-mass counter-rotating components that we are missing. Likewise, co-rotating components of the same mass fraction leave a more subtle signature on the observed stellar kinematics, and so can only be detected when the KDC population is rather young. Our sample is therefore too small to reliably put constraints on the importance of counter-rotation to recent star formation.

These are the first observations in an ongoing follow-up of \sauron\/ survey galaxies using high-spatial resolution integral field spectroscopy to study the central regions of these objects in detail. The empirical results presented in this paper are a first step towards understanding the intrinsic structure of the central few hundred parsecs of these objects, and already reveal new insights. Future application of dynamical models, using both \sauron\/ and \oasis\/ data will substantiate this further.

\section*{Acknowledgments}
The \sauron\/ team gratefully acknowledges the contribution to the \oasis\/ CFHT observing campaign by Pierre Martin, Pierre Ferruit, Yannick Copin, Fabien Wernli and Arlette P\'econtal. We thank Glenn van de Ven for useful discussion during the preparation of this manuscript. The \sauron\ project is made possible through grants 614.13.003, 781.74.203, 614.000.301 and 614.031.015 from the Netherlands Organization for Scientific Research (NWO) and financial contributions from the Institut National des Sciences de l'Univers, the Universit\'e Claude Bernard Lyon~I, the Universities of Durham, Leiden, and Oxford, the British Council, PPARC grant `Extragalactic Astronomy \& Cosmology at Durham 1998--2002', and the Netherlands Research School for Astronomy NOVA. RLD is grateful for the award of a PPARC Senior Fellowship (PPA/Y/S/1999/00854) and postdoctoral support through PPARC grant PPA/G/S/2000/00729. We are grateful to the Lorentz center at Leiden University for generous hospitality. MC acknowledges support from a VENI grant 639.041.203 awarded by the Netherlands Organisation for Scientific Research (NWO). JFB acknowledges support from the Euro3D Research Training Network, funded by the EC under contract HPRN-CT-2002-00305. This project made use of the HyperLeda and NED databases. Part of this work is based on data obtained from the ESO/ST-ECF Science Archive Facility.\looseness-2



\appendix

\section[]{Major-axis long-slit profile of NGC\,3489}
\label{appendix:a}

\begin{figure*}
 \begin{center}
  \includegraphics[width=16cm]{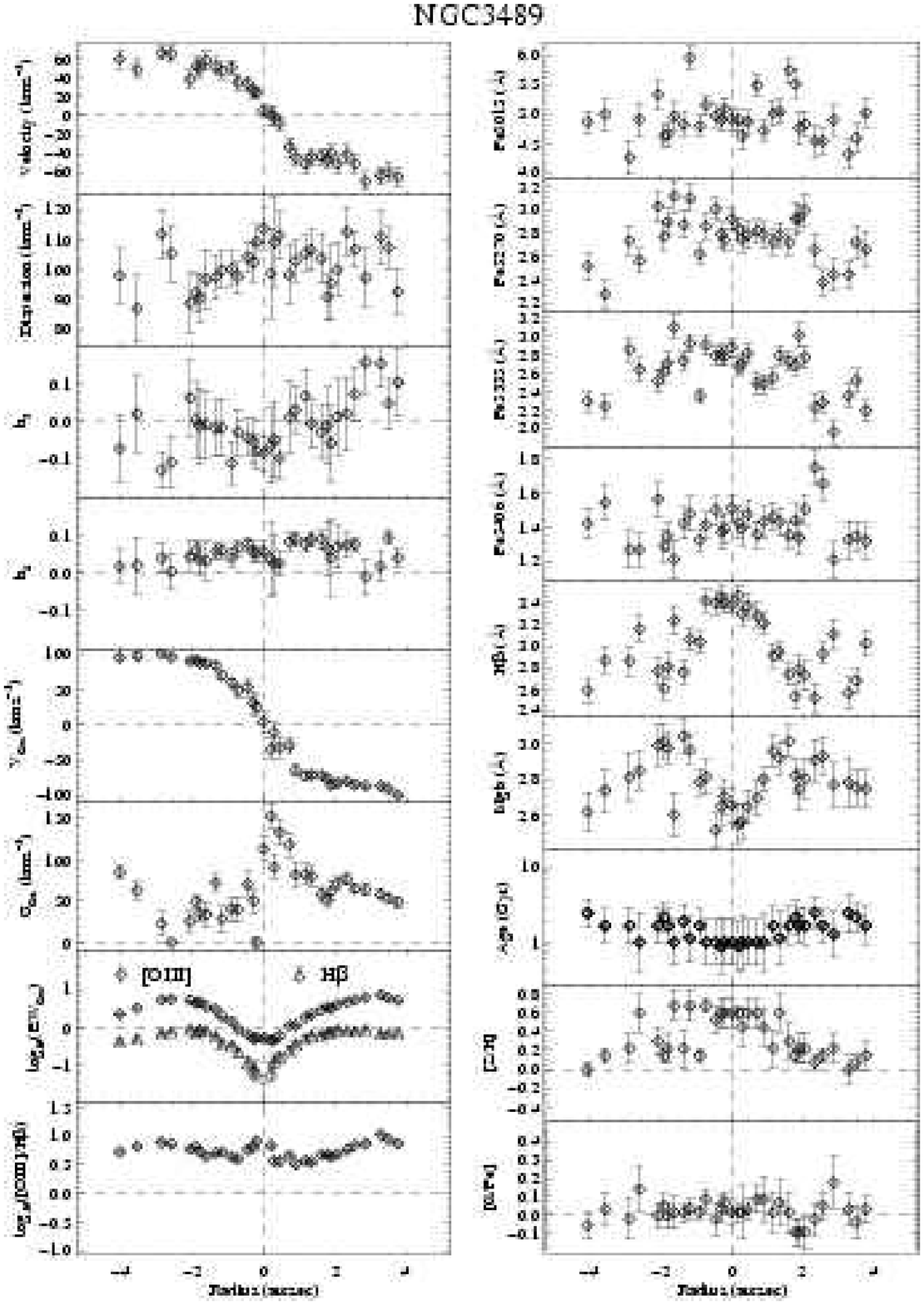}
 \end{center}
 \caption[]{Simulated major-axis slit aperture 0\farcsec2 wide for NGC\,3489. Each of the parameters derived from the two-dimensional maps are shown, with measurement errors, to illustrate the typical quality of our data and how the uncertainties relate to structures in the parameters.}
 \label{fig:ngc3489_profile}
\end{figure*}

Figure \ref{fig:ngc3489_profile} presents profiles of the measured stellar kinematics, gas properties, line-strength indices and stellar population parameters for NGC\,3489, extracted along a major-axis `slit' aperture 0\farcsec2 wide, collecting the value from each Voronoi bin whose centre falls inside. This illustrates the typical measurement error associated with each bin in the maps presented in Fig. \ref{fig:ngc1023}.

\section[]{Kinemetric profiles of stellar velocity}
\label{appendix:b}
\begin{figure*}
 \begin{center}
  \includegraphics[width=16cm,angle=0]{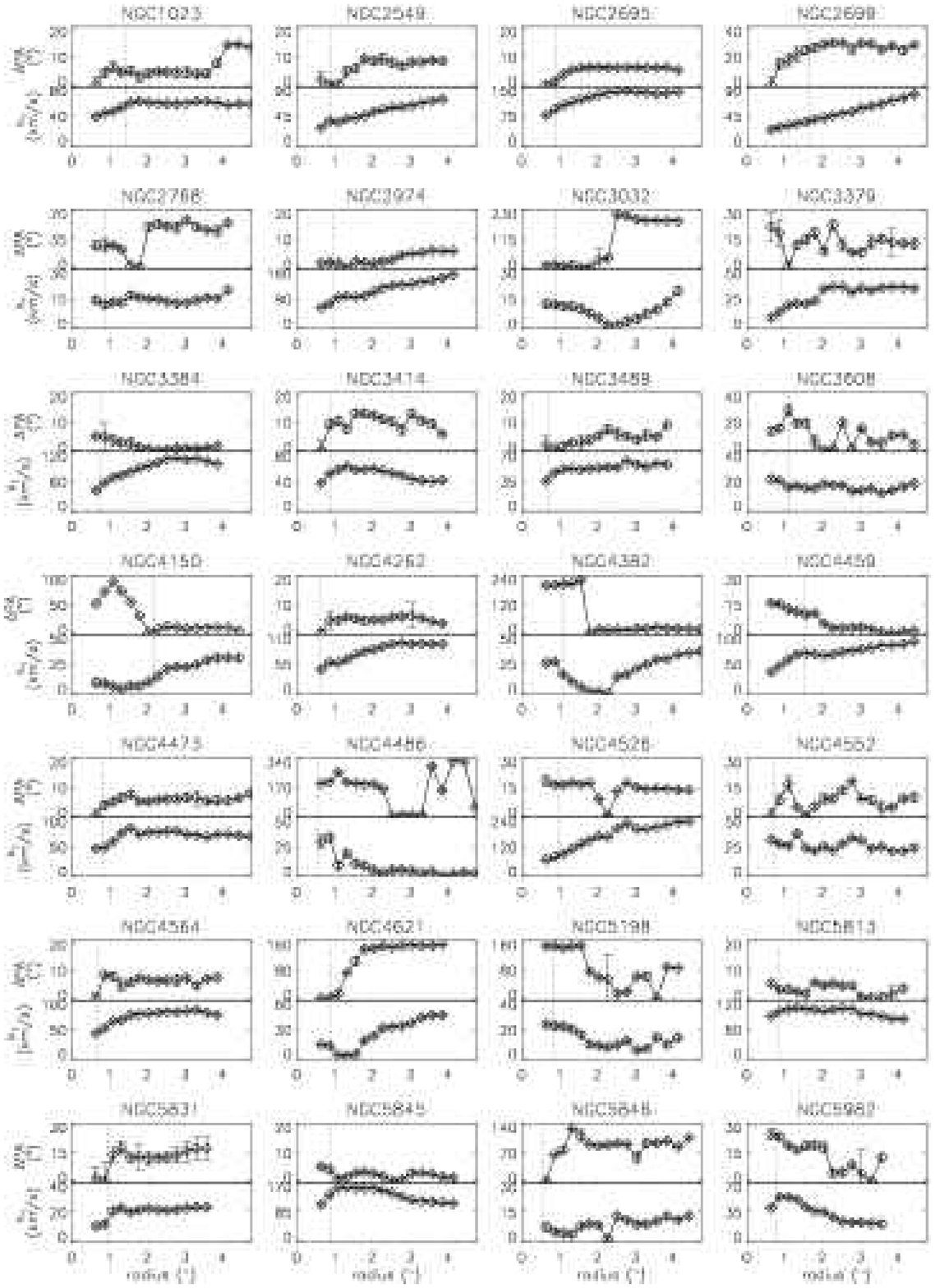}
 \end{center}
 \caption[]{Radial profiles of kinematic position angle ($\Delta$PA) and peak absolute rotation (\ki), measured using the kinemetry method of \citet{krajnovic06}.}
 \label{fig:all_kinemetry_profiles}
\end{figure*}

Fig.~\ref{fig:all_kinemetry_profiles} presents profiles of kinematic position angles and peak absolute rotation values for all 28 galaxies, measured using the kinemetry method of \citet{krajnovic06}. See Section \ref{sec:kinemetry} for details.

\section[]{Comparison of \oasis\/ and \sauron\/ spectra}
\label{appendix:c}

\begin{figure*}
 \begin{center}
  \includegraphics[width=12cm,angle=90]{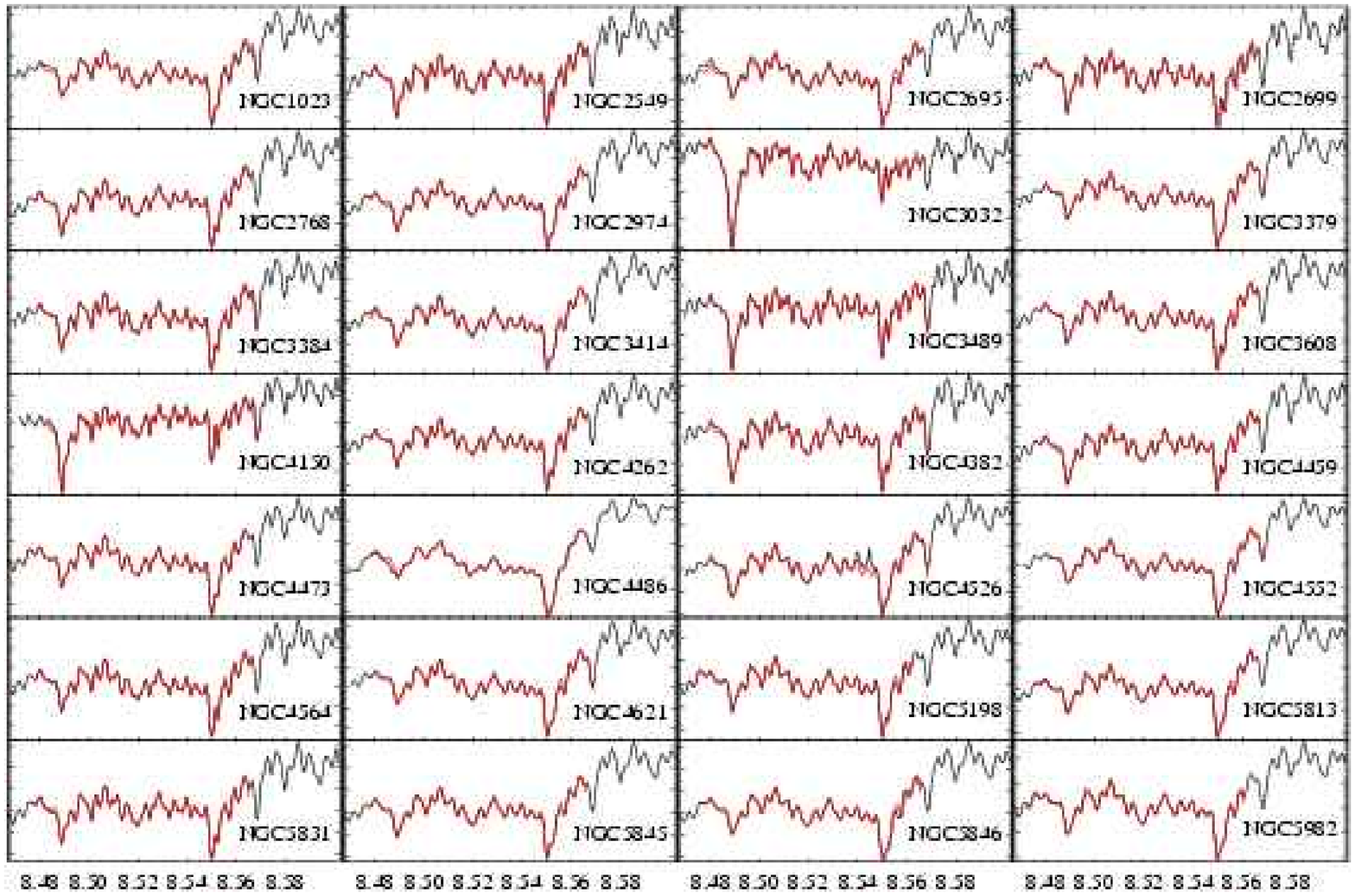}
 \end{center}
 \caption[]{Comparison of \oasis\/ (black) and \sauron\/ (red) spectra, plotted in $\ln\lambda$. The spectra from both instruments have been integrated within the same 4\arcsec\/ circular aperture centred on the galaxy. The spectra have been 
 de-redshifted, and normalized around 5100~\AA\/ for the purpose of this comparison. The \sauron\/ spectra have been broadened and resampled to match the \oasis\/ system. Both data sets have been independently corrected for emission line contamination. These spectra are used for the comparisons presented in Section \ref{sec:analysis}. The \sauron\/ spectra have been truncated to a wavelength range of 4825-5280~\AA, to ensure a common wavelength coverage within the aperture for all galaxies, without encountering gaps in the field coverage due to the filter cut-off \citep[see][]{kuntschner06}.}
 \label{fig:spectra_comp}
\end{figure*}

Fig.~\ref{fig:spectra_comp} presents fully calibrated \oasis\/ spectra (black lines) integrated within a 4\arcsec\/ circular aperture centred on the galaxy nucleus for all the objects in our sample. Overplotted in red are equivalent spectra from \sauron\/ integrated within the same 4\arcsec\/ aperture. Both sets of spectra are de-redshifted, and normalized around 5100~\AA. The \sauron\/ spectra have been broadened and resampled to match the \oasis\/ spectral resolution and sampling. Both sets of spectra have been `cleaned' of gas, by subtracting an emission line model, as described in Section \ref{sec:gas}.

\section[]{Comparison of \oasis\/ and \sauron\/ equivalent width limits}
\label{appendix:d}

\begin{figure*}
 \begin{center}
  \includegraphics[width=16cm,angle=0]{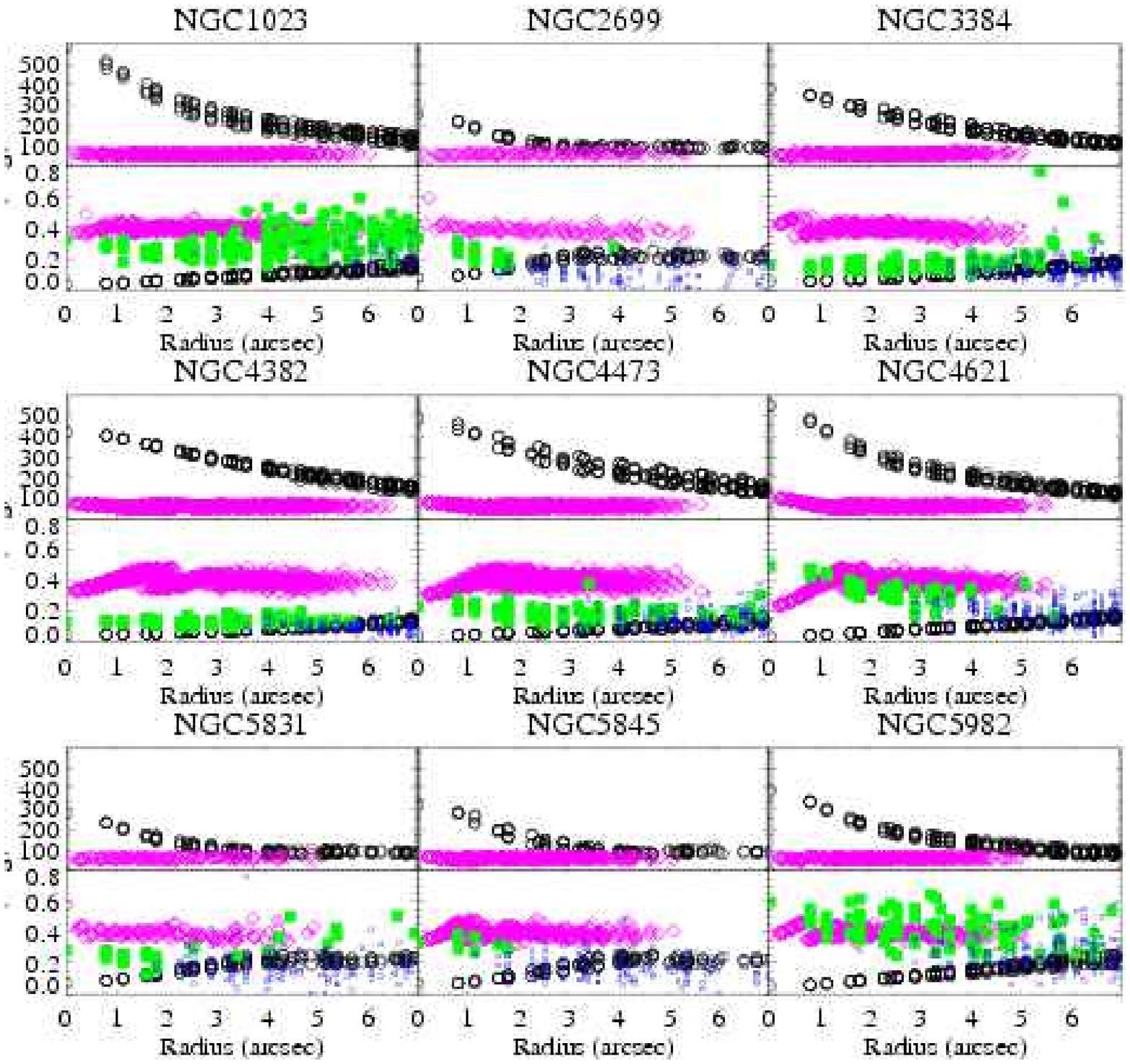}
 \end{center}
 \caption[]{{\it Top panels}: Comparison of $S/N$ as a function of circular radius for \oasis\/ (diamonds) and \sauron\/ (circles). {\it Bottom panels}: Corresponding equivalent width detection limits derived using eq.(1) for both data sets (same symbols as top panels). Overplotted with large filled (detections) and small open symbols (non-detections) are the \oiii\/ equivalent width values measured with \sauron\/ (from Paper V). This shows that the very high $S/N$ of the \sauron\/ data results in significantly lower detection limits than that obtained with \oasis\/ at higher spatial sampling. This explains why some objects have clearly detected gas with \sauron\/ but not with \oasis.}
 \label{fig:ew_limits_comp}
\end{figure*}

Fig.~\ref{fig:ew_limits_comp} illustrates the difference in equivalent width limits between the \sauron\/ data of Paper V and our \oasis\/ data, for the cases where \sauron\/ detects gas but \oasis\/ does not, or where the gas detection with \sauron\/ is stronger. For each galaxy, the upper panels show the estimated $S/N$ determined from the error spectrum propagated through the reduction process of both data sets (circles = \sauron, diamonds = \oasis). This estimator is effectively an upper limit, as correlation effects (e.g. seeing PSF) can be difficult to model accurately, resulting in slightly underestimated noise estimates, at the level of $\sim 5$\%.

The lower panels show how these $S/N$ values (same symbols) translate into equivalent width limits, using eq. (1), assuming an intrinsic emission-line velocity dispersion of 50~\kms. Overplotted on these limits are the measured emission-line equivalent widths of \oiii\/ from the \sauron\/ data, where filled square symbols indicate significant detections (i.e. above the $A/N$ threshold), and open square symbols are non-detections.

This clearly shows that the poor or non-detection of emission with \oasis\/ which is observed with \sauron\/ arises simply from the very different $S/N$ of the two data sets. The lower $S/N$ of the \oasis\/ data comes mainly from the much smaller spatial sampling on the sky, as well as generally shorter total integration times, smaller telescope aperture, and small difference in throughput of the two instruments.

\bsp

\label{lastpage}

\end{document}